 \documentclass[preprint,11pt,a4paper]{elsarticle}

\usepackage{graphicx}
\usepackage{amssymb}
\usepackage{amsmath}
\usepackage{lineno}

\biboptions{compress}

\usepackage[cm]{fullpage}
\usepackage[title]{appendix}
\usepackage{cases}
\usepackage{siunitx}
\usepackage{doi}
\usepackage{hyperref}
\hypersetup{
    colorlinks,
    linkcolor={red!50!black},
    citecolor={blue!50!black},
    urlcolor={blue!80!black}
}
\usepackage[font=small,labelfont=bf]{caption}
\usepackage[font=footnotesize,labelfont=bf,subrefformat=parens]{subcaption}
\usepackage{xcolor}
\usepackage{booktabs}  % provides commands such as \toprule[<wd>] to replace \hline in tables
\usepackage{multirow}
\usepackage{framed}

\journal{Physics of Fluids}

\begin{document}

\newcommand{\vf}[1]{\underline{#1}}
\newcommand{\tf}[1]{\underline{\underline{#1}}}
\newcommand{\pd}[2]{\frac{\partial #1}{\partial #2}}
\newcommand{\ucd}[1]{\overset{\scriptscriptstyle \triangledown}{\tf{#1}}}
% Dimensionless numbers
\newcommand\Rey{\operatorname{\mathit{Re}}}
\newcommand\Str{\operatorname{\mathit{Sr}}}
\newcommand\Wei{\operatorname{\mathit{Wi}}}
\newcommand\Deb{\operatorname{\mathit{De}}}

\begin{frontmatter}

%% Title, authors and addresses

%% use the tnoteref command within \title for footnotes;
%% use the tnotetext command for the associated footnote;
%% use the fnref command within \author or \address for footnotes;
%% use the fntext command for the associated footnote;
%% use the corref command within \author for corresponding author footnotes;
%% use the cortext command for the associated footnote;
%% use the ead command for the email address,
%% and the form \ead[url] for the home page:
%%
%% \title{Title\tnoteref{label1}}
%% \tnotetext[label1]{}
%% \author{Name\corref{cor1}\fnref{label2}}
%% \ead{email address}
%% \ead[url]{home page}
%% \fntext[label2]{}
%% \cortext[cor1]{}
%% \address{Address\fnref{label3}}
%% \fntext[label3]{}

%\title{Finite volume solution of transient Bingham flows}
\title{Theoretical study of the flow in a fluid damper containing high viscosity silicone oil: 
effects of shear-thinning and viscoelasticity}

%% use optional labels to link authors explicitly to addresses:
%% \author[label1,label2]{<author name>}
%% \address[label1]{<address>}
%% \address[label2]{<address>}

\author[up]{Alexandros Syrakos\corref{cor1}}
\ead{syrakos@upatras.gr, alexandros.syrakos@gmail.com}

\author[up]{Yannis Dimakopoulos}
\ead{dimako@chemeng.upatras.gr}

\author[up]{John Tsamopoulos}
\ead{tsamo@chemeng.upatras.gr}

\cortext[cor1]{Corresponding author}

\address[up]{Laboratory of Fluid Mechanics and Rheology, Department of Chemical Engineering, 
University of Patras, 26500 Patras, Greece}

\begin{abstract}
%% Text of abstract
The flow inside a fluid damper where a piston reciprocates sinusoidally inside an outer casing 
containing high-viscosity silicone oil is simulated using a Finite Volume method, at various 
excitation frequencies. The oil is modelled by the Carreau-Yasuda (CY) and Phan-Thien \& Tanner 
(PTT) constitutive equations. Both models account for shear-thinning but only the PTT model accounts 
for elasticity. The CY and other generalised Newtonian models have been previously used in 
theoretical studies of fluid dampers, but the present study is the first to perform full 
two-dimensional (axisymmetric) simulations employing a viscoelastic constitutive equation. It is 
found that the CY and PTT predictions are similar when the excitation frequency is low, but at 
medium and higher frequencies the CY model fails to describe important phenomena that are predicted 
by the PTT model and observed in experimental studies found in the literature, such as the 
hysteresis of the force-displacement and force-velocity loops. Elastic effects are quantified by 
applying a decomposition of the damper force into elastic and viscous components, inspired from LAOS 
(Large Amplitude Oscillatory Shear) theory. The CY model also overestimates the damper force 
relative to the PTT, because it underpredicts the flow development length inside the piston-cylinder 
gap. It is thus concluded that (a) fluid elasticity must be accounted for and (b) theoretical 
approaches that rely on the assumption of one-dimensional flow in the piston-cylinder gap are of 
limited accuracy, even if they account for fluid viscoelasticity. The consequences of using 
lower-viscosity silicone oil are also briefly examined.
\end{abstract}

% \begin{keyword}
% %% keywords here, in the form: keyword \sep keyword
% %% MSC codes here, in the form: \MSC code \sep code
% %% or \MSC[2008] code \sep code (2000 is the default)
%  fluid damper \sep viscoelasticity \sep Phan-Thien \& Tanner fluid \sep Carreau -- Yasuda fluid 
% \sep hysteresis \sep finite volume method
% \end{keyword}

\end{frontmatter}

\begin{framed} \centering
\noindent This is the accepted version of the article published in:

Physics of Fluids 30, 030708 (2018); \doi{10.1063/1.5011755}
\end{framed}

%%
%% Start line numbering here if you want
%%
%% \linenumbers
% \begin{linenumbers}

%% main text

\section{Introduction}
\label{sec: introduction}

Viscous dampers are devices that dissipate mechanical energy into heat through the action of 
viscous stresses in a fluid; such stresses develop because the fluid is forced to flow though a 
constriction. A common design involves the motion of a piston in a cylinder filled with a fluid, 
such that large velocity gradients develop in the narrow gap between the piston head and the 
cylinder, resulting in viscous and pressure forces that resist the piston motion. Viscous dampers 
are used in a range of applications, including the protection of buildings and bridges from damage 
induced by seismic and wind excitations \cite{Constantinou_1993, Konstantinidis_2014}, and the 
absorption of shock and vibration energy in vehicles \cite{Yao_2002, Nguyen_2009b} and aerospace 
structures \cite{Rittweger_2002}.

A common mathematical model used to describe the behaviour of fluid dampers assumes that the 
instantaneous damper force $F$ depends only on the instantaneous piston velocity $U$ according to
%^b
\begin{equation} \label{eq: viscous damper}
 F \;=\; C \, U^n
\end{equation}
%^a
where $C$ is the damping constant and $n$ is an exponent usually in the range $0.3$ to $2$ 
\cite{Seleemah_1997}. Dampers that conform to the above relationship are characterised as purely 
viscous dampers. Also, they are classified as linear if $n = 1$ or nonlinear if $n \neq 1$. 
Although Eq.\ \eqref{eq: viscous damper} often describes well the behaviour of fluid dampers in a 
range of operating conditions, dampers in many cases also exhibit a spring-like behaviour, 
especially at high frequencies, where the damper force has a component that is in phase with the 
piston displacement. For linear dampers, a simplifying analysis \cite{Constantinou_1993, 
Seleemah_1997} shows that if a sinusoidal piston displacement $X = X_0 \sin(\omega t)$ is forced 
upon the damper and the resulting force is at a phase lag with the piston velocity then this force 
can be expressed as
%^b
\begin{equation} \label{eq: linear viscoelastic damper}
 F  \;=\;  C \, \dot{X}  \;+\; K \, X
\end{equation}
%^a
where the dot denotes differentiation with respect to time ($\dot{X} \equiv dX/dt \equiv U$) and 
$K$ is called the storage stiffness. Equation \eqref{eq: linear viscoelastic damper} essentially 
assumes that the actual damper can be modelled as an ideal linear purely viscous damper connected 
in parallel with an ideal spring. Alternatively, one can assume these two components to be 
connected in series, which results in the macroscopic Maxwell model \cite{Constantinou_1993}:
%^b
\begin{equation} \label{eq: macroscopic Maxwell}
 F \;+\; \lambda \dot{F} \;=\; C \, \dot{X}
\end{equation}
%^a
where $\lambda \equiv C/K$ is called the relaxation time of the model. Macroscopic models of the 
damper behaviour such as \eqref{eq: viscous damper}, \eqref{eq: linear viscoelastic damper} and 
\eqref{eq: macroscopic Maxwell} are needed for the prediction of the dynamic response to 
excitations of structures that incorporate dampers. These and many more such models that have been 
proposed over time (see e.g.\ \cite{Makris_1991, Wang_2011, Lu_2012} and references therein) are 
mostly phenomenological, i.e.\ they are not derived from first principles but are empirical 
relationships. Each model incorporates a number of parameters such as $C$ and $\lambda$ whose 
values have to be obtained by fitting the model to experimental data for each particular damper.

The mechanical behaviour of a fluid damper depends on several factors (damper design and 
dimensions, fluid properties, operating conditions) and to obtain insight into the role of each of 
these one has to consider the damper operation from first principles. Energy dissipation occurs as 
the mechanical excitation causes parts of the damper to move relative to each other, forcing fluid 
through an orifice. Therefore the equations governing fluid flow must be considered. The flow is in 
general multidimensional and time-dependent, but in several studies simplifications have been made 
to make it amenable to solution by analytical means or by simple numerical methods. The 
simplifications commonly made are: (a) that the only important region is that of the orifice; (b) 
that the orifice is narrow and long so that the flow there can be assumed one-dimensional; and (c) 
that the flow is quasi-steady, i.e.\ the effect of the time derivatives in the governing equations 
can be neglected. These reduce the problem to planar or annular Couette-Poiseuille flow that can be 
solved to obtain the wall shear stress and pressure gradient, and hence the viscous and pressure 
force on the piston.

A fluid very commonly employed in dampers, especially those used in civil engineering applications, 
is silicone oil (poly(dimethylsiloxane) (PDMS) \cite{Currie_1950, Longin_1998, Kokuti_2014}). 
Studies concerning silicone oil dampers include \cite{Constantinou_1993, Seleemah_1997, 
Rittweger_2002, Black_2007, Hou_2008, Yun_2008, Jiao_2016}. Silicone oil is a polymeric fluid whose 
long molecules become aligned under shear resulting in a drop of viscosity; this property is called 
shear-thinning, and is more pronounced in silicone oils of high molecular weight. Due to 
shear-thinning, the force with which the damper reacts to excitations is significantly lower than 
what it would be if the fluid viscosity was constant. Therefore, shear-thinning must be accounted 
for in theoretical studies, and a popular constitutive equation that can describe this sort of 
behaviour is the Carreau-Yasuda (CY) \cite{Yasuda_1981} model. The CY model has been used in 
\cite{Jiao_2016} for a theoretical study of the flow in a silicone oil damper under the 
aforementioned simplifying assumptions of one-dimensional flow that render the problem amenable to 
analytical solution. Two dimensional (axisymmetric) simulations of the whole flow have also been 
performed, assuming silicone oil to behave as either a Newtonian \cite{Hou_2008} or a CY fluid 
\cite{Sun_2014}, using commercial Computational Fluid Dynamics (CFD) solvers.

Silicone oil, like most polymeric fluids, possesses another property, namely elasticity, which is a 
tendency to partially recover from its deformation. The CY constitutive equation, being of 
generalised Newtonian type, relates the stress tensor only to the instantaneous rate of strain 
tensor and hence cannot describe elastic effects, which depend on the flow history. Constitutive 
equations that predict elastic behaviour are more difficult to solve than generalised Newtonian 
ones and hence theoretical analyses usually assume elasticity to be negligible and generalised 
Newtonian models such as CY to be sufficient. Yet, the CY model fails to predict the stiffness 
aspect of the damper's behaviour mentioned above, and it is natural to wonder whether this stiffness 
is due to the elasticity of silicone oil. This has been investigated in a couple of theoretical 
studies employing the one-dimensional flow simplifying assumption \cite{Makris_1996, Hou_2008} which 
suggest that indeed fluid elasticity is a cause of damper stiffness, although this has been disputed 
in favour of fluid compressibility as the cause \cite{Wang_2007, Jiao_2016}. A better picture would 
be obtained by performing simulations of the whole flow using a viscoelastic constitutive equation, 
and to the best of the authors' knowledge the present study is the first to attempt this. A possible 
reason for the lack of such prior studies is that viscoelastic simulations are much more expensive 
than those involving generalised Newtonian fluids such as the CY, as additional differential 
equations have to be solved (one per stress component), which are unwieldy at high elasticity (the 
notorious high-Weissenberg number problem -- see e.g.\ \cite{Fattal_2005}), while also much smaller 
time steps are needed in order to capture the elastic phenomena (whereas if the flow is assumed 
generalised Newtonian, it is also quasi-steady due to the high viscosity of typical damper fluids 
and one can employ large time steps without loss of accuracy -- see \cite{Syrakos_2016}). 

In the present paper we present flow simulations in a fluid damper containing high-viscosity 
silicone oils, modelled as single-mode linear Phan-Thien \& Tanner \cite{Phan_Thien_1977} (PTT)
viscoelastic liquids. The simulations are performed with an in-house Finite Volume solver. The PTT 
constitutive equation accounts for both shear-thinning and elasticity, and its parameters were 
selected so as to represent the rheological behaviour of high-viscosity silicone oils as recorded 
in the literature \cite{Currie_1950, Longin_1998, Kokuti_2014}). The viscoelastic simulations are 
complemented by simulations using the Carreau-Yasuda model which only accounts for shear-thinning 
but not elasticity, so as to investigate whether this inexpensive and easy to implement alternative 
suffices to produce realistic results, as has been assumed. The present results show that the PTT 
and CY models produce almost identical results at low oscillation frequencies, but at medium and 
high frequencies the PTT model does predict significant damper stiffness behaviour whereas the CY 
model does not. Furthermore, the PTT model, contrary to the CY, predicts significant variation of 
the fluid velocity along the piston-cylinder gap, i.e.\ that the flow in the gap is 
two-dimensional. This means that the theoretical approach based on the one-dimensional viscoelastic 
flow assumption, employed in \cite{Makris_1996, Hou_2008}, is of limited accuracy. The study 
concludes with a brief investigation of flow with silicone oil of lower viscosity, to expose the 
main differences with the high-viscosity case.

Before ending this introductory section, it is worth mentioning that damper stiffness is observed 
experimentally also in other types of fluid damper, including magnetorheological (MR) 
\cite{Snyder_2001, Yao_2002, Wang_2007, Wang_2011, Parlak_2012} and extrusion dampers 
\cite{Rodgers_2007}. The experimental hysteresis loops obtained for these kinds of dampers are also 
qualitatively very similar to the present PTT results (but not the CY results), indicating that 
fluid elasticity is an important factor for these dampers as well. The fluids employed in MR and 
extrusion dampers additionally possess a yield stress, i.e.\ a limiting value of stress below which 
they do not flow but behave as solids; theoretical studies have thus far concentrated on the effect 
of yield stress, employing the Bingham \cite{Wereley_1998, Nguyen_2009, Parlak_2012, Syrakos_2016} 
or Herschel-Bulkley \cite{Wang_2007} constitutive models to describe the viscoplasticity of these 
fluids, but neglecting the elasticity. These models can explain the non-linear force-velocity 
relationship such as that expressed by Eq.\ \eqref{eq: viscous damper}, but not the damper stiffness 
(the spring-like property) such as expressed by Eqs.\ \eqref{eq: linear viscoelastic damper} and 
\eqref{eq: macroscopic Maxwell}. The present results suggest that elasticity must be incorporated 
into these viscoplastic constitutive equations in order to obtain more realistic predictions.

\section{Problem definition and governing equations}
\label{sec: definition}

\begin{figure}[t]
    \centering
    \begin{subfigure}[b]{0.47\textwidth}
        \centering
        \includegraphics[width=0.95\linewidth]{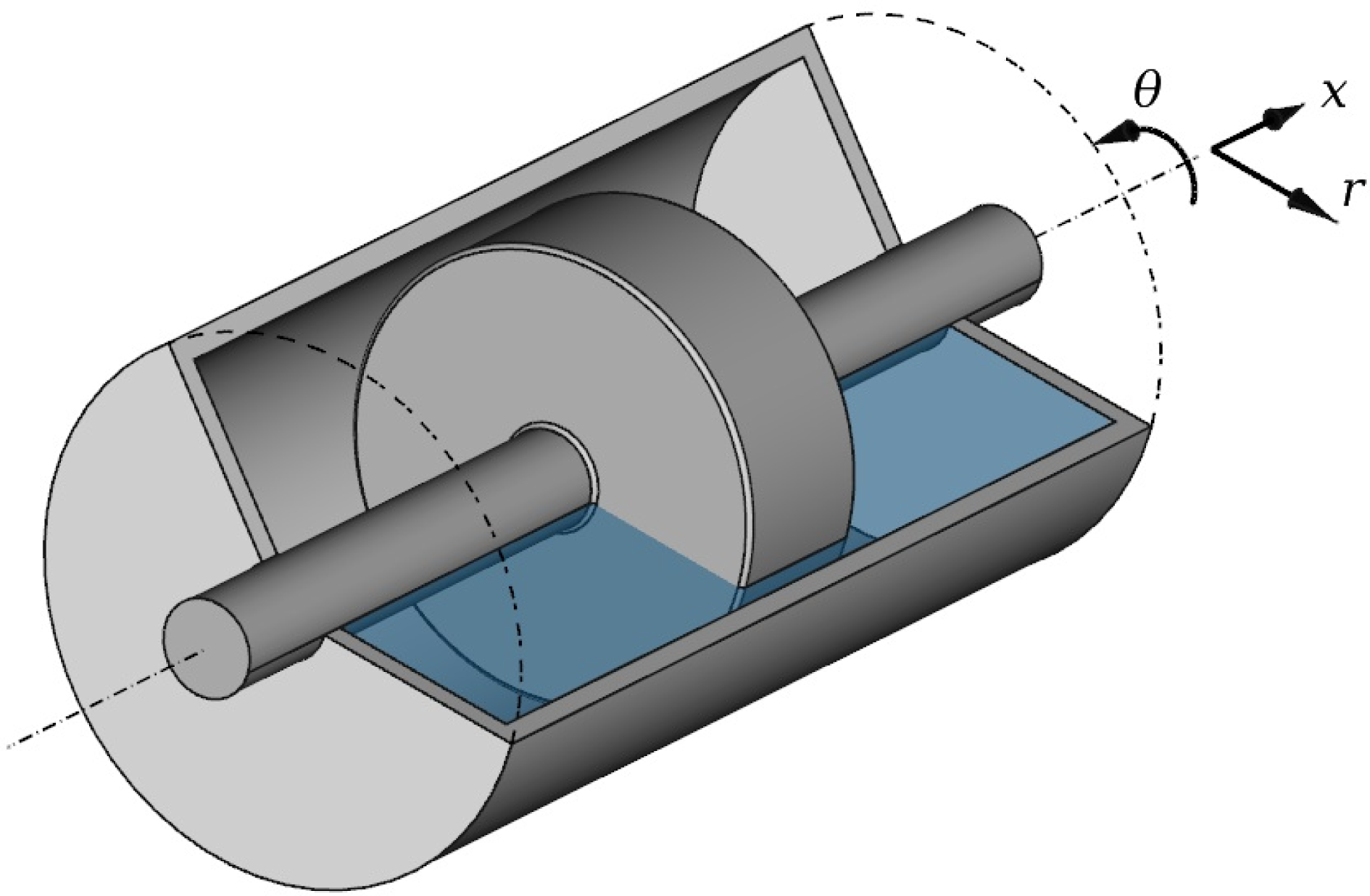}
        \caption{3-D model}
        \label{sfig: geometry 3D}
    \end{subfigure}
    \begin{subfigure}[b]{0.52\textwidth}
        \centering
        \includegraphics[width=\linewidth]{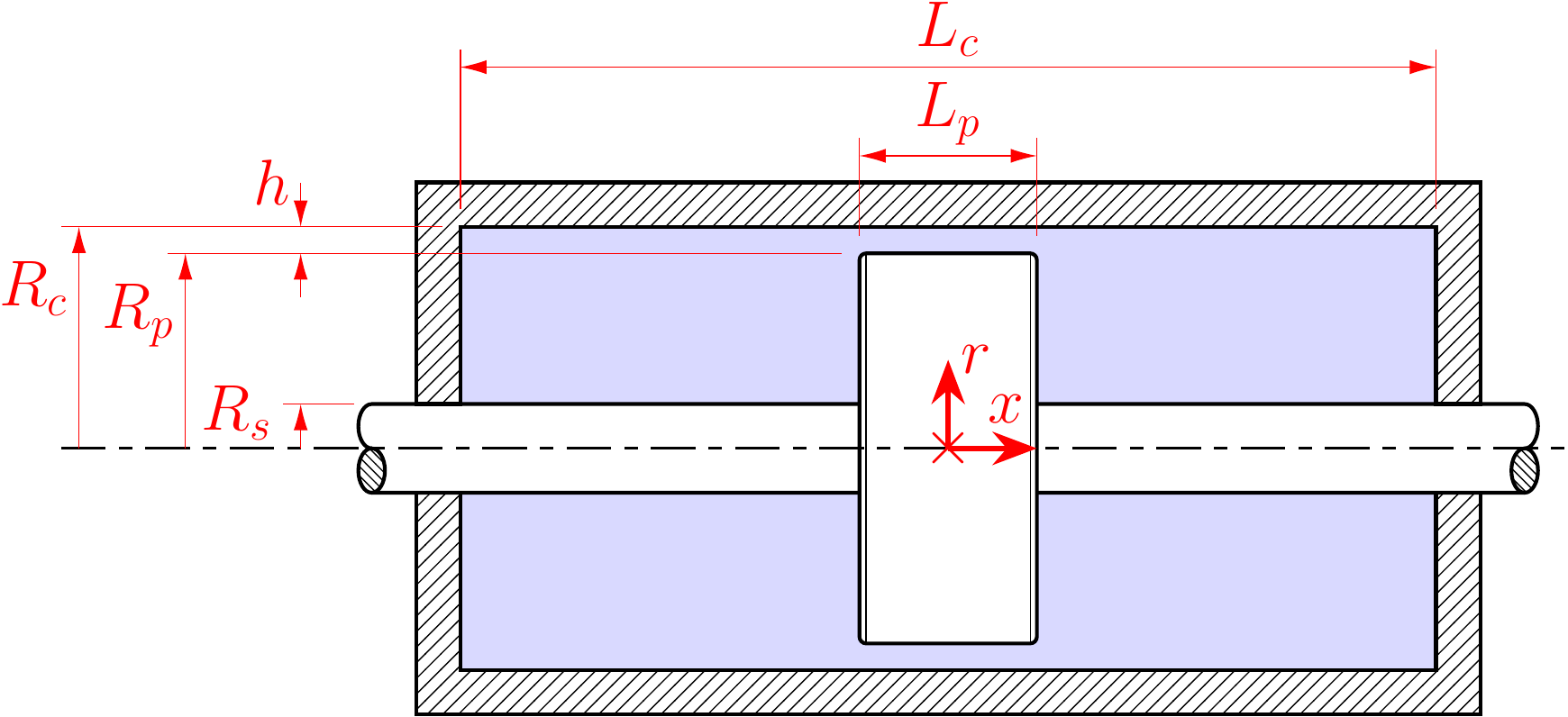}
        \caption{2-D section}
        \label{sfig: geometry 2D} 
    \end{subfigure}
    \caption{Layout of the damper. In \subref{sfig: geometry 3D} part of the cylinder is removed to 
reveal the shaft and piston. The flow is axisymmetric and can be solved on a single plane $\theta = 
0$ (coloured in \subref{sfig: geometry 3D}).}
  \label{fig: geometry}
\end{figure}

The geometry of the damper under consideration is shown in Fig.\ \ref{fig: geometry}. It is of a 
simple and common design, consisting of a cylindrical piston of radius $R_p$ and length $L_p$ that 
reciprocates in a cylinder of radius $R_c$, forcing the fluid to flow through the annular gap of 
width $h = R_c - R_p$. The piston is fixed on a shaft of radius $R_s$ that extends on both sides of 
the piston so that, as the shaft and piston oscillate, the fluid volume remains constant and 
compressibility effects are minimised; in fact in the present study the fluid is assumed to be 
incompressible in order to isolate the effects of the fluid viscoelasticity from other possible 
sources of damper stiffness such as fluid compressibility. The dimensions of the damper are listed 
in Table \ref{table: damper dimensions}; they are in the range of sizes of silicone oil dampers used 
in experimental studies such as  \cite{Hou_2008, Sun_2014}. In the literature the relative gap width 
$h/R_c$ varies from small values for low viscosity oils (e.g.\ $0.022$ for $1$ \si{Pa.s} oil in 
\cite{Hou_2008}) to large values for high viscosity oils (e.g.\ $0.412$ for $630$ \si{Pa.s} oil in 
\cite{Jiao_2016}), since the maximum damper force increases by increasing the fluid viscosity but 
decreases by increasing the gap width. In the present work the chosen model fluids have nominal 
viscosities of 100 and 500 \si{Pa.s} and the relative gap width is set to $h/R_c = 0.120$.

\begin{table}[b]
\caption{Dimensions of the damper, in millimetres (see Fig.\ \ref{sfig: geometry 2D}).}
\label{table: damper dimensions}
\begin{center}
\begin{small}   % to make the font smaller
\renewcommand\arraystretch{1.25}   % Adjust row height (default=1)
\begin{tabular}{ l | r }
\toprule
 Cylinder radius, $R_c$      &  25 %\si{mm}
\\
 Cylinder length, $L_c$      & 110 %\si{mm}
\\
 Piston radius, $R_p$        &  22 %\si{mm}
\\
 Piston length, $L_p$        &  20 %\si{mm}
\\
 Gap width, $h = R_c - R_p$  &   3 %\si{mm}
\\
 Shaft radius, $R_s$         &   5 %\si{mm}
\\
\bottomrule
\end{tabular}
\end{small}
\end{center}
\end{table}

This geometry is best fitted by a system of cylindrical polar coordinates $(x, r, \theta)$ (Fig.\ 
\ref{sfig: geometry 3D}), with $\vf{e}_x$, $\vf{e}_r$, $\vf{e}_{\theta}$ denoting the unit vectors 
along the coordinate directions. The flow is assumed to be axisymmetric, so that the solution is 
independent of $\theta$ and the problem is reduced to two dimensions. The fluid velocity is denoted 
by $\vf{u}$, and its components are denoted by $u = \vf{u} \cdot \vf{e}_x$ (axial component) and $v 
= \vf{u} \cdot \vf{e}_r$ (radial component). The azimuthal velocity component, $\vf{u} \cdot 
\vf{e}_{\theta}$, is zero. The shaft is set to reciprocate along the axial direction such that the 
$x-$coordinate of the midpoint of the piston changes in time as
%^b
\begin{equation} \label{eq: shaft position}
x_p(t) = \alpha \cos(\omega t)
\end{equation}
%^a
where $x = 0$ is midway along the cylinder (point marked with $\times$ in Fig.\ \ref{sfig: geometry 
2D}), $\alpha$ is the amplitude of oscillation, and $\omega$ is the angular frequency related to 
the frequency $f$ by $\omega = 2\pi f$. The period of oscillation is $T = 1/f = 2\pi/\omega$. For 
civil engineering applications the relevant frequencies are typically less than 4 \si{Hz} 
\cite{Constantinou_1993} although short, stiff structures may have natural frequencies of the order 
of 10 \si{Hz} \cite{Makris_1996}; in other applications, higher frequencies may be relevant. In the 
present work, simulations were performed for frequencies of $f =$ 0.5, 2, 8 and 32 \si{Hz}, while 
the amplitude of oscillation was held fixed at $\alpha =$ 12 \si{mm}. At the start of the 
simulation, at time $t = 0$, the fluid is assumed to be at rest. Once the shaft starts to move, 
after a transient period the flow will reach a periodic state where the flow field at time $t$ will 
be identical to that at time $t-T$. In the present work we are interested in this periodic state, 
and due to viscoelastic effects this is attained faster if the initial piston velocity is zero, 
hence the imposed motion \eqref{eq: shaft position} contains a cosine term rather than a sine term. 
This means that at $t = 0$ the piston is located at its extreme right position. The velocity of the 
shaft and piston, $dx_p/dt$, is therefore
%^b
\begin{equation} \label{eq: piston velocity}
  u_{p}(t) \;=\; -\alpha \omega \sin(\omega t) \;=\; -U_p \sin(\omega t) 
\end{equation}
%^a
where $U_p = \omega \alpha$ is the maximum piston velocity. 

\begin{figure}[thb]
  \centering
  \includegraphics[scale=1.00]{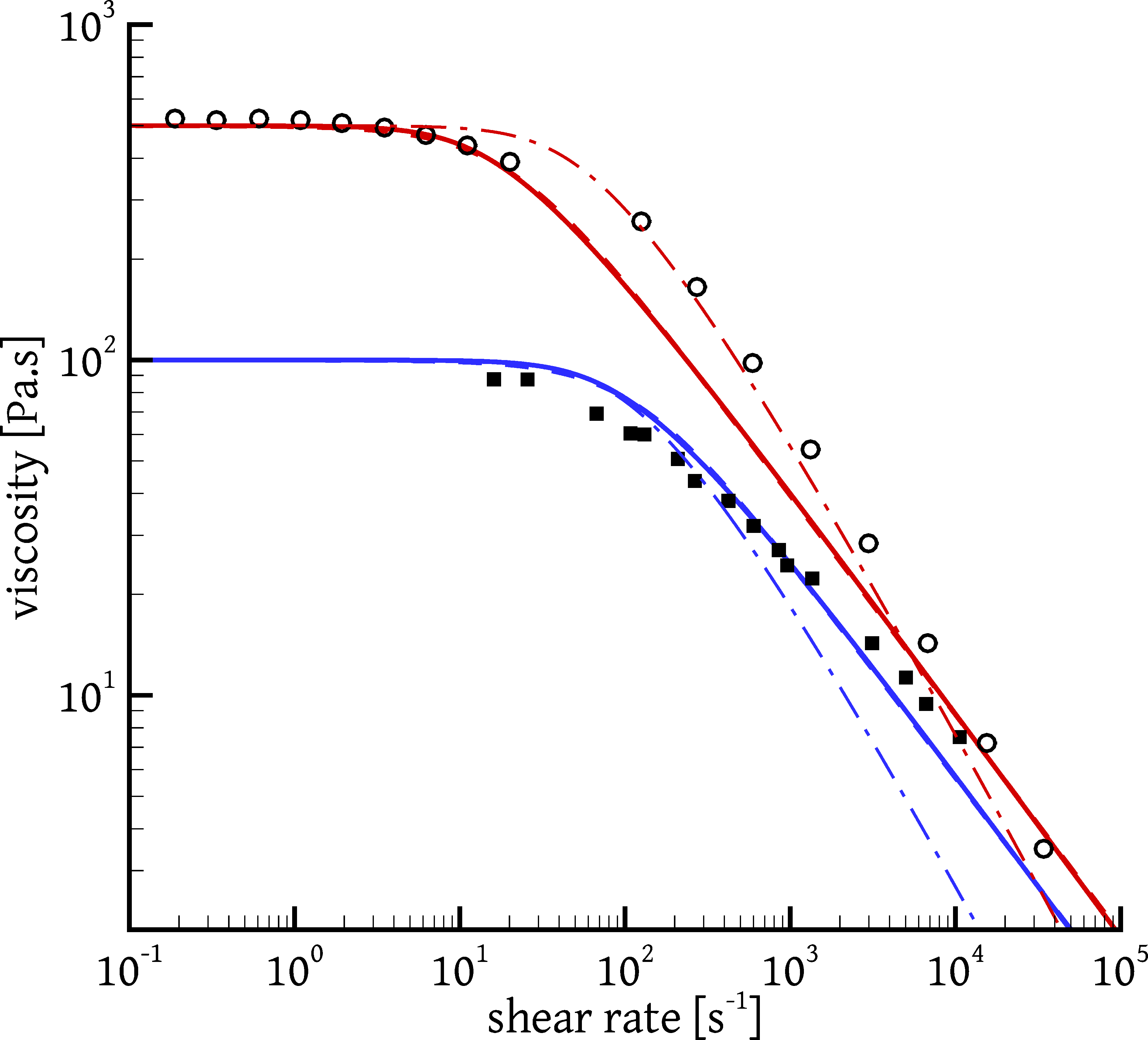}
  \caption{Variation of steady shear viscosity with shear rate, as measured for two different 
silicone oils (symbols) and predicted by constitutive models (lines). Experimental data are from 
\cite{Currie_1950} (${\scriptstyle\blacksquare}$) and \cite{Kokuti_2014} ($\circ$). The continuous 
lines correspond to the linear PTT fluids lPTT-100 (blue) and lPTT-500 (red) of Table \ref{table: 
fluids}. Dashed lines correspond to the Carreau-Yasuda fluids CY-100 (blue) and CY-500 (red) of 
Table \ref{table: fluids}, but are nearly indistinguishable from the lPTT-100 and lPTT-500 lines.
Dash-dot lines correspond to exponential PTT fluids with parameters \{$\eta=100$ \si{Pa.s}, 
$\lambda=0.01$ \si{s}, $\epsilon=0.25$\} (blue) and \{$\eta=500$ \si{Pa.s}, $\lambda=0.06$ \si{s}, 
$\epsilon=0.025$\} (red).}
  \label{fig: steady shear viscosity}
\end{figure}

The flow inside the damper is governed by the following equations that express the mass and 
momentum balance of the fluid, respectively, at a microscopic (but continuum) level:
%^b
\begin{equation} \label{eq: continuity}
% \pd{\rho}{t} \;+\; \nabla \cdot \left( \rho \vf{u} \right) \;=\; 0
 \nabla \cdot \vf{u} \;=\; 0
\end{equation}
%^b
\begin{equation} \label{eq: momentum}
 \pd{(\rho\vf{u})}{t} \;+\; \nabla \cdot \left( \rho \vf{u} \vf{u} \right) \;=\; -\nabla p \;+\;
 \nabla\cdot\tf{\tau}
\end{equation}
%^a
where $t$ is time, $\vf{u}$ is the velocity vector, $p$ is the pressure, $\rho$ is the density 
(which is constant, set to 1000 \si{kg/m^3}), and $\tf{\tau}$ is the stress tensor. To simplify the 
analysis the flow will be assumed incompressible and isothermal\footnote{The temperature rise of the 
fluid due to viscous dissipation can be important and deserves a separate study. It is investigated 
theoretically in \cite{Makris_1998, Makris_1998b} and experimentally in \cite{Black_2007}. In 
\cite{Black_2007}, experiments performed with a damper of much larger size than the present model, 
showed that if the amplitude of oscillation is significantly greater than the piston diameter then 
the oil temperature in the neighbourhood of the piston can rise by 50 or more degrees Celsius during 
6 oscillation cycles (the temperature rise is smaller farther away from the piston), which leads to 
a 10\% drop in the maximum damper force during the same period. A force drop is expected, because 
such a temperature rise would reduce the oil viscosity to half its room temperature value 
\cite{Currie_1950, Barlow_1964}. However, other experiments (again in \cite{Black_2007}) with a 
smaller damper showed negligible effect of the temperature rise on the force magnitude, so this 
issue requires further investigation.}. These equations must be complemented by a constitutive 
equation that relates the stress tensor to the kinematics of the flow. The simplest such equation is 
the Newtonian one, $\tf{\tau} = \eta \dot{\tf{\gamma}}$, where $\eta$ is the fluid viscosity and 
$\dot{\tf{\gamma}} \equiv \nabla \vf{u} + (\nabla \vf{u})^{\mathrm{T}}$ (T denoting the transpose) 
is the rate-of-strain tensor. However, this does not accurately describe the behaviour of polymeric 
liquids for all flow conditions. Such liquids consist of long molecular chains that interact with 
each other in a complex manner that typically results in a shear-thinning behaviour, i.e.\ the 
viscosity drops as the shear rate increases. This can be seen clearly in Fig.\ \ref{fig: steady 
shear viscosity} where the symbols denote viscosity measurements at different shear rates in steady 
shear experiments, reported in \cite{Currie_1950, Kokuti_2014}, for two different silicone oils with 
respective zero-shear viscosities of 90 and 525 \si{Pa.s}. Steady shear experiments 
\cite{Morrison_2001} are typically performed in capillary or rotational rheometers where a velocity 
$\vf{u} \equiv (u_1, u_2, u_3) = (\dot{\gamma}_{12} x_2, 0, 0)$ is imposed on the fluid (as in 
Couette flow between parallel plates), $\dot{\gamma}_{12}$ being the only non-zero component of 
$\dot{\tf{\gamma}}$, and the $\tau_{12}$ component of the stress tensor $\tf{\tau}$ is measured; the 
viscosity is then defined as $\eta(\dot{\gamma}_{12}) \equiv \tau_{12} / \dot{\gamma}_{12}$. A 
simple and popular way to construct a shear-thinning constitutive equation is then to generalise the 
Newtonian one by allowing $\eta$ to vary, $\tf{\tau} = \eta(\dot{\gamma}) \dot{\tf{\gamma}}$ where 
$\dot{\gamma} \equiv (\frac{1}{2} \dot{\tf{\gamma}} : \dot{\tf{\gamma}})^{1/2}$ is the magnitude of 
$\dot{\tf{\gamma}}$ and the function $\eta(\dot{\gamma})$ is fitted to steady shear data such as 
those plotted in Fig.\ \ref{fig: steady shear viscosity}.

Figure \ref{fig: steady shear viscosity} shows that the viscosity of silicone oil is nearly 
constant up to a critical shear rate beyond which it drops continually. A generalised Newtonian 
model capable of describing this kind of behaviour is the Carreau-Yasuda model \cite{Yasuda_1981}, 
with four parameters\footnote{The full Carreau-Yasuda model includes a fifth parameter, the 
viscosity limit $\eta_{\infty}$ as $\dot{\gamma} \rightarrow \infty$. The experimental data for 
silicone oil \cite{Currie_1950, Kokuti_2014} such as those plotted in Fig.\ \ref{fig: steady shear 
viscosity} do not reveal a non-zero such limit; therefore, we set $\eta_{\infty} = 0$, in which case 
the Carreau-Yasuda model reduces to Eq.\ \eqref{eq: constitutive Carreau-Yasuda}.} ($\eta_0$, 
$\dot{\gamma}_0$, $n$, $m$):
%^b
\begin{equation} \label{eq: constitutive Carreau-Yasuda}
 \eta(\dot{\gamma}) \;=\; \eta_0 \left[ 1 \;+\; \left( \frac{\dot{\gamma}}{\dot{\gamma}_0} 
\right)^m \right]^{\frac{n-1}{m}}
\end{equation}
%^a
At low shear rates ($\dot{\gamma} \ll \dot{\gamma}_0$) the model predicts Newtonian behaviour with 
viscosity $\eta = \eta_0$ while at high shear rates ($\dot{\gamma} \gg \dot{\gamma}_0$) the model 
predicts Power-Law behaviour $\eta = k \dot{\gamma}^{n-1}$ with shear-thinning index $n$ and 
consistency index $k = \eta_0 / \dot{\gamma}_0^{n-1}$; the exponent $m$ determines how rapid the 
transition is from Newtonian to Power-Law behaviour as $\dot{\gamma}$ increases. As mentioned in 
Section \ref{sec: introduction}, this model has been used to approximate the behaviour of damper 
fluids in the literature.

On the other hand, generalised Newtonian models do not account for fluid elasticity, which can 
affect the behaviour of the fluid in substantial ways. A popular viscoelastic model capable of 
describing the rheology of shear-thinning polymeric liquids is the Phan-Thien \& Tanner (PTT) model 
\cite{Phan_Thien_1977, Phan_Thien_1978}; it has been used successfully to simulate several 
important processes of scientific and industrial interest (e.g.\ \cite{Foteinopoulou_2004, 
Dimakopoulos_2009, Papaioannou_2014, Pettas_2015, Fraggedakis_2016}). The simplified linear version 
of this model \cite{Phan_Thien_1977} can be expressed as
%^b
\begin{equation} \label{eq: constitutive lPTT}
 \underbrace{\left( 1 \;+\; \frac{\lambda \epsilon}{\eta_0} 
\mathrm{tr}(\tf{\tau})\right)}_{Y\left(\mathrm{tr}(\tf{\tau})\right)}
\, \tf{\tau} \;+\; \lambda \ucd{\tau} \;=\; \eta_0 \dot{\tf{\gamma}}
\end{equation}
%^a
where $\mathrm{tr}(\tf{\tau}) = \sum_i \tau_{ii}$ is the trace of the stress tensor, $\lambda$ and 
$\eta_0$ are the fluid relaxation time and viscosity, respectively, and the triangle denotes the 
upper-convected derivative
%^b
\begin{equation} \label{eq: upper convected derivative}
 \ucd{\tau} \;\equiv\; \pd{\tf{\tau}}{t} \;+\; \vf{u} \cdot \nabla \tf{\tau} 
            \;-\;      \left( (\nabla \vf{u})^{\mathrm{T}} \cdot \tf{\tau} \;+\; 
                              \tf{\tau} \cdot \nabla \vf{u} 
\right)
\end{equation}
%^a
The PTT model \eqref{eq: constitutive lPTT} uses also a non-negative dimensionless parameter 
$\epsilon$ which is related to shear-thinning. A feel for the physical significance of the model 
can be obtained by noting that for $\epsilon = 0$, and viewing the upper convected derivative as a 
form of time derivative, the PTT model is analogous to the macroscopic Maxwell model \eqref{eq: 
macroscopic Maxwell} albeit applied to a microscopic fluid element\footnote{The ``microscopic'' 
fluid element is still much larger than the molecular dimensions so that the fluid can be regarded 
as a continuum.} instead of to the whole damper. In fact, for $\epsilon = 0$ the model \eqref{eq: 
constitutive lPTT} is called the Upper Convected Maxwell (UCM) model. For $\epsilon > 0$ the 
function $Y(\mathrm{tr}(\tf{\tau}))$ causes shear-thinning. This can be seen in Fig.\ \ref{fig: 
steady shear viscosity} where the shear-thinning behaviour of the two linear PTT model fluids of 
Table \ref{table: fluids} in steady shear flow is plotted (the solution of eq.\ \eqref{eq: 
constitutive lPTT} for steady shear flow can be found in \cite{Azaiez_1996}; see also Appendix 
\ref{appendix: steady shear flow}).

\begin{table}[tb]
\caption{Model fluids used in the simulations. In every case the density is $\rho = 1000$ 
\si{kg/m^3}.}
\label{table: fluids}
\begin{center}
\begin{small}   % to make the font smaller
\renewcommand\arraystretch{1.25}   % Adjust row height (default=1)
\begin{tabular}{ l | l }
\toprule
 N-100:    & 
 Newtonian fluid of viscosity $\eta = 100$ \si{Pa.s}
\\
 lPTT-100  & 
 Eq.\ \eqref{eq: constitutive lPTT}:
 $\eta_0 = 100$ \si{Pa.s}, $\lambda = 0.01$ \si{s}, $\epsilon = 0.25$
\\
 CY-100    &
 Eq.\ \eqref{eq: constitutive Carreau-Yasuda}:
 $\eta_0 = 100$ \si{Pa.s}, $\dot{\gamma}_0 = 119$ \si{s^{-1}}, $n = 0.353$, $m = 1.433$
\\
 lPTT-500  &
 Eq.\ \eqref{eq: constitutive lPTT}:  
 $\eta_0 = 500$ \si{Pa.s}, $\lambda = 0.06$ \si{s}, $\epsilon = 0.25$
\\
 CY-500    &
 Eq.\ \eqref{eq: constitutive Carreau-Yasuda}:
 $\eta_0 = 500$ \si{Pa.s}, $\dot{\gamma}_0 = 20.7$ \si{s^{-1}}, $n = 0.346$, $m = 1.381$
\\
\bottomrule
\end{tabular}
\end{small}
\end{center}
\end{table}

There is another version of the PTT model commonly in use, the exponential version 
\cite{Phan_Thien_1978}, which differs from Eq.\ \eqref{eq: constitutive lPTT} in the definition of 
the function $Y(\mathrm{tr}(\tf{\tau})) = \exp( (\lambda \epsilon / \eta) \, 
\mathrm{tr}(\tf{\tau}))$. The behaviour of a couple of such fluids is also plotted in Fig.\ 
\ref{fig: steady shear viscosity} with dash-dot lines (see Appendix \ref{appendix: steady shear 
flow} for calculation details). It was noticed that for viscosities around $100$ \si{Pa.s} the 
exponential model predicts excessive shear thinning whereas the linear PTT model was closer to the 
experimentally observed behaviour of silicone oil. For viscosities around $500$ \si{Pa.s} the 
experimentally observed shear thinning is between that predicted by the two models. In the present 
work it was decided to use the linear PTT model, and in particular the fluids lPTT-100 and lPTT-500 
of Table \ref{table: fluids}.  The experimental data of the 90 \si{Pa.s} and 525 \si{Pa.s} silicone 
oils plotted in Fig.\ \ref{fig: steady shear viscosity} were used as guides in selecting the PTT 
parameters, although a precise fitting to the data was not performed and the parameters were given 
nice, rounded values. In particular, as shown in Appendix \ref{appendix: steady shear flow}, the 
PTT parameter $\eta_0$ is equal to the viscosity at $\dot{\gamma} \rightarrow 0$ and values of 
$\eta_0 = 100$ \si{Pa.s} and $500$ \si{Pa.s} were selected. On the other hand, as also shown in 
Appendix \ref{appendix: steady shear flow}, the data of Fig.\ \ref{fig: steady shear viscosity} are 
not sufficient for obtaining realistic values for $\lambda$ and $\epsilon$ individually because the 
curves of all PTT fluids for which the product $\epsilon \lambda^2$ is the same coincide. 
Therefore, the value of one of these parameters must come from another type of rheological 
experiment. The PTT parameter $\lambda$ is intended to represent the fluid relaxation time, i.e.\ 
the time needed for stresses in the fluid to relax after the imposed shear rate is removed 
($\dot{\tf{\gamma}} = 0$). Longin et al.\ \cite{Longin_1998} provide experimental measurements of 
the storage and loss moduli of a 100 \si{Pa.s} silicone oil, from which they derive a discrete 
spectrum of $N=10$ relaxation moduli $g_i$, $i = 1, \ldots, 10$, and associated relaxation times 
$\lambda_i$. From these we calculated the viscosity-averaged relaxation time as
%^b
\begin{equation} \label{eq: average relaxation time}
 \lambda \;=\; \frac{\sum_{i=1}^N g_i \lambda_i^2}{\sum_{i=1}^N g_i \lambda_i}
\end{equation}
%^a
which gives a value of $\lambda = 0.01$ \si{s}. Assigning this value to $\lambda$ for the lPTT-100 
fluid, the value $\epsilon = 0.25$ was then chosen in order to obtain a nice viscosity versus shear 
rate curve compared to the data for a 90 \si{Pa.s} silicone oil given in \cite{Currie_1950} (Fig.\ 
\ref{fig: steady shear viscosity}). A similar procedure was followed for the lPTT-500 fluid, using 
data from \cite{Kokuti_2014}.

Flow simulations with a viscoelastic constitutive equation such as \eqref{eq: constitutive lPTT} 
are much more expensive than those with a generalised Newtonian constitutive equation such as 
\eqref{eq: constitutive Carreau-Yasuda} because in the latter the stress tensor is given by 
explicit expressions whereas in the former it has to be obtained through the solution of a partial 
differential equation for each relevant stress component. Thus it is tempting to assume that 
elasticity is not important and treat the flow as generalised Newtonian, accounting only for 
shear-thinning. To investigate how realistic such an assumption is for the present flow, we 
performed simulations also with the Carreau-Yasuda CY-100 and CY-500 fluids defined in Table 
\ref{table: fluids}, whose parameters were chosen such that they match the lPTT-100 and lPTT-500 
fluids, respectively, in the plot of Fig.\ \ref{fig: steady shear viscosity}. This was achieved by 
fixing $\eta_0$ to 100 (for CY-100) and 500 (for CY-500) \si{Pa.s} and selecting the rest of the 
parameters, $\dot{\gamma}_0$, $m$, $n$, such that they minimise the functional
%^b
\begin{equation*}
\int_{ \dot{\gamma}_1}^{\dot{\gamma}_2} \left[ 
\ln \left( \eta_{\mathrm{CY}}(\dot{\gamma},\dot{\gamma}_0,m,n) \right) - 
\ln \left( \eta_{\mathrm{PTT}}(\dot{\gamma}) \right)
\right]^2 \mathrm{d}(\ln \dot{\gamma})
\end{equation*}
%^a
where $\dot{\gamma}_1 = 1$ (CY-100) or $0.1$ (CY-500) and $\dot{\gamma}_2 = 10^5$ \si{s^{-1}} are 
the lower and upper limits of the shear-rate range of Fig.\ \ref{fig: steady shear viscosity}, 
$\eta_{\mathrm{CY}}$ is the Carreau-Yasuda viscosity function \eqref{eq: constitutive 
Carreau-Yasuda}, and $\eta_{\mathrm{PTT}}$ is the steady shear viscosity of the linear PTT fluids 
plotted in Fig.\ \ref{fig: steady shear viscosity}. The resulting $\eta_{\mathrm{CY}}$ functions 
are plotted in dashed lines in Fig.\ \ref{fig: steady shear viscosity} but are not discernible 
because they nearly coincide with the linear PTT viscosities. A few Newtonian simulations were also 
performed.

To some degree, whether inertia, shear-thinning, and elasticity of the fluid are important for this 
particular flow can be assessed a priori. This depends on the fluid properties, damper geometry, 
and operating conditions, and is expressed in terms of dimensionless numbers. Whether 
shear-thinning is important or not depends on the shear rates encountered. The highest shear rates 
will occur at the critical region of the gap between the piston and cylinder, when the piston 
velocity is maximum, i.e.\ when $|u_p| = U_p = \omega \alpha$ (Eq.\ \eqref{eq: piston velocity}) at 
$t = T/4$, $3T/4$, etc. At these times, the mean fluid velocity at the gap $U_f$ can be found from 
the fact that the rate of fluid volume pushed towards one side by the piston, $U_p (\pi R_p^2 - \pi 
R_s^2)$, must equal the rate of fluid volume crossing the gap towards the other side, $U_f (\pi 
R_c^2 - \pi R_p^2)$, as follows from mass conservation and incompressibility. Therefore, $U_f = U_p 
(R_p^2 - R_s^2) / (R_c^2 - R_p^2)$. Assuming that the fluid velocity roughly varies from $U_p$ at 
the piston surface to $-U_f$ at half-way between the piston and cylinder, a characteristic shear 
rate can be defined as $\dot{\gamma}_c \equiv U / (h/2)$ where $U = U_p + U_f$ is the relative 
velocity between the fluid and the piston. Table \ref{table: operating conditions} lists the values 
of $\dot{\gamma}_c$ for the selected frequencies; from Eq.\ \eqref{eq: constitutive Carreau-Yasuda} 
or Fig.\ \ref{fig: steady shear viscosity} we can deduce the viscosity that corresponds to each 
$\dot{\gamma}_c$, and compare it to the nominal $\eta_0$. Table \ref{table: operating conditions} 
also lists the ratios $\eta(\dot{\gamma}_c) / \eta_0$, and from these values it is evident that 
shear-thinning is expected to play a very significant role, even at low frequencies.

\begin{table}[tb]
\caption{Operating conditions for which simulations were performed, and associated dimensionless 
numbers. The oscillation amplitude is $\alpha$ = 12 \si{mm} in every case.}
\label{table: operating conditions}
\begin{center}
\begin{small}   % to make the font smaller
\renewcommand\arraystretch{1.25}   % Adjust row height (default=1)
\begin{tabular}{ l | c c c c | c c }
\toprule
                  &  \multicolumn{4}{c |}{ 100 \si{Pa.s} } & \multicolumn{2}{c}{ 500 \si{Pa.s} }
\\ \midrule
 $f$  [\si{Hz}]   &    0.5  &   2     &   8      &   32     &  0.5     &  2
\\
 $U_p$ [\si{m/s}] &   0.038 &  0.151  &   0.603  &   2.413  &  0.038    &  0.151  
\\
 $U$ [\si{m/s}]   &   0.16  &  0.64   &   2.57   &   10.27  &  0.16    &  0.64  
\\
 $\dot{\gamma}_c$ [\si{s^{-1}}] &
                   107    &  428    &   1711   &   6845   &   107    &  428
\\
 $\eta(\dot{\gamma}_c) / \eta_0$ &
                  0.759   &  0.399  &  0.178   &  0.078   &  0.315   &  0.137
\\
 $\Rey$           &  0.002  &  0.010  &  0.039   &  0.154   &  0.001   &  0.002 
\\
 $\Rey_c$         &  0.003  &  0.024  &  0.216   &  1.984   &  0.002   &  0.014
\\
 $\Wei$           &  1.07   &  4.28   &  17.1    &  68.5    &  6.42    &  25.7
\\
 $\Deb$           &  0.005  &  0.02   &  0.08    &  0.32    &  0.03    &  0.12
\\
 $\Str$           &  214    &  214    &  214     &  214     &  214     &  214 
\\
\bottomrule
\end{tabular}
\end{small}
\end{center}
\end{table}

To assess the importance of inertia and elasticity, it is expedient to express the governing 
equations in dimensionless form. To this end, we normalise lengths by half the gap width, $H \equiv 
h/2$, time by the period of oscillation $T$, velocities by $U = U_p + U_f$, and stresses and 
pressure by $\eta_0 U / H$. Whence substituting $x = \tilde{x} H$, $t = \tilde{t} T$, $\vf{u} = 
\tilde{\vf{u}} U$ etc.\ Eqs.\ \eqref{eq: momentum} and \eqref{eq: constitutive lPTT} -- \eqref{eq: 
upper convected derivative} can be expressed as
%^b
\begin{equation} \label{eq: momentum nd}
  \Rey \left( 
  \frac{1}{\Str} \pd{\tilde{\vf{u}}}{\tilde{t}}
  \;+\;
  \tilde{\nabla} \cdot \left( \tilde{\vf{u}} \tilde{\vf{u}} \right)
  \right)
  \;=\;
  -\tilde{\nabla} \tilde{p}
  \;+\;
  \tilde{\nabla} \cdot \tilde{\tf{\tau}}
\end{equation}
%^a
%^b
\begin{equation} \label{eq: constitutive lPTT nd}
 \left( 1 \;+\; \epsilon \Wei \, \mathrm{tr}(\tilde{\tf{\tau}}) \right) \tilde{\tf{\tau}}
 \;\;+\;\;
 \left(
     \Deb \pd{\tilde{\tf{\tau}}}{\tilde{t}}
     \;+\;
     \Wei 
     \left(
         \tilde{\vf{u}} \cdot \tilde{\nabla} \tilde{\tf{\tau}}
         \:-\: (\tilde{\nabla} \tilde{\vf{u}})^{\mathrm{T}} \cdot \tilde{\tf{\tau}}
         \:-\: \tilde{\tf{\tau}} \cdot \tilde{\nabla} \tilde{\vf{u}}
     \right)
 \right)
 \;\;=\;\;
 \tilde{\tf{\dot{\gamma}}}
\end{equation}
%^a
where $\Rey \equiv \rho U H / \eta_0$ is the Reynolds number, $\Str \equiv T/(H/U)$ is the 
Strouhal number, $\Wei \equiv \lambda U / H$ is the Weissenberg number, $\Deb \equiv \lambda / T$ 
($= \Wei / \Str$) is the Deborah number, and the tilde (\textasciitilde) denotes dimensionless 
quantities, with $\tilde{\tf{\dot{\gamma}}} \equiv \tf{\dot{\gamma}} / (U/H)$.

The Reynolds and Strouhal numbers appear in the convection terms of the momentum equation. For this 
particular problem, it follows from $T = 2\pi/\omega$ and $U = U_p + U_f = \alpha \omega (1 + 
(R_p^2 - R_s^2) / (R_c^2 - R_p^2))$ that $\Str = \tilde{\alpha} \, 2\pi (R_c^2-R_s^2) / (R_c^2 - 
R_p^2)$ where $\tilde{\alpha} \equiv \alpha / H$ is the dimensionless oscillation amplitude. Thus, 
the inherent inversely proportional relationship between the velocity $U$ and the period $T$ 
removes the dependence of $\Str$ on $f$ and makes it a function of only the oscillation amplitude 
and the damper geometry, which are held constant in the present study. The Reynolds number is an 
indicator of the ratio of fluid inertia to viscous forces, and its values for each fluid / 
frequency combination are listed in Table \ref{table: operating conditions}. This definition of 
Reynolds number does not account for shear-thinning and overestimates the viscous forces, so the 
Table includes also a Reynolds number based on the viscosity at the characteristic shear rate, 
$\Rey_c \equiv \rho U H / \eta(\dot{\gamma}_c)$. The values of both Reynolds numbers can be seen to 
be small, making the left-hand side of the momentum equation \eqref{eq: momentum nd} small compared 
to the two terms appearing in the right-hand side. Thus the role of fluid inertia is expected to be 
limited in the present flow; nevertheless, the inertia terms were retained in the equations solved 
and it will be seen in Sec.\ \ref{ssec: results; inertia} that inertia does play a minor role for 
this flow, at high frequencies.

The Weissenberg number is an indicator of the ratio of elastic to viscous forces, while the Deborah 
number is the ratio of the fluid relaxation time to a time scale characterising the flow, which in 
our case is the oscillation period $T$ \cite{Poole_2012}. Both numbers are indicative of the 
importance of elastic phenomena in the flow, and it may be seen that if they tend to zero then 
the constitutive equation \eqref{eq: constitutive lPTT nd} tends to the Newtonian constitutive 
equation. However, the values listed in Table \ref{table: operating conditions} are not small, and 
suggest significant elastic effects.

Before ending this section, one more issue must be discussed. Calculation of the damper force is of 
crucial importance, and is performed by integrating the stress and pressure over the surface of the 
piston and the shaft. At the points where the moving shaft meets the stationary cylindrical casing 
the wall velocity varies discontinuously. If the fluid is Newtonian and the no-slip boundary 
condition is assumed then it can be shown \cite{Batchelor_2000} that the shear stress on the shaft 
varies as $(\delta x)^{-1}$, where $\delta x$ is the distance from the singularity point. The force 
on the shaft is equal to the integral of the shear stress over the shaft surface, i.e.\ it is 
proportional to the integral of $(\delta x)^{-1}$ from $\delta x = 0$ to the length of the shaft, 
which is infinite. Clearly, an infinite force is not a realistic result. In our previous work 
\cite{Syrakos_2016} this hurdle was overcome by assuming a Navier slip boundary condition, which 
was shown to limit the force to finite values \cite{He_2009}, and is a realistic assumption 
according to molecular simulations studies \cite{Koplik_1995, Qian_2005} even for Newtonian flows. 
In non-Newtonian flows slip phenomena are much more pronounced \cite{Hatzikiriakos_2015, 
Hatzikiriakos_2012, Archer_2005, Karapetsas_2006}. To be consistent with our previous approach, in 
the present study we also employed Navier slip boundary conditions, in fact on all solid walls; 
thus, if $\vf{u}$ and $\vf{u}_w$ are the fluid and wall velocities at the boundary then
%^b
\begin{equation} \label{eq: Navier slip}
 (\vf{u}-\vf{u}_w) \cdot \vf{s} \;=\; \beta (\vf{n} \cdot \tf{\tau}) \cdot \vf{s}
\end{equation}
%^a
where $\vf{n}$ is the unit vector normal to the wall and $\vf{s}$ the unit vector tangential to the 
wall within the plane in which the equations are solved. The slip coefficient $\beta$ is assigned 
here values of $5 \times 10^{-7}$ \si{m/Pa.s} (lPTT-100 and CY-100 fluids) and $10^{-6}$ 
\si{m/Pa.s} (lPTT-500 and CY-500 fluids). These values are significantly lower than those used in 
our previous work \cite{Syrakos_2016}. A feel for the effect of slip in the critical region of the 
gap can be obtained as follows. Assume that the flow in that region is one-dimensional and the 
stress on the cylinder, $(\vf{n} \cdot \tf{\tau}) \cdot \vf{s} = \tau_{rx}$, is typically of order 
$\eta(\dot{\gamma}_c) U / H$ (Table \ref{table: operating conditions}) when the cylinder moves with 
maximum velocity. Then from Eq.\ \eqref{eq: Navier slip} it follows that
%^b
\begin{equation*}
 u - u_w \;=\; \beta \, \eta(\dot{\gamma}_c) \frac{U}{H} \;\Rightarrow\;
 \frac{u - u_w}{U} \;=\; \frac{\beta \eta(\dot{\gamma}_c)}{H}
\end{equation*}
%^a
(the product $\beta\eta$ is called the \textit{slip length}). The right-hand side has values 
ranging from 2.5\% ($f$ = 0.5 \si{Hz}) to 0.26\% ($f$ = 32 \si{Hz}) for the 100 \si{Pa.s} fluids, 
and from 10.5\% ($f$ = 0.5 \si{Hz}) to 4.6\% ($f$ = 2 \si{Hz}) for the 500 \si{Pa.s} fluids. Thus 
the slip velocity $u-u_w$ is small, but not negligible, compared to the velocity scale $U$ and the 
flow is not much affected by the wall slip (this is confirmed by the velocity profiles shown in 
Fig.\ \ref{fig: velocity profiles}, to be discussed in Sec.\ \ref{ssec: results; elasticity}). We 
note that the $(\delta x)^{-1}$ stress variation of Newtonian fluids is barely non-integrable; a 
variation $(\delta x)^{-a}$ would be integrable for any $a < 1$ and result in finite force. 
Therefore, it is reasonable to expect that for shear-thinning fluids whose viscosity tends to zero 
as the shear rate tends to infinity, such as those presently employed, the force would be finite 
even with the no-slip boundary condition. However, this issue was not investigated further.

\section{Numerical method}
\label{sec: method}

The equations given in the previous Section were solved using a finite volume method, which was 
developed on the foundation of an existing method for generalised Newtonian and viscoplastic flows 
\cite{Syrakos_06a,Syrakos_06b,Syrakos_2016}. In the present Section the method will only be 
summarised, while the extensions pertaining to viscoelastic flow will be described in detail in a 
separate publication.

\begin{figure}[thb]
    \centering
    \begin{minipage}[b]{0.6\textwidth}
        \begin{subfigure}[b]{\textwidth}
            \centering
            \includegraphics[width=0.95\linewidth]{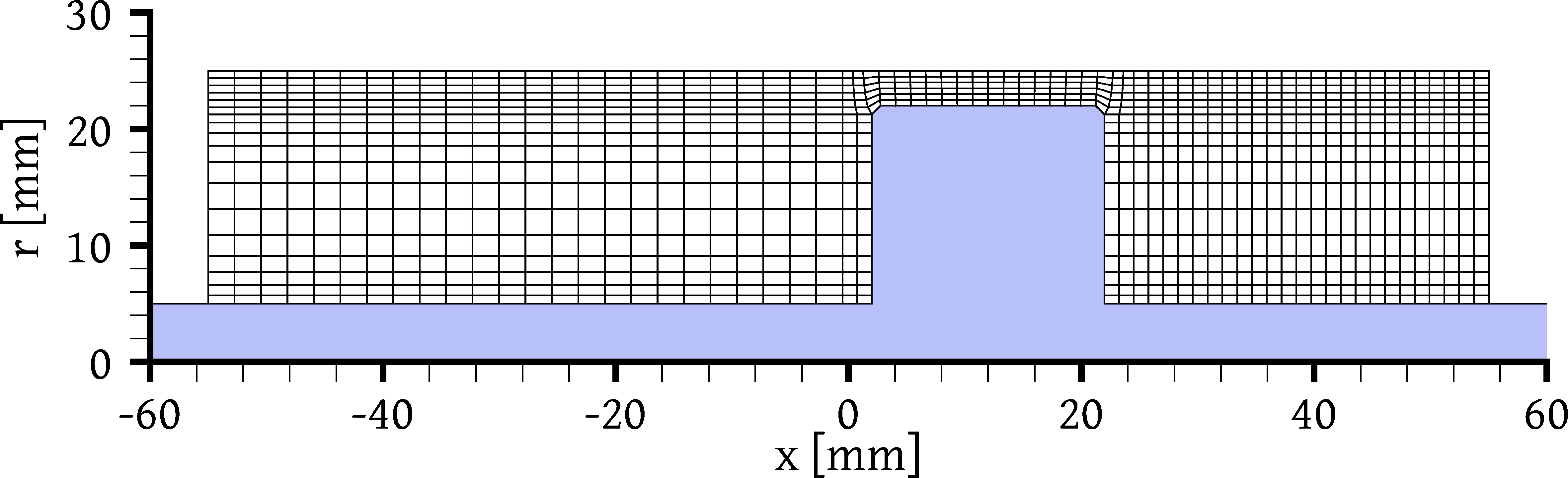}
            \caption{}
            \label{sfig: grid offset}
        \end{subfigure}
        \begin{subfigure}[b]{\textwidth}
            \centering
            \includegraphics[width=0.95\linewidth]{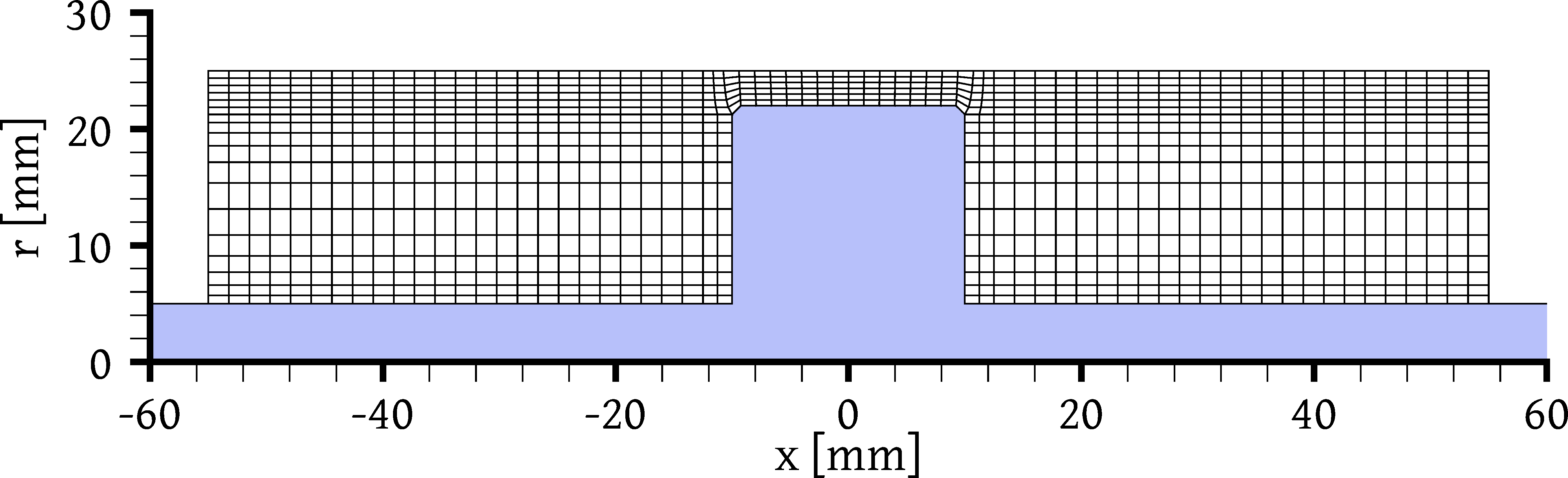}
            \caption{}
            \label{sfig: grid centred}
        \end{subfigure}
    \end{minipage}
    \begin{minipage}[b]{0.35\textwidth}
        \begin{subfigure}[b]{\textwidth}
            \centering
            \includegraphics[width=0.95\linewidth]{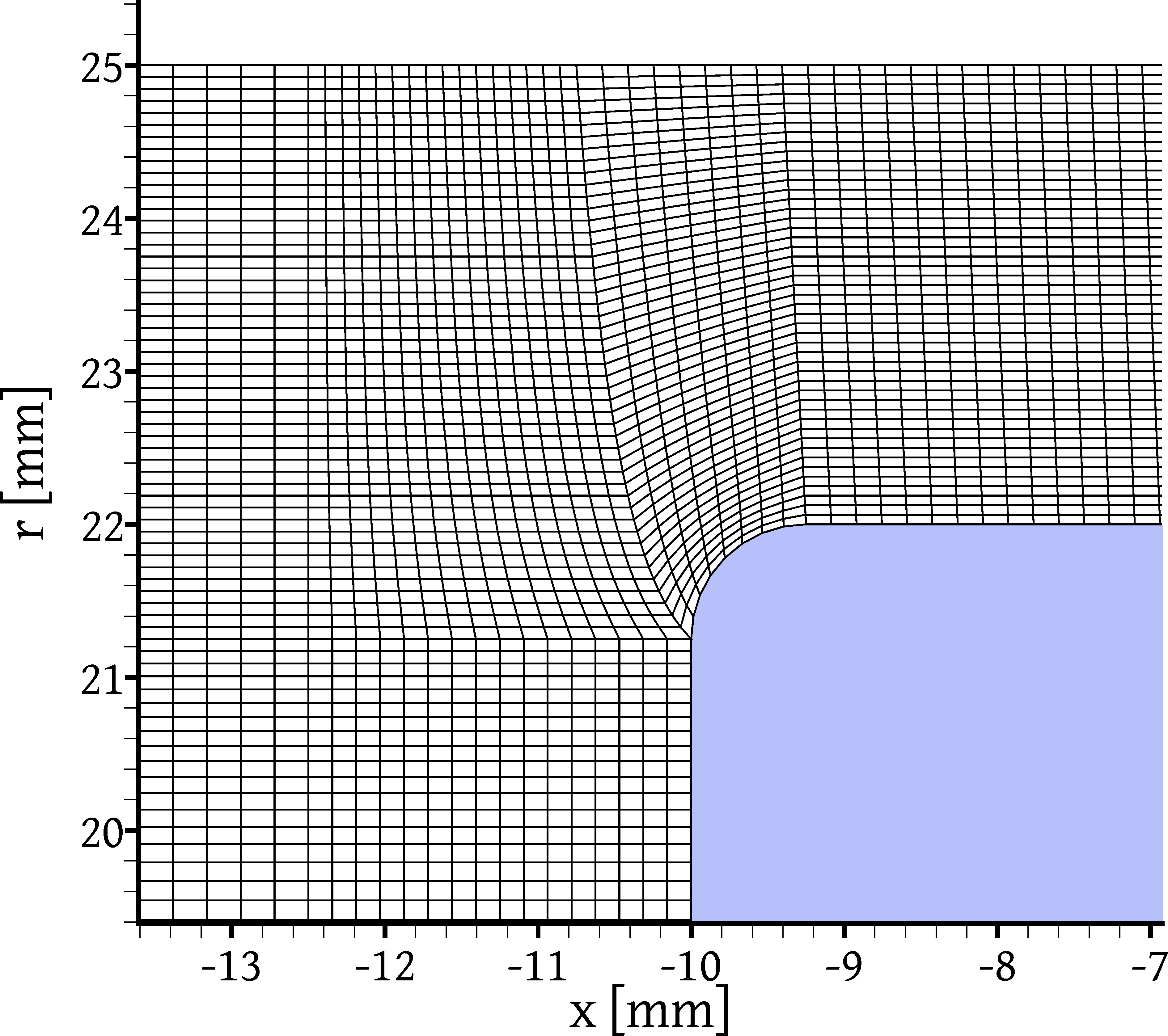}
            \vspace{1cm}
            \caption{}
            \label{sfig: grid corner}
        \end{subfigure}
    \end{minipage}
    \caption{\subref{sfig: grid offset} View of a coarse grid at a time instance when the piston is 
at the extreme right position. \subref{sfig: grid centred} View at a time instance when the piston 
is at the middle position. \subref{sfig: grid corner} Close-up of grid 2 (Table \ref{table: grids}) 
near a piston corner.}
    \label{fig: grid}
\end{figure}

The $x-r$ plane was discretised by a series of block-structured grids of increasing fineness, a 
coarse one of which is shown in Figs.\ \ref{sfig: grid offset} and \ref{sfig: grid centred}. The 
grid changes dynamically in time in order to follow the motion of the piston: the cells surrounding 
the piston up to $2.5$ \si{mm} on either side translate along with it without deforming, while the 
rest of the grid cells contract or expand accordingly (Figs.\ \ref{sfig: grid offset}, \ref{sfig: 
grid centred}). Since the solution of the set of discretised equations was accelerated using a 
multigrid algorithm, grids of varying fineness were created. Solutions were obtained on the three 
finest grids, the sizes of which are listed in Table \ref{table: grids}. The same Table lists the 
number of cells in the radial direction between the piston and the cylinder, which are of equal 
radial width. Preliminary simulations suggested that using sharp piston corners does not create 
additional numerical difficulties, but to be on the safe side \cite{Pavlidis_2010} it was decided to 
use rounded corners, of radius $0.75$ \si{mm}, as shown in Fig.\ \ref{sfig: grid corner}. The blocks 
in contact with these corners were constructed using an elliptic grid generation technique 
\cite{Dimakopoulos_2003, Chatzidai_2009}.

\begin{table}[thb]
\caption{Sizes of the grids used in the simulations.}
\label{table: grids}
\begin{center}
\begin{small}   % to make the font smaller
\renewcommand\arraystretch{1.25}   % Adjust row height (default=1)
\begin{tabular}{ c | c c }
\toprule
 Grid  &  total cells          &   cells across gap  
\\
\midrule
 1     &   16,512              &          24                
\\
 2     &   66,048              &          48                
\\
 3     &  264,192              &          96                
\\
\bottomrule
\end{tabular}
\end{small}
\end{center}
\end{table}

Concerning the discretisation of the momentum and continuity equations, we used the same method as 
in our previous work \cite{Syrakos_2016}, which employs central differencing with least-squares 
gradients \cite{Syrakos_2017} in space and a three time level implicit scheme in time, which are 
second-order accurate. A complication arises in the viscoelastic case because the viscous forces on 
the cell faces are not an explicit function of the velocity but are calculated by linear 
interpolation of the stresses at the cells on either side of the face, which are stored as separate 
variables. This results in a lack of second derivatives of velocity in the momentum equation 
\eqref{eq: momentum} and leaves only the first derivatives of the convection terms, which makes the 
solution susceptible to spurious velocity oscillations, in exactly the same way that the appearance 
of only first derivatives of pressure in the same equation gives rise to the well-known problem of 
spurious pressure oscillations. In the present work, the pressure oscillations were suppressed by 
using a variant, proposed in \cite{Syrakos_06a}, of the renowned ``momentum interpolation'' 
technique \cite{Rhie_1983} while the velocity oscillations were suppressed in a similar manner, by 
devising a scheme for interpolation of the stresses at cell faces that is inspired from the one 
proposed in \cite{Oliveira_1998, Matos_2009} and re-introduces (discrete) second derivatives of 
velocity into the momentum equation. The details of this interpolation scheme, which follows the 
philosophy of our momentum interpolation scheme \cite{Syrakos_06a}, will be given in a separate 
publication. The momentum convection terms were discretised with plain central differences; the 
aforementioned stress interpolation scheme proved effective in eliminating the velocity oscillations 
such that a high-resolution scheme (e.g.\ CUBISTA -- see below) was not necessary for these terms.

The viscoelastic constitutive equations were discretised along similar lines as in 
\cite{Oliveira_1998, Afonso_2012}. To reduce their stiffness, these equations were expressed 
in terms of the logarithm of the conformation tensor instead of the stress tensor itself, as 
suggested in \cite{Fattal_2004, Fattal_2005, Afonso_2009}. The viscoelastic constitutive equation 
\eqref{eq: constitutive lPTT} contains no diffusion terms, and therefore again a spurious stress 
oscillation issue may arise; to avoid it, its convection terms were discretised with the CUBISTA 
high resolution scheme \cite{Alves_2003}. At domain boundaries, all of which are solid walls, the 
pressure and stresses were linearly extrapolated from the interior. The system of non-linear 
algebraic equations that arise from the finite volume discretisation was solved with an extended 
SIMPLE algorithm, each iteration of which includes the solution of one linear system per component 
of the log-conformation tensor. The SIMPLE solver was used as a smoother in a multigrid framework 
\cite{Syrakos_06b}. For one of the test cases, the lPTT-500 fluid at the $f = 2$ frequency, the 
SIMPLE / multigrid solver exhibited convergence difficulties that were overcome by applying a 
vector extrapolation technique \cite{Sidi_2012} where each vector contains the estimate of all 
log-conformation tensor components ($rr$, $xx$, $rx$, and $\theta\theta$) at all grid cells after a 
multigrid cycle.

The code was validated against available data on viscoelastic lid-driven cavity flows, which 
resemble the present flow configuration in that they are bounded all around by solid walls and the 
flow is induced by a moving wall. Comparison of results obtained with our code against results 
presented in the literature shows very good agreement. Indicative results concerning the location 
and strength of the main vortex are shown in Table \ref{table: validation}. The model fluids 
include a Newtonian solvent contribution to the stress tensor: $\tf{\tau} = \tf{\tau}_p + 
\tf{\tau}_s$ where the polymeric component $\tf{\tau}_p$ is given by Eq.\ \eqref{eq: constitutive 
lPTT} and the solvent component by $\tf{\tau}_s = \eta_s \tf{\dot{\gamma}}$. Then the flow is 
determined by an additional dimensionless number, the viscosity ratio $B \equiv \eta_s / (\eta_s + 
\eta_p)$ where $\eta_p$ is the viscosity of the polymeric component, denoted as $\eta_0$ in Eq.\ 
\eqref{eq: constitutive lPTT}.

\begin{table}[thb]
\caption{$\tilde{x}-$ and $\tilde{y}-$ coordinates (first and second columns) of the centre of the 
main vortex for viscoelastic flow in a square cavity of side $H$, and associated value of the 
streamfunction there (third column). The top wall moves in the positive $x-$direction with variable 
velocity $u(x) = 16 U (x/H)^2(1-x/H)^2$ (for the $\epsilon = 0$ case), or with uniform velocity $u 
= U$ (for the $\epsilon = 0.25$ cases); The coordinates are normalised by the cavity side $H$ and 
the streamfunction by $U H$. The Weissenberg and Reynolds numbers are defined as $\Rey \equiv \rho 
U H/(\eta_s + \eta_p)$ and $\Wei \equiv \lambda U / H$, respectively.}
\label{table: validation}
\begin{center}
\begin{small}   % to make the font smaller
\renewcommand\arraystretch{1.25}   % Adjust row height (default=1)
\begin{tabular}{ l | c c c }
\toprule
%                                   &  $x_c$  &  $y_c$  &  $\phi_c$
% \\
% \midrule
\multicolumn{4}{c}{ $\epsilon = 0$, $\Wei = 1$, $\Rey = 0$, $B = 0.5$ }
\\
\midrule
 Saramito  \cite{Saramito_2014}   &  0.429  &  0.818  &  -0.0619   \\
 Sousa et al. \cite{Sousa_2016}   &  0.434  &  0.814  &  -0.0619   \\
 Present results                  &  0.434  &  0.818  &  -0.0619
\\
\midrule
\multicolumn{4}{c}{ $\epsilon = 0.25$, $\Wei = 4$, $\Rey = 1$, $B = 1/9$ }
\\
\midrule
 Dalal et al.\ \cite{Dalal_2016}  &  0.429  &  0.812  &  -0.0694   \\
 Present results                  &  0.430  &  0.813  &  -0.0693
\\
\midrule
\multicolumn{4}{c}{ $\epsilon = 0.25$, $\Wei = 4$, $\Rey = 100$, $B = 1/9$ }
\\
\midrule
 Dalal et al.\ \cite{Dalal_2016}  &  0.783  &  0.767  &  -0.0594   \\
 Present results                  &  0.782  &  0.760  &  -0.0600
\\
\bottomrule
\end{tabular}
\end{small}
\end{center}
\end{table}

\begin{figure}[!tb]
    \centering
    \begin{subfigure}[b]{0.32\textwidth}
        \centering
        \includegraphics[width=0.95\linewidth]{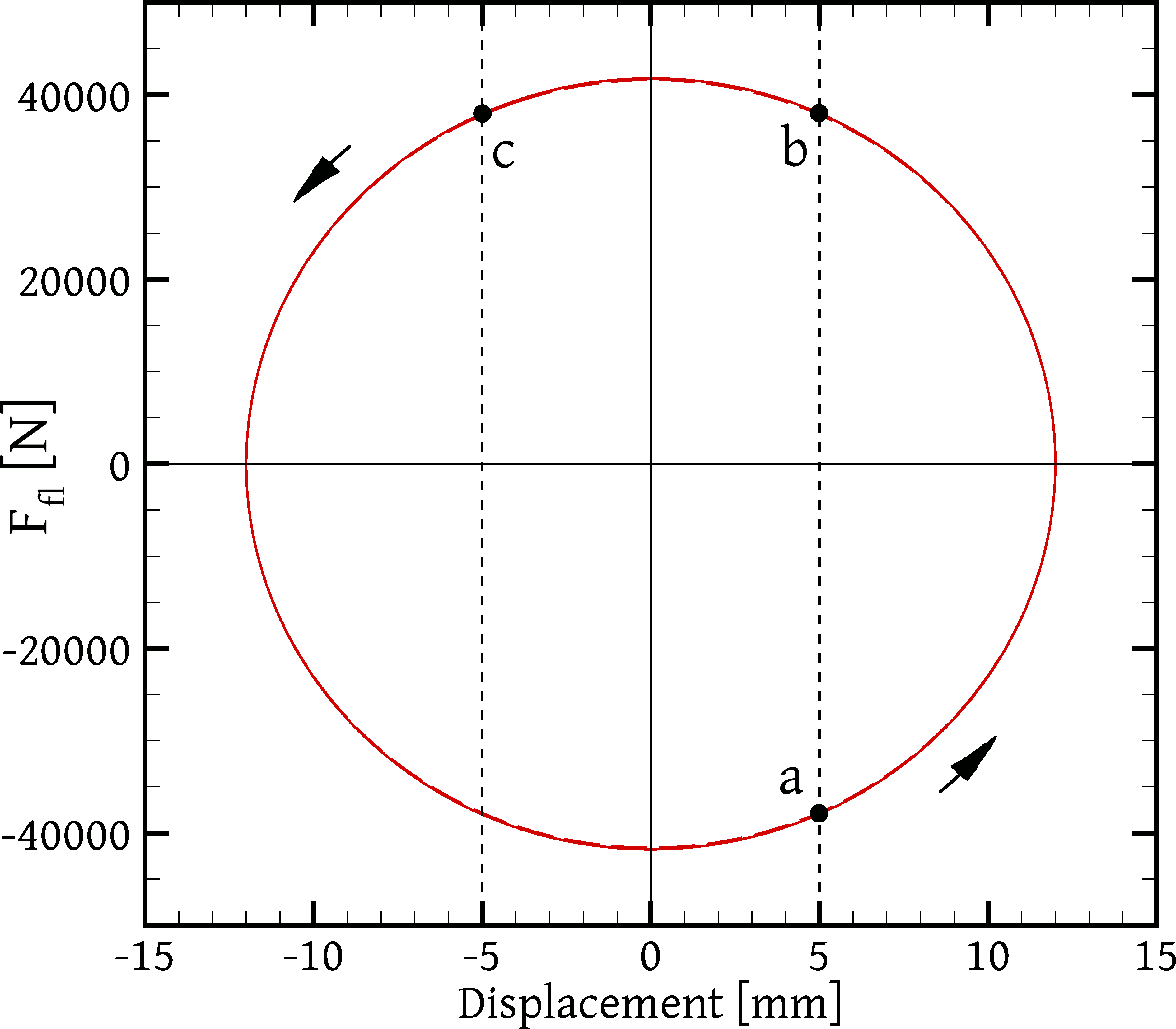}
        \caption{N-100, 32 \si{Hz} (0.41\%, 0.08\%)}
        \label{sfig: validation Newtonian}
    \end{subfigure}
    \begin{subfigure}[b]{0.32\textwidth}
        \centering
        \includegraphics[width=0.95\linewidth]{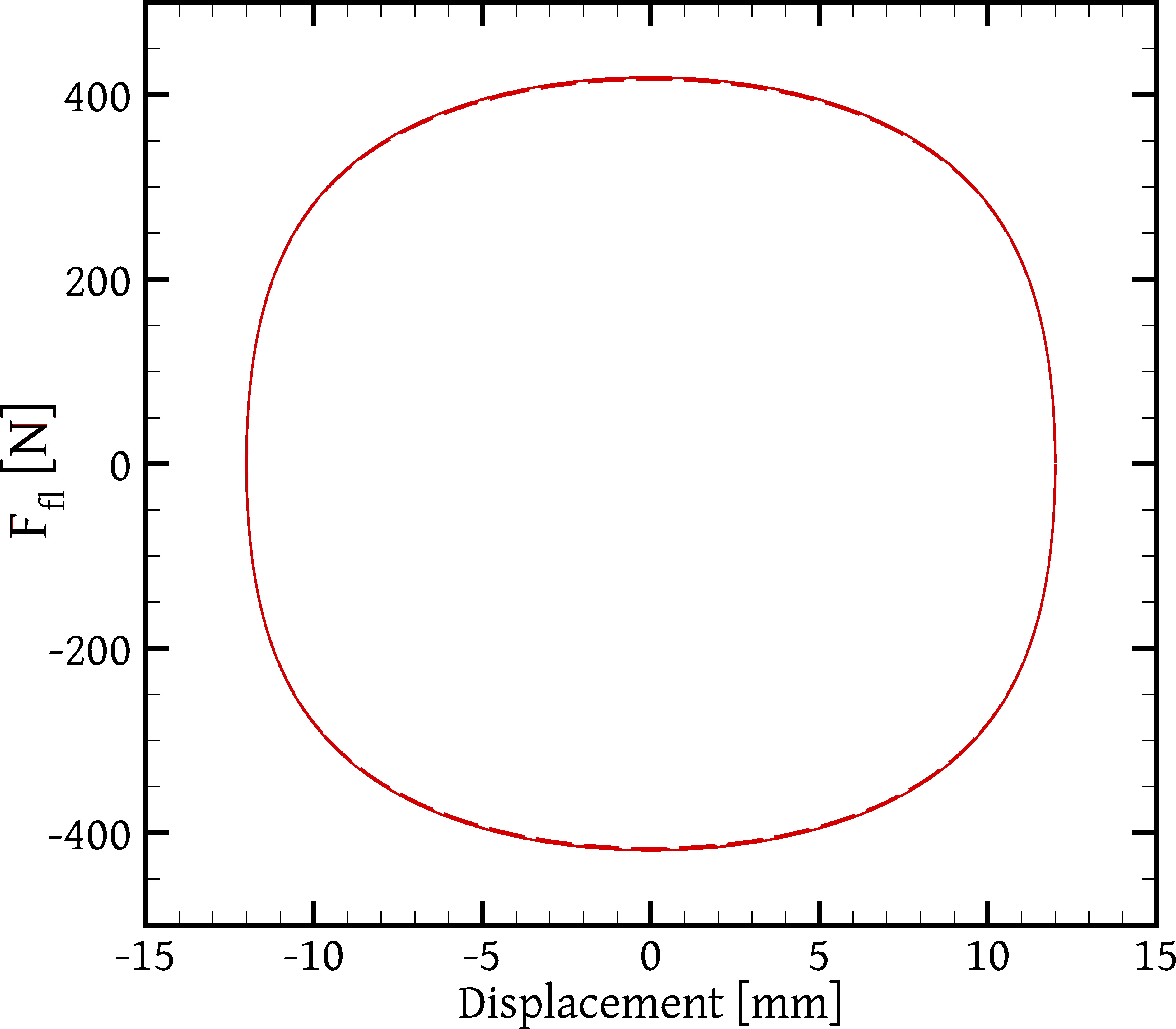}
        \caption{CY-100, 0.5 \si{Hz} (0.59\%, 0.14\%)}
        \label{sfig: validation CY-100 f=0.5}
    \end{subfigure}
    \begin{subfigure}[b]{0.32\textwidth}
        \centering
        \includegraphics[width=0.95\linewidth]{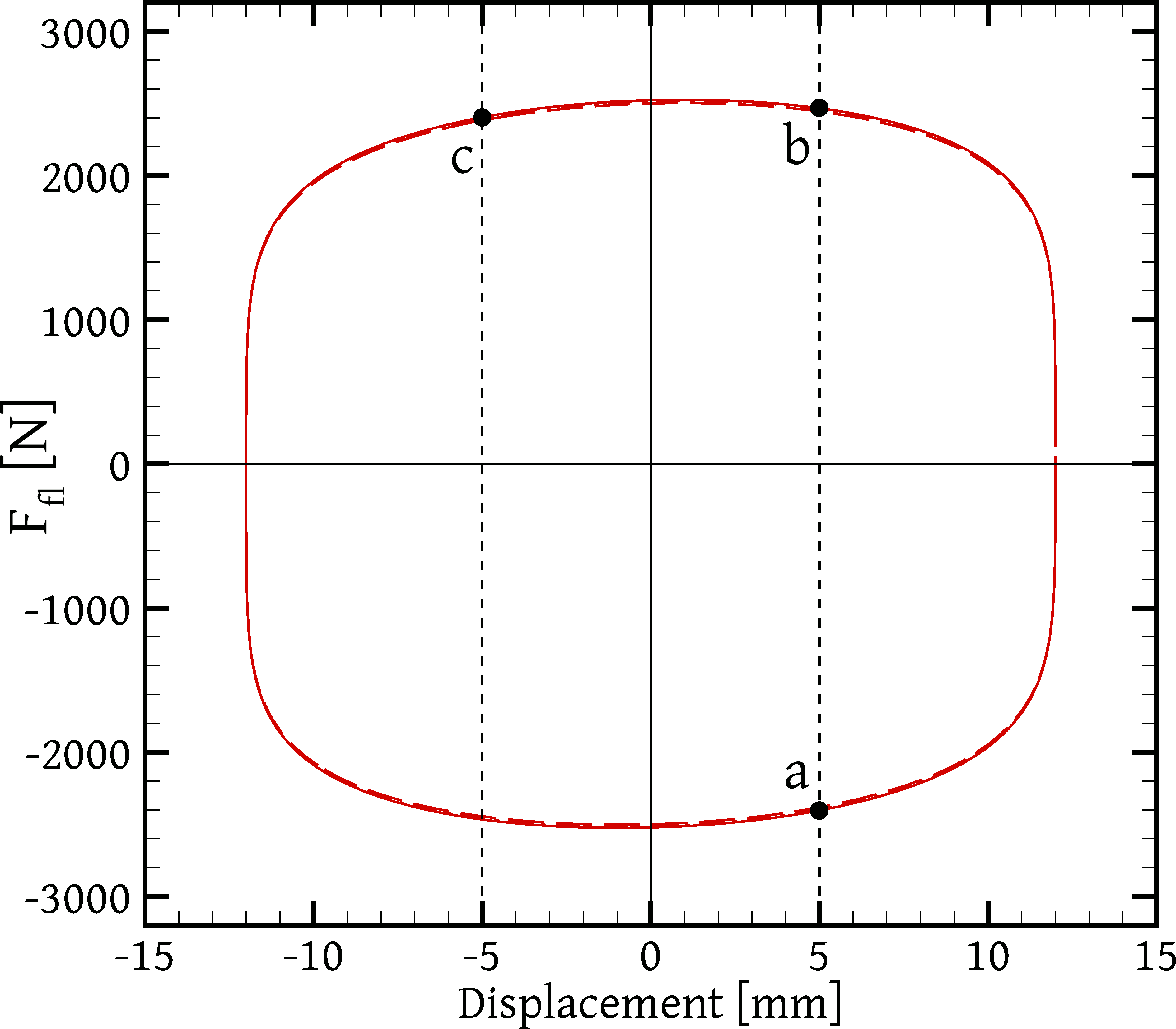}
        \caption{CY-100, 32 \si{Hz} (0.90\%, 0.23\%)}
        \label{sfig: validation CY-100 f=32}
    \end{subfigure}

\vspace{0.2cm}    

    \begin{subfigure}[b]{0.32\textwidth}
        \centering
        \includegraphics[width=0.95\linewidth]{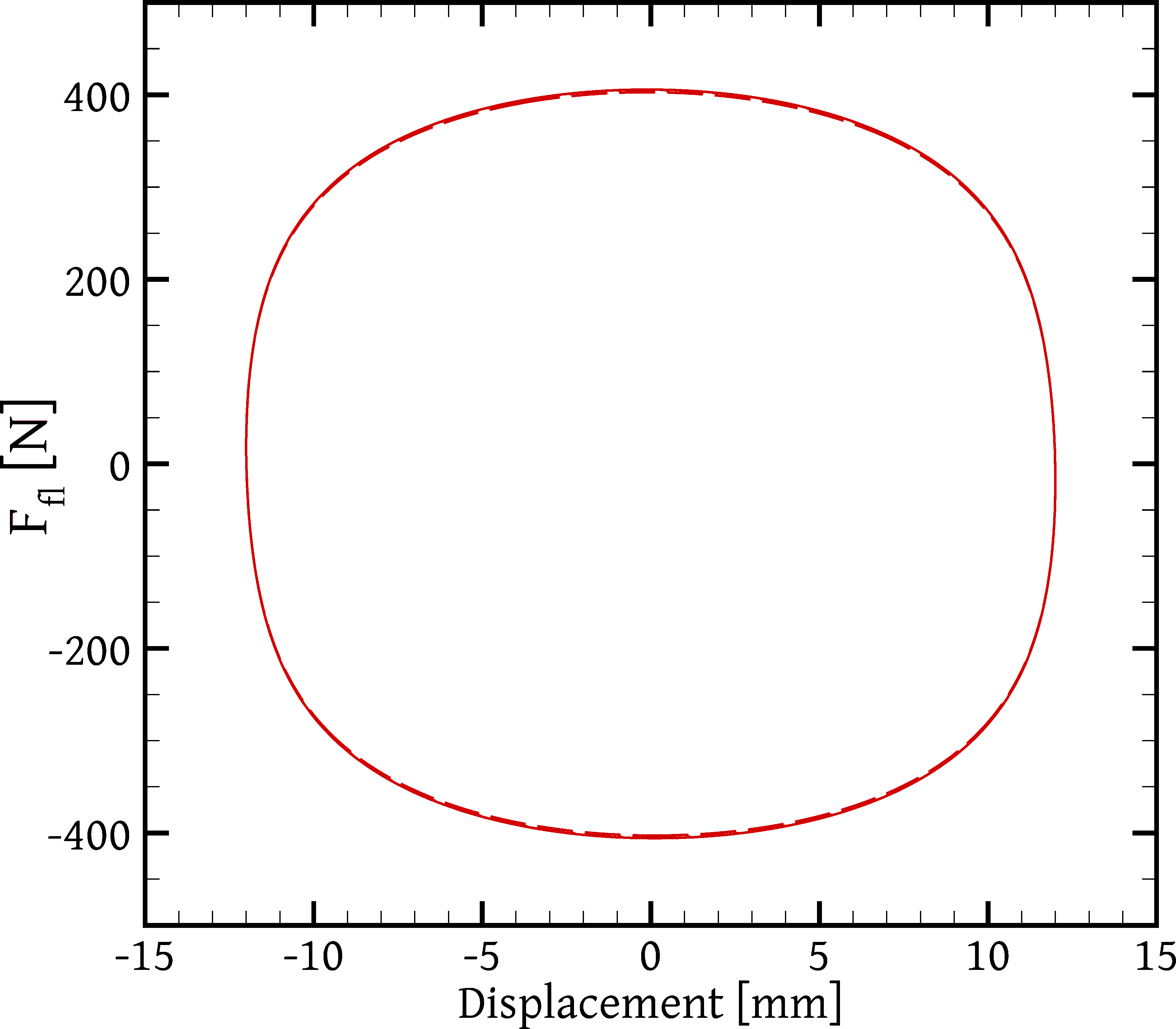}
        \caption{lPTT-100, 0.5 \si{Hz} (0.72\%, 0.14\%)}
        \label{sfig: validation lPTT-100 f=0.5}
    \end{subfigure}
    \begin{subfigure}[b]{0.32\textwidth}
        \centering
        \includegraphics[width=0.95\linewidth]{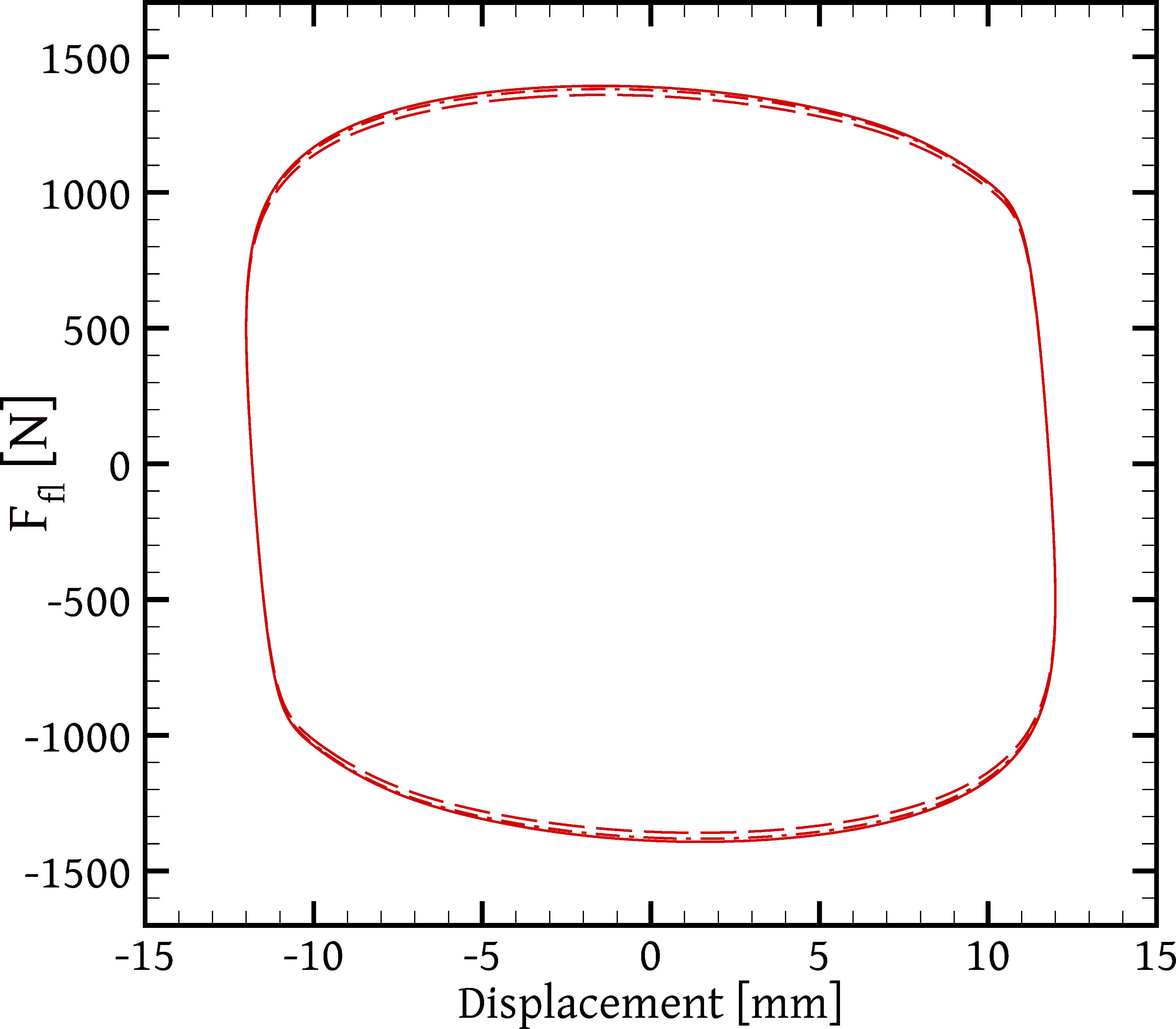}
        \caption{lPTT-100, 8 \si{Hz} (2.35\%, 0.81\%)}
        \label{sfig: validation lPTT-100 f=8}
    \end{subfigure}
    \begin{subfigure}[b]{0.32\textwidth}
        \centering
        \includegraphics[width=0.95\linewidth]{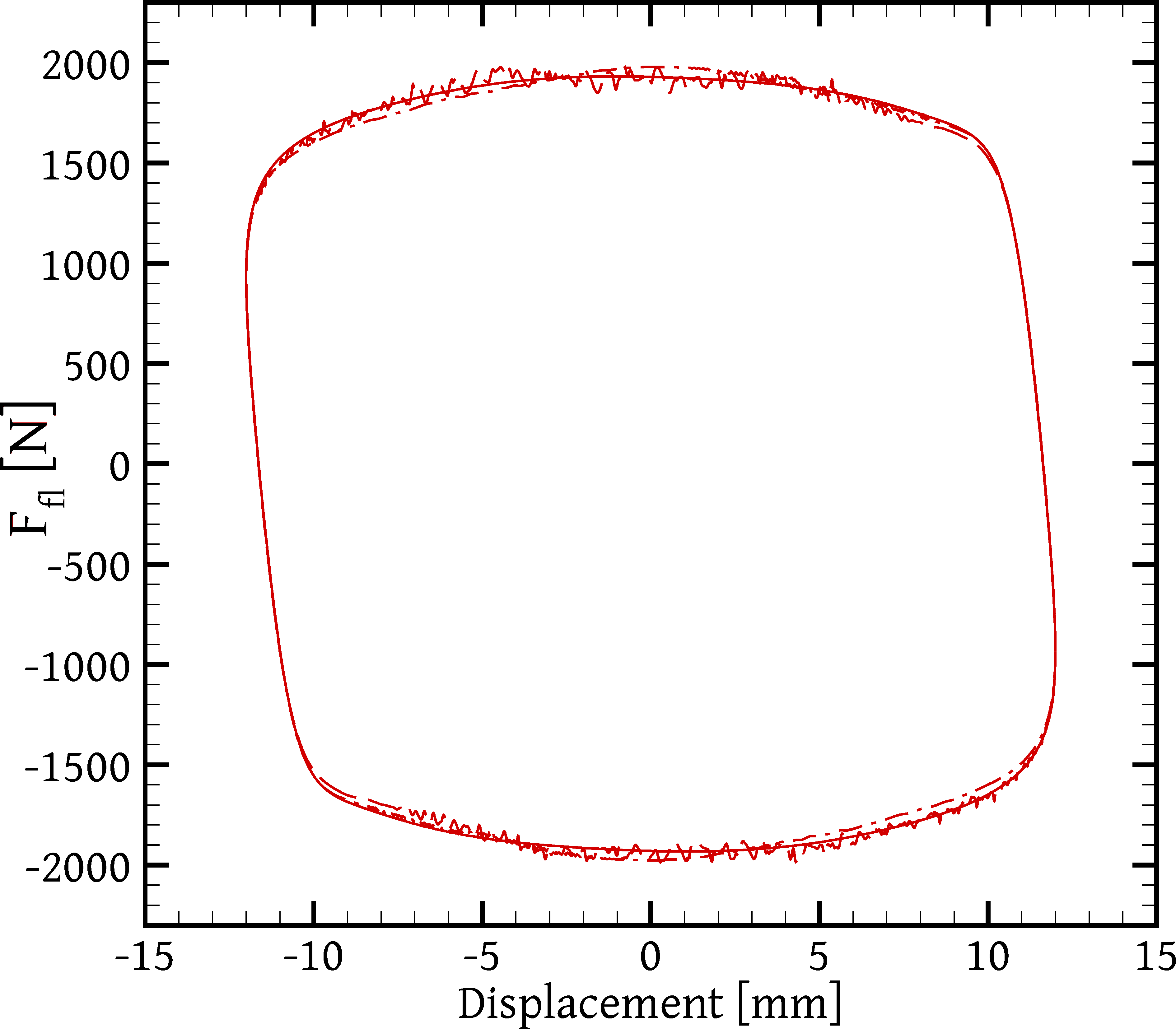}
        \caption{lPTT-100, 32 \si{Hz} (1.36\%, -2.63\%)}
        \label{sfig: validation lPTT-100 f=32}
    \end{subfigure}

\vspace{0.2cm}    

    \begin{subfigure}[b]{0.32\textwidth}
        \centering
        \includegraphics[width=0.95\linewidth]{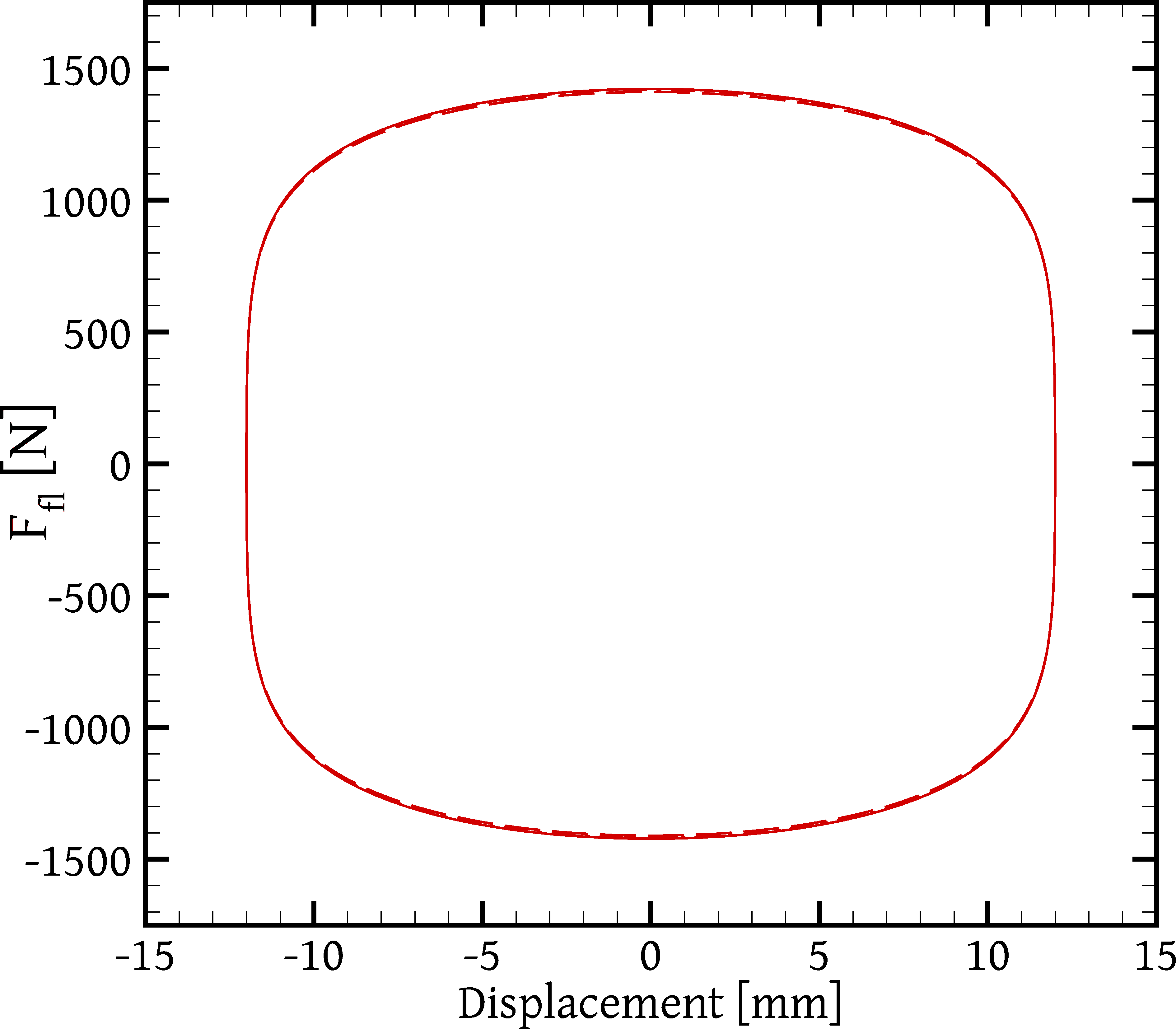}
        \caption{CY-500, 2 \si{Hz} (0.81\%, 0.20\%)}
        \label{sfig: validation CY-500 f=2}
    \end{subfigure}
    \begin{subfigure}[b]{0.32\textwidth}
        \centering
        \includegraphics[width=0.95\linewidth]{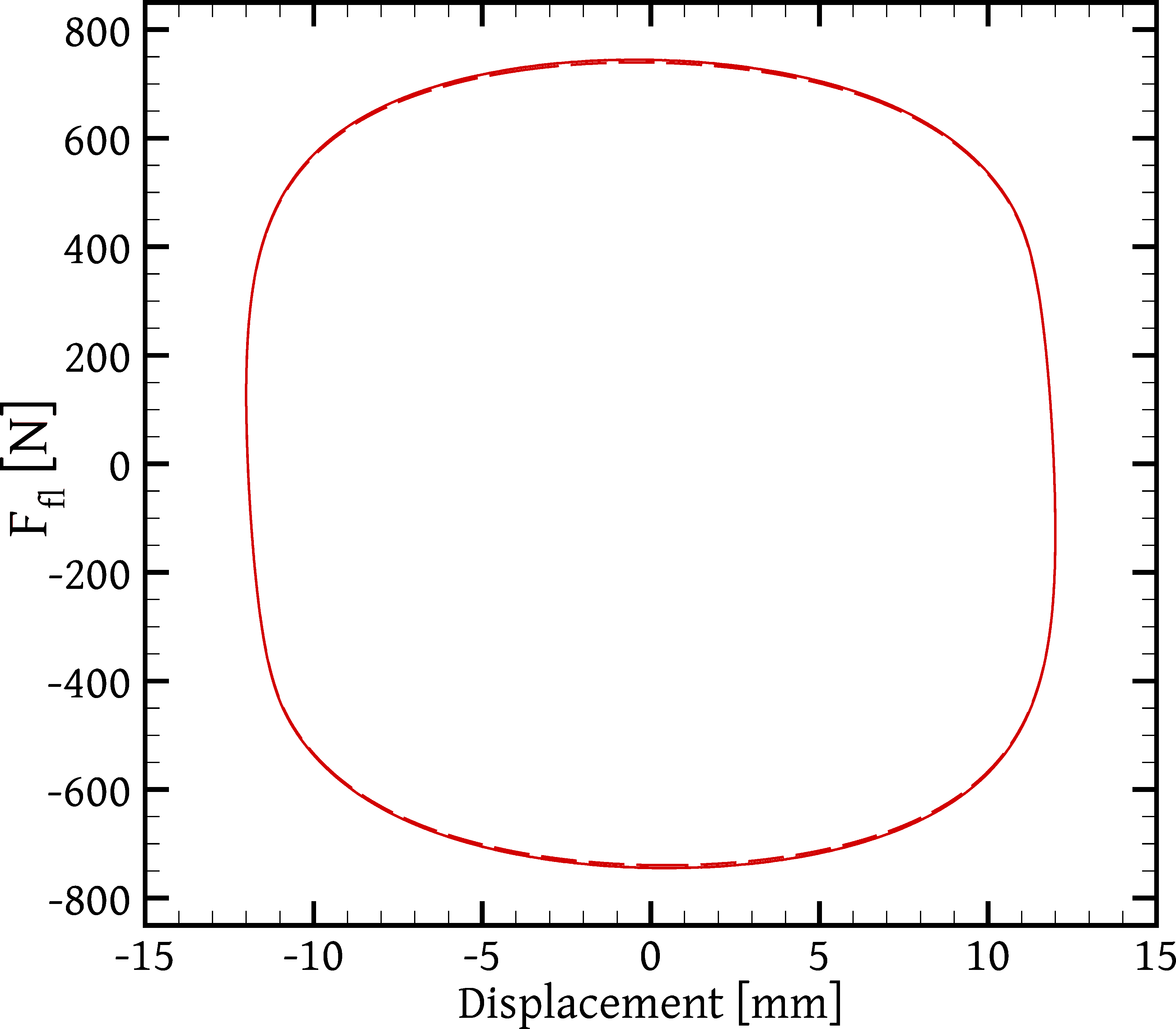}
        \caption{lPTT-500, 0.5 \si{Hz} (0.67\%, -0.03\%)}
        \label{sfig: validation lPTT-500 f=0.5}
    \end{subfigure}
    \begin{subfigure}[b]{0.32\textwidth}
        \centering
        \includegraphics[width=0.95\linewidth]{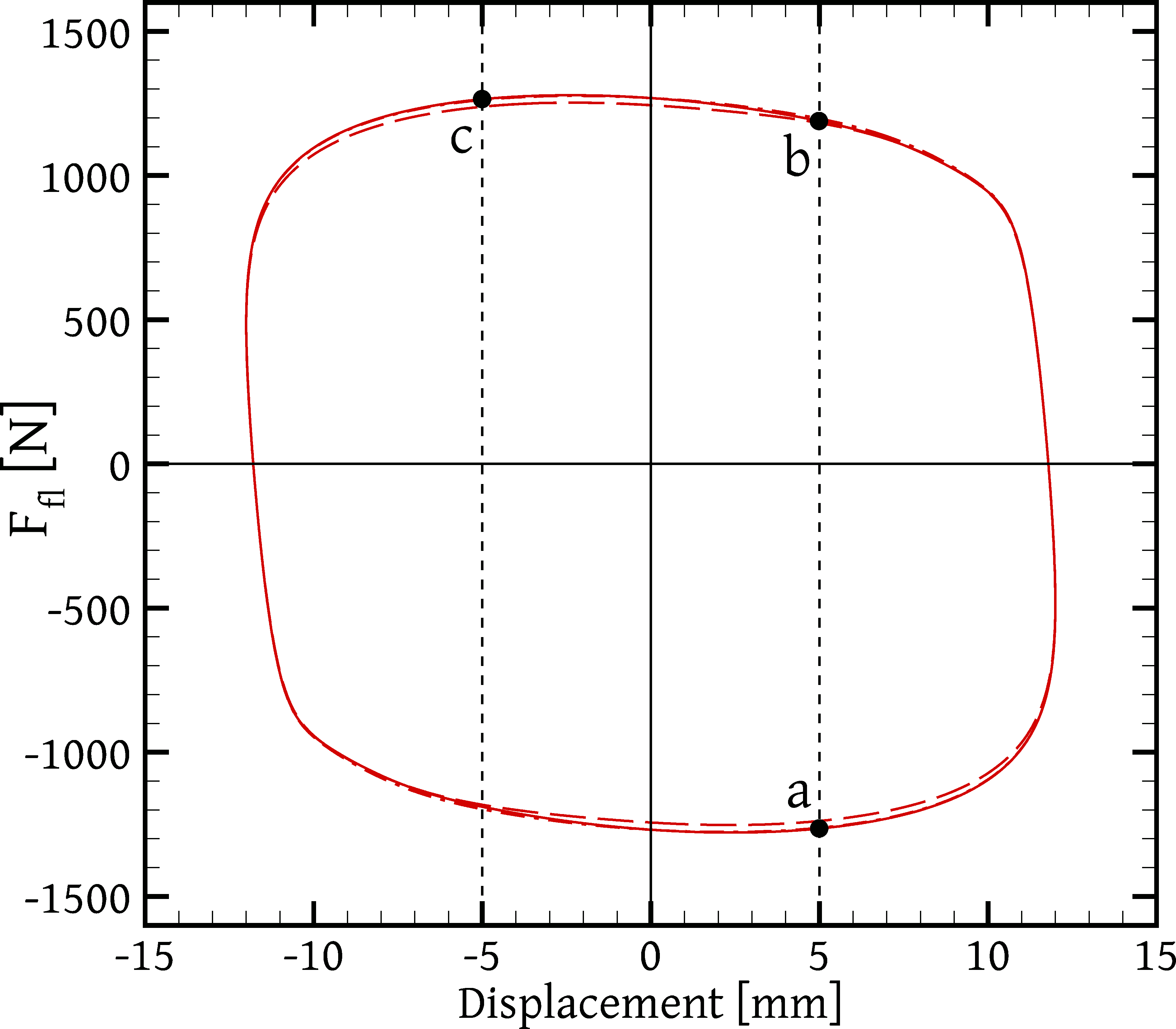}
        \caption{lPTT-500, 2 \si{Hz} (1.95\%, -0.01\%)}
        \label{sfig: validation lPTT-500 f=2}
    \end{subfigure}
    
\caption{Diagrams of the force $F_{fl}$, Eq.\ \eqref{eq: fluid force}, versus displacement of the 
piston midpoint compared to the $x = 0$ position, for various fluids and frequencies. Dashed lines: 
period 2, calculated on grid 1; dash-dot lines: period 4, calculated on grid 2; continuous lines: 
period 6, calculated on grid 3. The numbers in parentheses are the percentage difference in force 
at zero displacement between grid 3 and either grid 1 (first number) or grid 2 (second number). The 
loops are traversed in the anticlockwise sense as time progresses.}
  \label{fig: validation}
\end{figure}

Each damper flow simulation had a total duration of six periods, of which the first two were 
calculated on grid 1, the following two on grid 2, and the last two on grid 3 (Table \ref{table: 
grids}). To ensure that the last (sixth) period represents the periodic state, we compared force 
versus shaft displacement diagrams for periods 2, 4 and 6, with indicative results shown in Fig.\ 
\ref{fig: validation}. As can be seen from the figures, the force changes very little between 
periods 2, 4 and 6 which suggests that the periodic state is reached quickly, and also that grid 
convergence has been achieved on grid 3 (because periods 2, 4 and 6 are calculated on different 
grids). We note that as viscoelasticity ($\Wei$) increases the force difference between different 
grids grows, which shows that increasing viscoelasticity necessitates the use of finer grids to 
maintain a given level of accuracy. The results suggest that the error on the finest grid is in 
each case less than 1 \%. The force calculated on grid 2 for the lPTT-100 fluid at $f = 32$ \si{Hz} 
(Fig.\ \ref{sfig: validation lPTT-100 f=32}) has a component that oscillates in time (which is 
responsible for the relatively large zero-displacement force difference of 2.63\% between grids 2 
and 3) but this vanishes on the finest grid.

The time step used in the simulations was $\Delta t = T / 4000$ for the CY and Newtonian fluids and 
$\Delta t = T / 8000$ for the l-PTT fluids. Since the Reynolds numbers are small, the inertial terms 
in the momentum Eq.\ \eqref{eq: momentum nd}, which include the temporal term, are not very 
significant and thus for the Newtonian and CY simulations the flow is quasi-steady 
\cite{Syrakos_2016} (or nearly so) and the choice of time step is not crucial (the chosen $\Delta t 
= T/4000$ is most probably unnecessarily small). On the other hand, for the viscoelastic simulations 
a temporal term appears also in the constitutive Eq.\ \eqref{eq: constitutive lPTT nd} which, 
according to the values of $\Wei$ and $\Deb$ listed in Table \ref{table: operating conditions}, may 
be significant; hence the small chosen time step $\Delta t = T / 8000$ for the viscoelastic 
simulations. This is 20 times smaller than that used in our previous study \cite{Syrakos_2016}, and 
was found to offer adequate accuracy by tests where the time steps used on the coarse grids 2 and 1 
were, respectively, two ($\Delta t = T / 4000$) and four ($\Delta t = T / 2000$) times larger than 
this value, which was used on grid 3. For example, this procedure for the lPTT-100, $f$ = 8 \si{Hz} 
case yielded force differences at zero displacement of 1.85\% and 0.68\% between grids 3 and 1, and 
3 and 2, respectively. These differences are even smaller than those reported in Fig.\ \ref{sfig: 
validation lPTT-100 f=8}, which suggests that the spatial and temporal discretisation contributions 
to the discretisation error are of opposite sign. The present choice of time step, which is coupled 
to the oscillation period, leads to variable resolutions of the relaxation time depending on the 
oscillation frequency, namely to ratios $\lambda / \Delta t$ of 40, 160, 640 and 2560 at $f$ = 0.5, 
2, 8 and 32 \si{Hz}, respectively, for the lPTT-100 fluid and of 240 and 960 at $f$ = 0.5 and 2 
\si{Hz}, respectively, for the lPTT-500 fluid. At low frequencies this ratio is small, but the 
$\Wei$ and $\Deb$ numbers are small as well (Table \ref{table: operating conditions}) and the 
importance of viscoelasticity diminishes, as the results of Sec.\ \ref{sec: results} will confirm, 
so that a high resolution of the viscoelastic phenomena is not necessary in order to accurately 
simulate the flow. Finally, we note that the Courant-Friedrichs-Lewy number (CFL) on the finest grid 
can be approximated as $U \Delta t/ (h/96)$ where $h/96$ is the grid spacing across the gap, which 
gives CFL values of 2.56 and 1.28 for the CY and l-PTT fluids, respectively (independent of the 
frequency $f$). The CFL number is used in explicit temporal discretisation schemes to control their 
stability; while the present method is implicit and a CFL $< 1$ is not a prerequisite for stability, 
usually a CFL number smaller than unity indicates adequate temporal resolution in relation to the 
spatial resolution. In \cite{Oliveira_2001}, where the same temporal discretisation scheme was used, 
good accuracy was reported with CFL numbers as high as 3--6.

\section{Results}
\label{sec: results}

In an experimental study, the damper under investigation would be subjected to a predetermined 
displacement history, and the required force that causes this displacement would be recorded using a 
load cell. This applied force, $F_{ap}$, causes the acceleration of the shaft-piston assemblage, and 
is counteracted by the force $F_{fl}$ exerted by the fluid on the shaft-piston and by the friction 
force $F_{fr}$ at the damper bearings and seals. Therefore, Newton's second law can be expressed as
%^b
\begin{equation} \label{eq: force balance}
 F_{ap} \;+\; F_{fl} \;+\; F_{fr} \;=\; M_p a_p
\end{equation}
%^a
where $M_p$ is the mass of the shaft-piston assemblage and $a_p$ is its acceleration. The latter is 
obtained by differentiating Eq.\ \eqref{eq: piston velocity}:
%^b
\begin{equation} \label{eq: shaft acceleration}
 a_p(t) \;=\; -\alpha \omega^2 \cos(\omega t) \;=\; - A_p \cos(\omega t)
\end{equation}
%^a
where $A_p = \alpha \omega^2$ is the maximum acceleration, occurring when the piston reaches its 
extreme positions.

The present numerical simulations allow prediction of the reaction force exerted by the fluid on 
the piston and shaft, by integrating the stress and pressure over the piston surface and on the 
part of the shaft's surface that is immersed in the fluid:
%^b
\begin{equation} \label{eq: fluid force}
 F_{fl} \;=\; \iint_S \left( -p \vf{n} \;+\; \vf{n} \cdot \tf{\tau} \right) \cdot \vf{e}_x \, 
              \mathrm{d}s
\end{equation}
%^a
where $S$ is the shaft-piston surface, $\mathrm{d}s$ is an infinitesimal element of that surface, 
and $\vf{n}$ is the unit vector normal to that surface at each point, directed towards the fluid. 
It is mostly on this force that the present study focuses. If the friction force $F_{fr}$ and the 
piston inertia term $M_p a_p$ are assumed negligible, then Eq.\ \eqref{eq: force balance} reduces to
%^b
\begin{equation}
 F_{ap} \approx -F_{fl}
\end{equation}
%^a
Therefore, the plots of $F_{fl}$ presented in this study, such as in Fig.\ \ref{fig: validation}, 
resemble mirror images of experimentally derived plots of $F_{ap}$ provided in the literature. 
However, in Section \ref{ssec: results; Fap} plots of $F_{ap}$ itself will also be presented, 
derived from Eq.\ \eqref{eq: force balance} with the shaft inertia included but $F_{fr}$ neglected.

Customarily, the damper behaviour is explored through the examination of plots of force versus 
piston displacement and piston velocity, which are easily obtainable in experimental studies. The 
present study includes such plots but is not restricted to these. One of the advantages of 
numerical simulation is that it provides estimates of all the flow variables at every point in the 
domain and at every time instance, thus providing a complete picture of the flow. This is exploited 
in order to go beyond the force-displacement/velocity plots and obtain additional insight. Since we 
are interested in the periodic state, the plots presented in this study correspond to the sixth 
oscillation period simulated ($t \in [5T, 6T]$), unless otherwise stated, in order to avoid any 
initial transient phenomena (although it was noticed that the periodic state is attained fairly 
quickly, with transient phenomena not persisting beyond the first period).

The area enclosed by the loops of force-displacement plots is equal to the energy absorbed by 
the damper in a single cycle and converted into heat by viscous action in the fluid\footnote{The 
instantaneous energy balance for the damper fluid includes the rate of work of the force $F_{fl}$, 
the rate of viscous dissipation of energy into heat, and the rate of change of energy stored in the 
fluid in the form of elastic energy. However, considering a full period of oscillation, if the flow 
has reached the periodic state, the stored elastic energy at time $t$ is exactly equal to that at 
time $t + T$. Therefore, the work of $F_{fl}$ during a complete cycle is equal to the amount of 
energy dissipated to heat by viscous action during the same period.}. A look at Fig.\ \ref{fig: 
validation} shows that the elliptical shape of the Newtonian loop \ref{sfig: validation Newtonian} 
is distorted towards a rectangular shape when the fluid is non-Newtonian, stretched in a direction 
that forms either a positive (counterclockwise) angle, for CY fluids (e.g.\ Fig.\ \ref{sfig: 
validation CY-100 f=32}), or a negative (clockwise) angle, for PTT fluids (e.g.\ Fig.\ \ref{sfig: 
validation lPTT-500 f=2}), compared to the horizontal axis. The non-Newtonian characteristics become 
more prominent as the frequency and/or zero-shear viscosity ($\eta_0$) increase. In 
force-displacement diagrams, this tilting of the loop at an oblique angle is a definite sign of 
\textit{hysteresis}, i.e.\ of dependence of the current state of the flow on its history.

\begin{figure}[!tb]
    \centering

    \begin{subfigure}[b]{0.49\textwidth}
        \centering
        \includegraphics[width=0.65\linewidth]{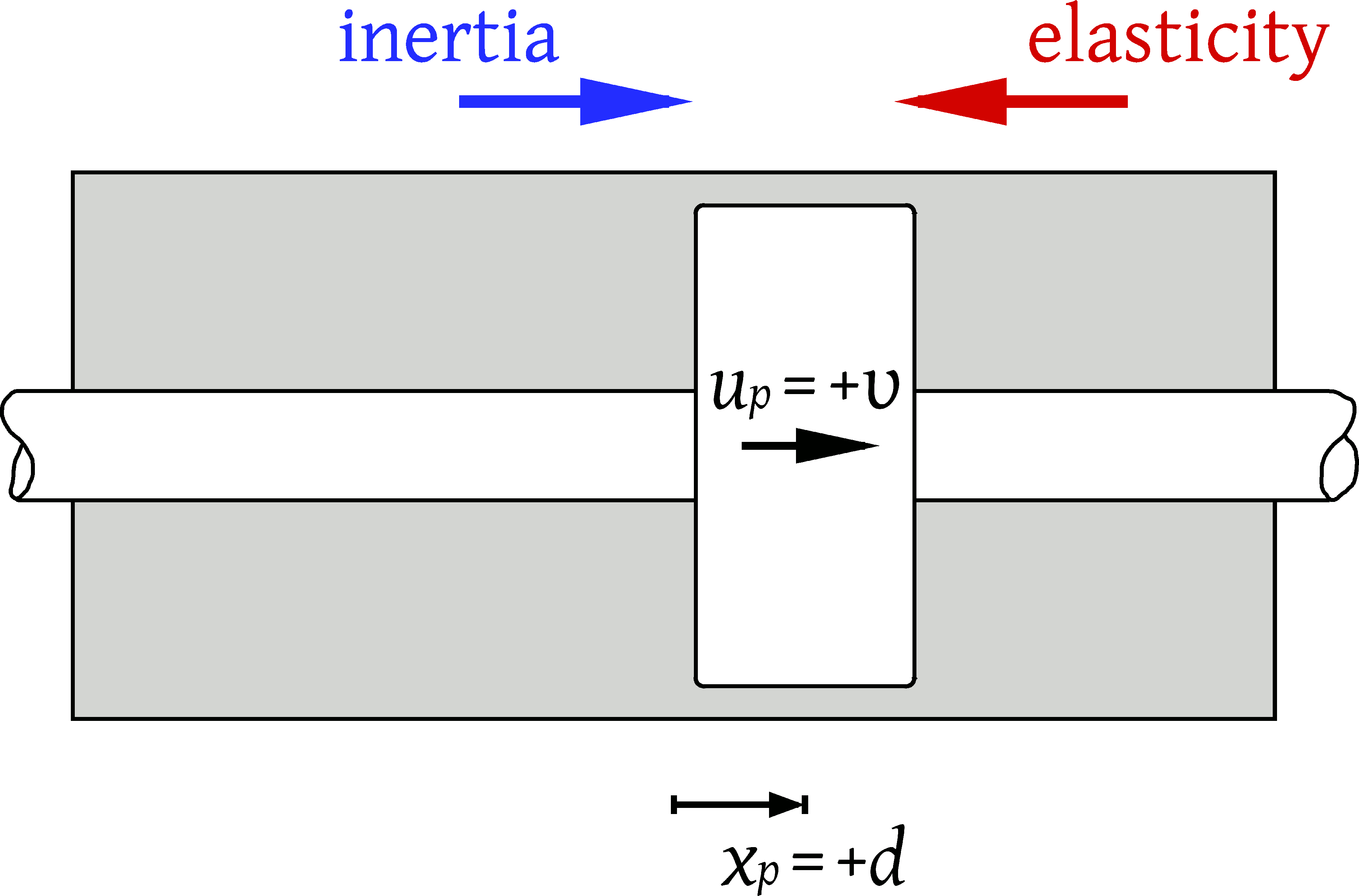}
        \caption{piston approaching the extreme right}
        \label{sfig: sketch piston Dp Up}
    \end{subfigure}

    \vspace{0.5cm}
    
    \begin{subfigure}[b]{0.49\textwidth}
        \centering
        \includegraphics[width=0.65\linewidth]{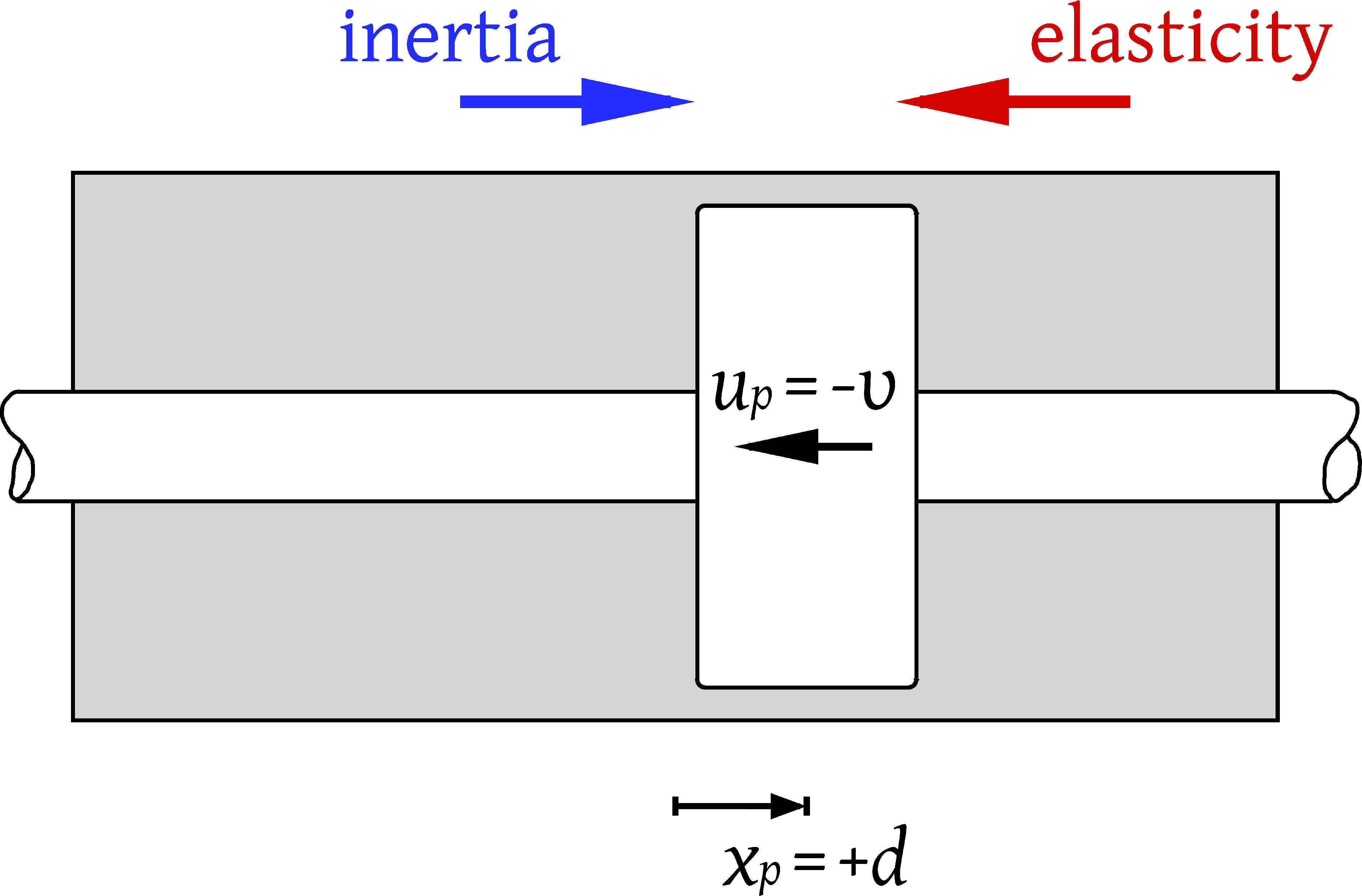}
        \caption{piston leaving the extreme right}
        \label{sfig: sketch piston Dp Um}
    \end{subfigure}

    \vspace{0.5cm}
    
    \begin{subfigure}[b]{0.49\textwidth}
        \centering
        \includegraphics[width=0.65\linewidth]{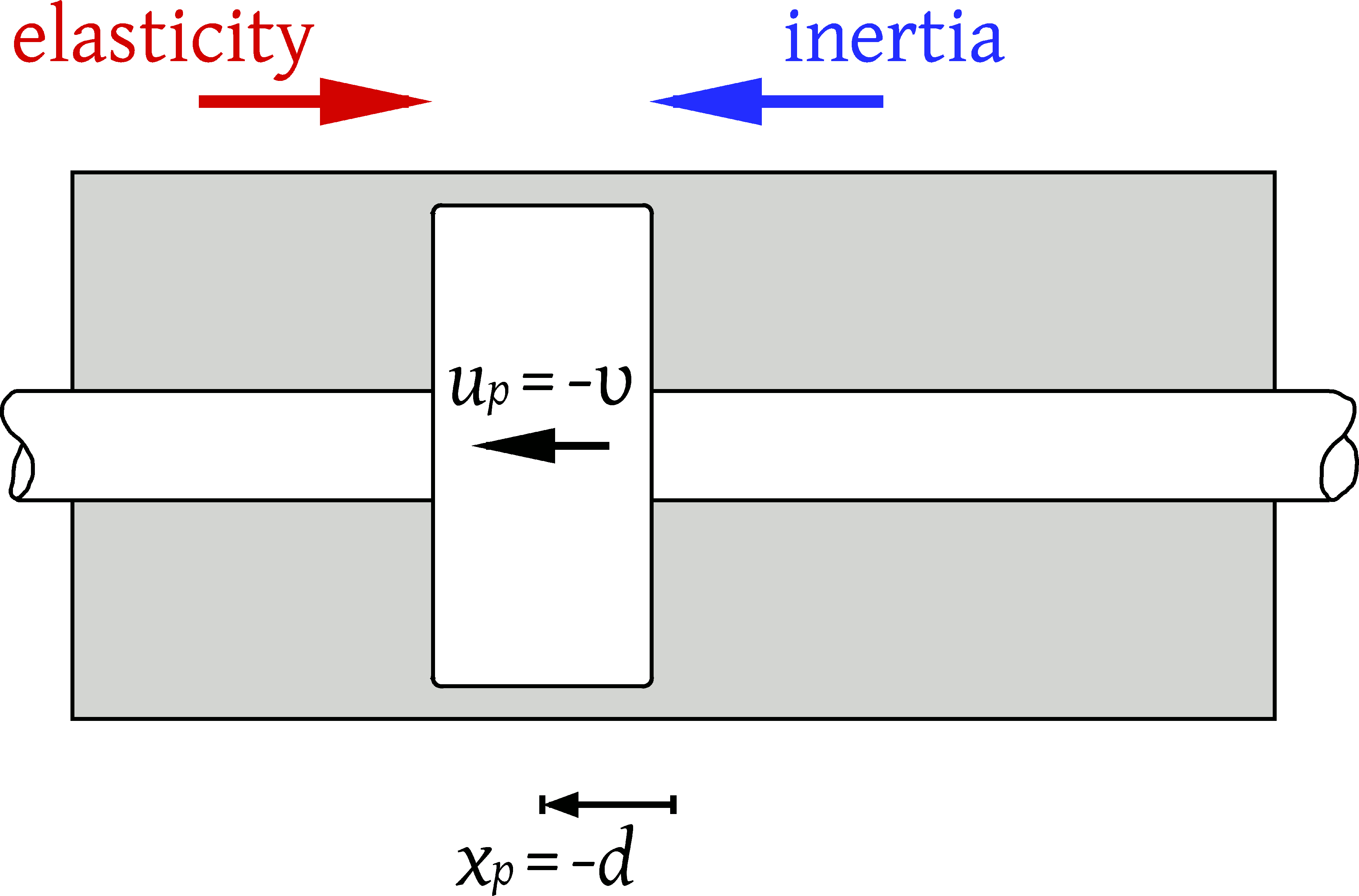}
        \caption{piston approaching the extreme left}
        \label{sfig: sketch piston Dm Um}
    \end{subfigure}

\caption{The damper at three different ``symmetric'' time instances: At \subref{sfig: sketch piston 
Dp Up} the piston is displaced towards the right and is moving towards the right. Later, at 
\subref{sfig: sketch piston Dp Um}, the piston has returned to the same position ($x_p(\mathrm{b}) 
= x_p(\mathrm{a})$) but is moving towards the left ($u_p(\mathrm{b}) = -u_p(\mathrm{a})$). Later
still, at \subref{sfig: sketch piston Dm Um}, the piston has moved to the opposite side 
($x_p(\mathrm{c}) = -x_p(\mathrm{b})$) and is moving towards the left ($u_p(\mathrm{c}) = 
u_p(\mathrm{b})$). The directions of the inertial and elastic components of the fluid force on the 
piston and shaft are shown in each case.}
  \label{fig: damper states}
\end{figure}

When the Reynolds number in Eq.\ \eqref{eq: momentum nd}, and, for PTT fluids, the $\Wei$ and 
$\Deb$ numbers in Eq.\ \eqref{eq: constitutive lPTT nd}, are very small then the corresponding 
terms become negligible and the governing equations become elliptic and quasi-steady state, so that 
the flow is not affected by its own history but is determined solely by the instantaneous boundary 
conditions. In this case, symmetry between two instantaneous damper states, such as any pair of the 
states (a), (b) and (c) depicted in Fig.\ \ref{fig: damper states}, results in identical force 
magnitudes. Such is the flow of the N-100 fluid even at the highest frequency of 32 \si{Hz} (Fig.\ 
\ref{sfig: validation Newtonian}) because of the very large viscosity of the fluid. In Fig.\ 
\ref{sfig: validation Newtonian} the three states (a), (b) and (c) are marked (the displacement $d$ 
of Fig.\ \ref{fig: damper states} is arbitrarily set equal to 5 \si{mm} in Fig.\ \ref{fig: 
validation}) and it can be seen that $|F_{fl}(\mathrm{a})| = |F_{fl}(\mathrm{b})| = 
|F_{fl}(\mathrm{c})|$. The N-100 loop is symmetric with respect to both the force = 0 and 
displacement = 0 axes.

On the other hand, for the CY and PTT fluids as the frequency increases the transient terms in 
Eqs.\ \eqref{eq: momentum nd} and \eqref{eq: constitutive lPTT nd} become more important and the 
flow history comes into play. Now, although damper states (a) and (b) are symmetric, they have 
different histories and therefore $|F_{fl}(\mathrm{a})| \neq |F_{fl}(\mathrm{b})|$ in both Fig.\ 
\ref{sfig: validation CY-100 f=32} and Fig.\ \ref{sfig: validation lPTT-500 f=2}. However, states 
(a) and (c) are not only symmetric, but the histories of the flows leading up to these states are 
also symmetric; therefore, $|F_{fl}(\mathrm{a})| = |F_{fl}(\mathrm{c})|$. Half a period suffices to 
describe the whole loop (the flow fields at times $t$ and $t + T/2$ are symmetric).

\begin{figure}[tb]
    \centering
    \begin{subfigure}[b]{0.49\textwidth}
        \centering
        \includegraphics[width=0.95\linewidth]{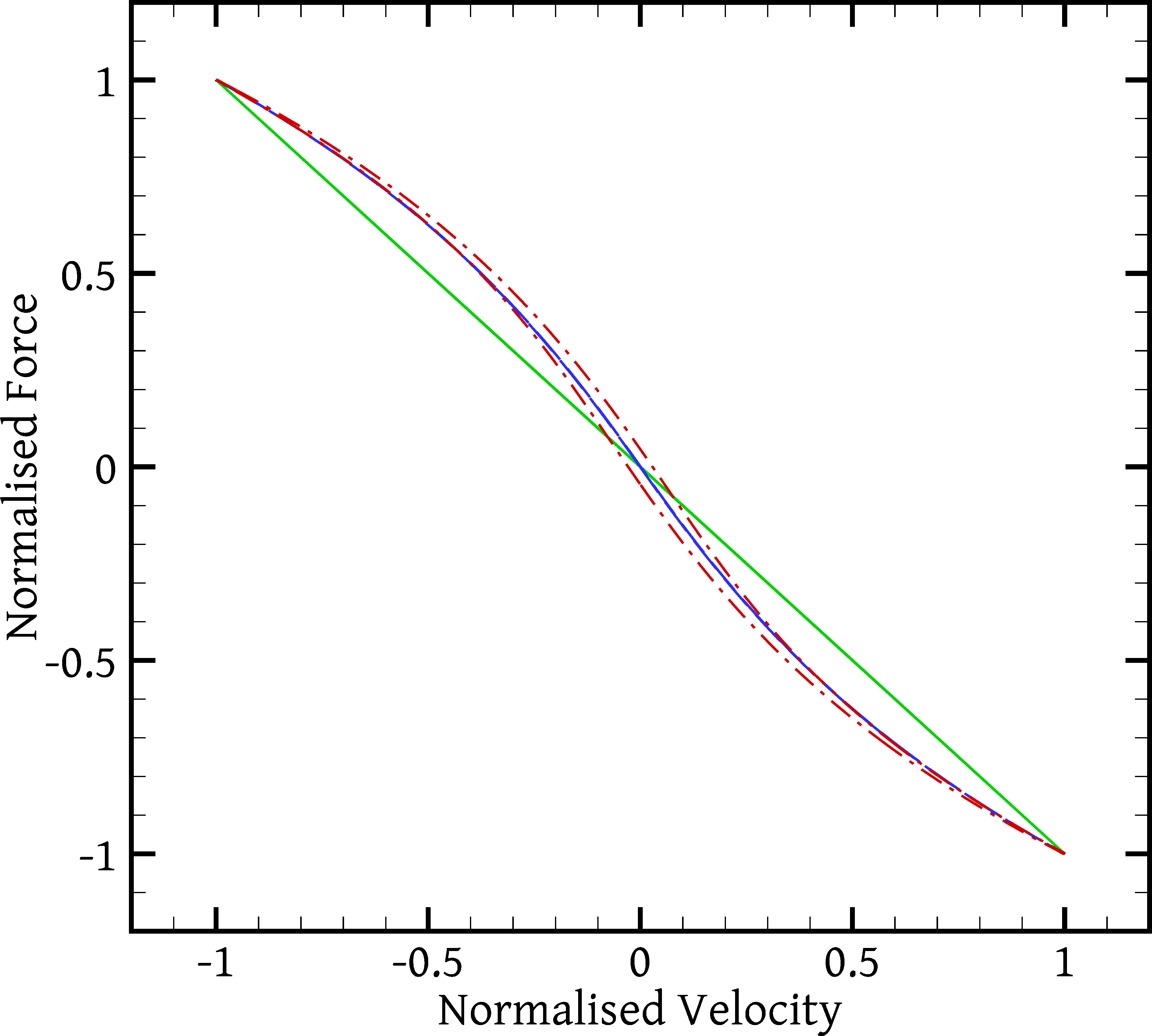}
        \caption{$f$ = 0.5 \si{Hz}}
        \label{sfig: force velocity 100 f=0.5 nd}
    \end{subfigure}
    \begin{subfigure}[b]{0.49\textwidth}
        \centering
        \includegraphics[width=0.95\linewidth]{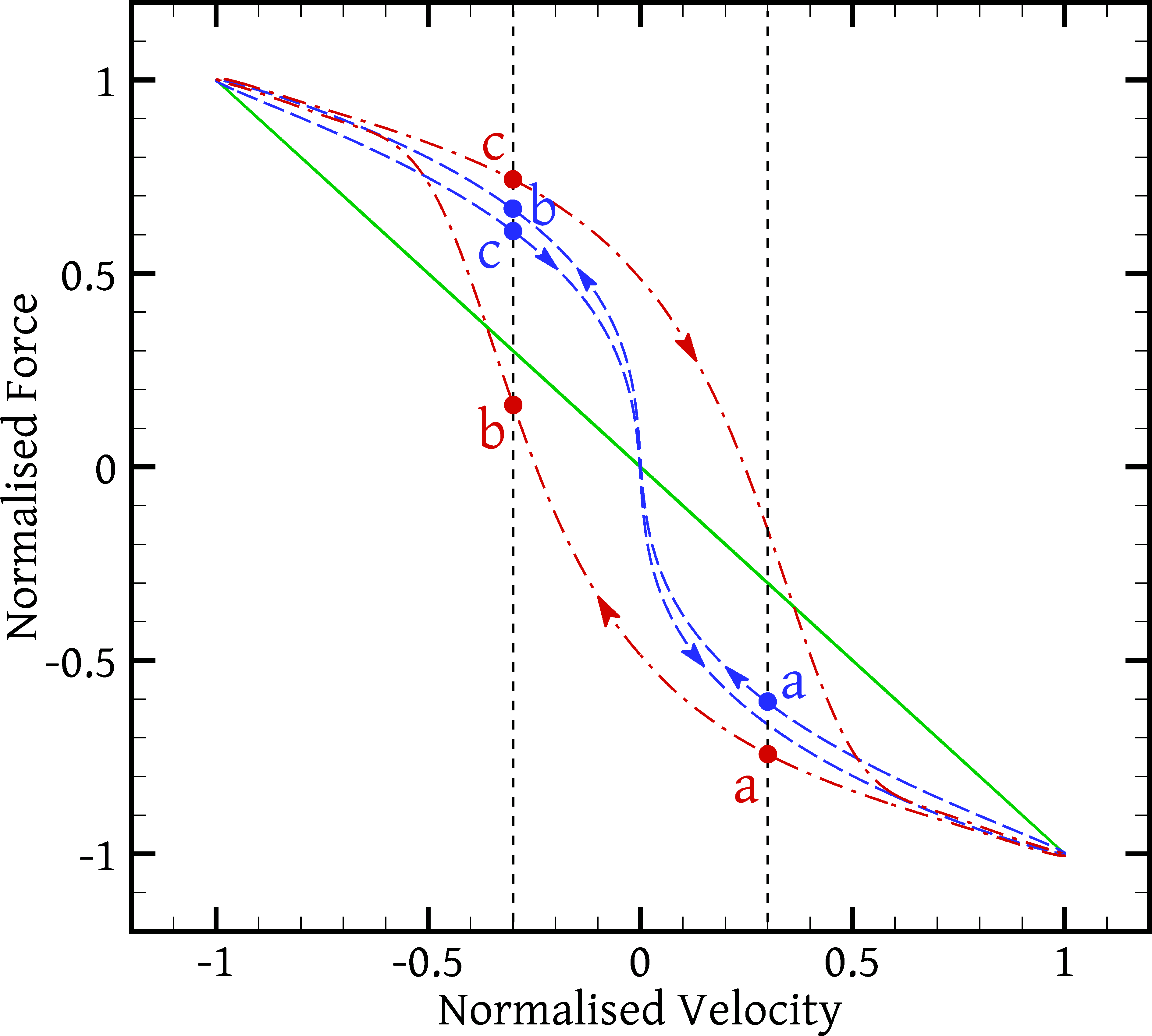}
        \caption{$f$ = 32 \si{Hz}}
        \label{sfig: force velocity 100 f=32 nd}
    \end{subfigure}
    
\caption{Force $F_{fl}$ normalised by its maximum value, against piston velocity normalised by 
its maximum value ($u_p / U_p$, Eq.\ \eqref{eq: piston velocity}), for the N-100 (solid green 
line), CY-100 (dashed blue line) and lPTT-100 (dash-dot red line) fluids at $f$ = 0.5 \si{Hz} 
(left) 
and 32 \si{Hz} (right). The CY-100 loops are traversed in the counterclockwise sense whereas the 
lPTT-100 loops in the clockwise sense.}
  \label{fig: ND force - ND velocity 100}
\end{figure}

It is customary to also plot the force versus the piston velocity, as in Fig.\ \ref{fig: ND force - 
ND velocity 100}. We shall use the normalised velocity $u_p / U_p$ (Eq.\ \eqref{eq: piston 
velocity}) for such plots, which allows to plot the loops of different frequencies on the same 
diagram (the maximum piston velocities $U_p$ for each frequency are listed in Table \ref{table: 
operating conditions}). The effects of the non-Newtonian character of the fluids are even more 
pronounced on force-velocity plots, especially at higher frequencies (Fig.\ \ref{sfig: force 
velocity 100 f=32 nd}) whereas at low frequencies the behaviour deviates mildly from the Newtonian 
one (Fig.\ \ref{sfig: force velocity 100 f=0.5 nd}). In force-velocity plots hysteresis is manifest 
in the plot having the form of a loop rather than of a single curve. This means that, by drawing a 
vertical line on the graph, for a given piston velocity there are two different values of force: 
one value is exhibited when the piston is accelerating, and a different value when it is 
decelerating. This can be observed in Fig.\ \ref{sfig: force velocity 100 f=32 nd} where 
$F_{fl}(\mathrm{b}) \neq F_{fl}(\mathrm{c})$ for both the CY and PTT fluids, whereas 
$F_{fl}(\mathrm{b}) = F_{fl}(\mathrm{c})$ for the Newtonian fluid (the states (a), (b), (c) have 
been arbitrarily set to correspond to a normalised velocity of $\pm 0.3$).

The force-displacement and force-velocity plots are very reminiscent of the Lissajous-Bowditch 
plots obtained in Large Amplitude Oscillatory Shear (LAOS) experiments \cite{Hyun_2011}, i.e.\ 
plots of shear stress versus strain or rate-of-strain, respectively, which are used for the 
characterisation of soft materials and complex fluids. A brief description of LAOS is given in 
Appendix \ref{appendix: LAOS flow}. This similarity allows us to borrow some tools from LAOS 
theory in the following sections. However, it must be noted that there are also significant 
differences between LAOS and the present flow: whereas in LAOS all the material undergoes the same 
deformation simultaneously in a Couette-type flow, the damper flow is two-dimensional and mostly 
pressure driven, with the stress and deformation rate varying along both the gap length and width, 
while each fluid particle remains in the critical gap region for only a limited amount of time 
during each period, provided that the oscillation amplitude is not very small. 

The precise shapes of the force-displacement and force-velocity plots will be examined in greater 
detail in the sections that follow, but first a general description of the flow field inside the 
damper will be given.

\subsection{General description of the flow field}

\begin{figure}[tb]
    \centering
    
    \includegraphics[width=0.75\linewidth]{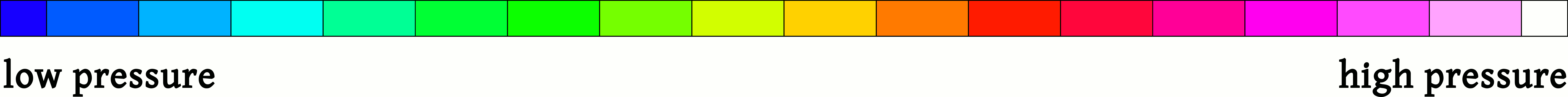}

\vspace{0.5cm}    

    \begin{subfigure}[b]{0.49\textwidth}
        \centering
        \includegraphics[width=0.95\linewidth]{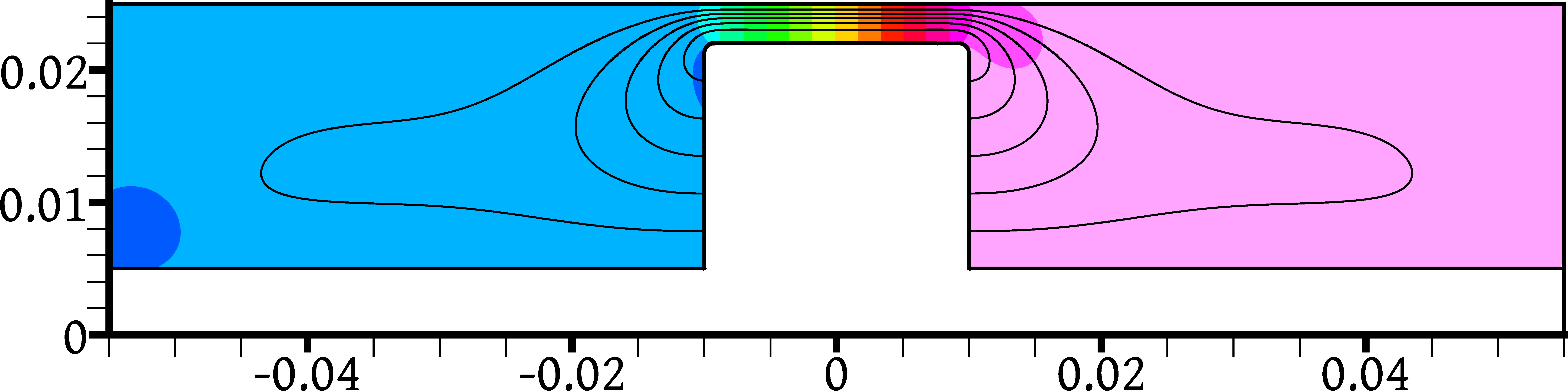}
        \caption{CY-100, 0.5 \si{Hz}; $\delta p$ = 18 \si{kPa}}
        \label{sfig: flow CY-100 f=0.5}
    \end{subfigure}
    \begin{subfigure}[b]{0.49\textwidth}
        \centering
        \includegraphics[width=0.95\linewidth]{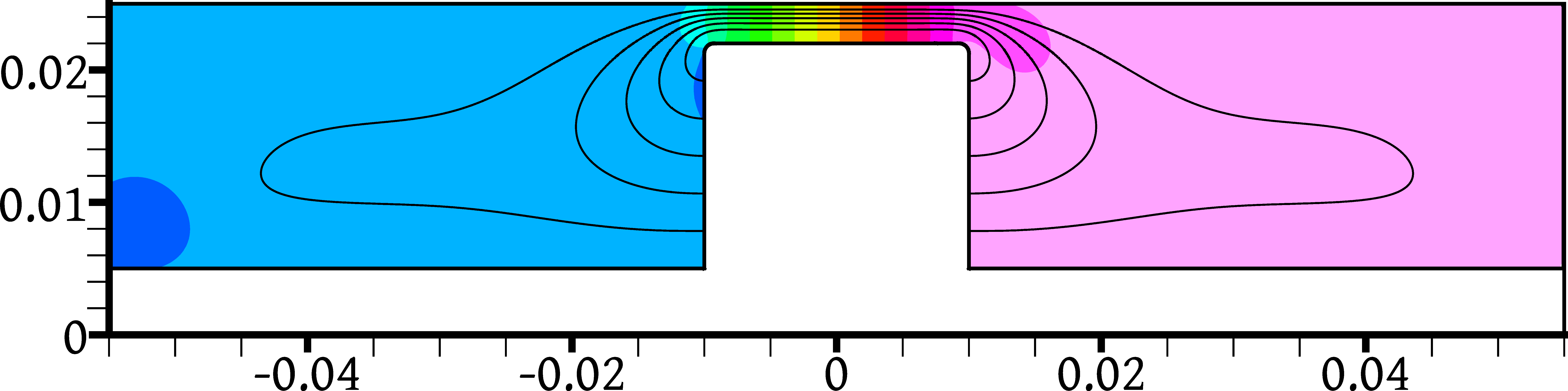}
        \caption{lPTT-100, 0.5 \si{Hz}; $\delta p$ = 17.5 \si{kPa}}
        \label{sfig: flow lPTT-100 f=0.5}
    \end{subfigure}

\vspace{0.2cm}    

    \begin{subfigure}[b]{0.49\textwidth}
        \centering
        \includegraphics[width=0.95\linewidth]{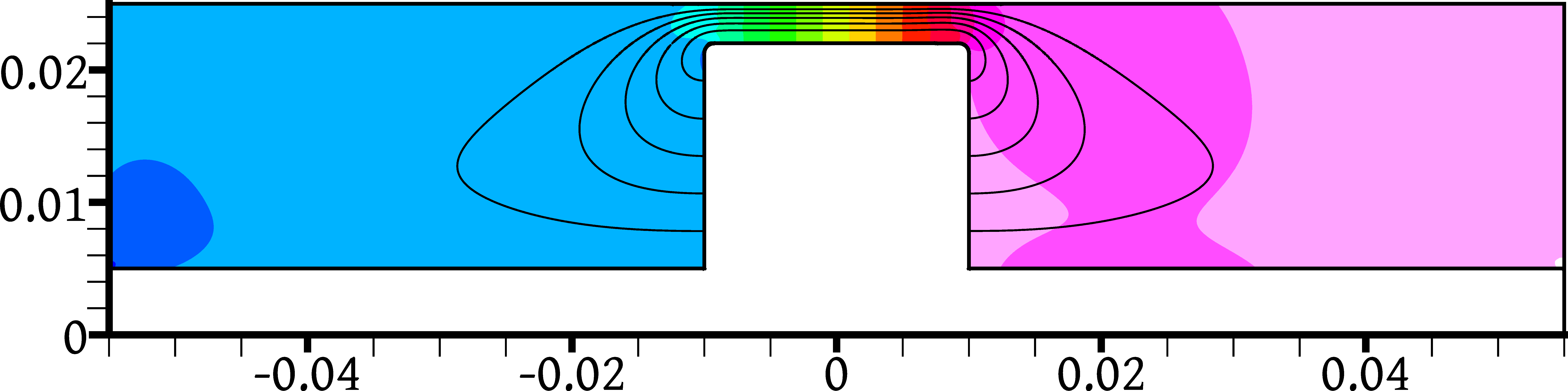}
        \caption{CY-100, 32 \si{Hz}; $\delta p$ = 110 \si{kPa}}
        \label{sfig: flow CY-100 f=32}
    \end{subfigure}
    \begin{subfigure}[b]{0.49\textwidth}
        \centering
        \includegraphics[width=0.95\linewidth]{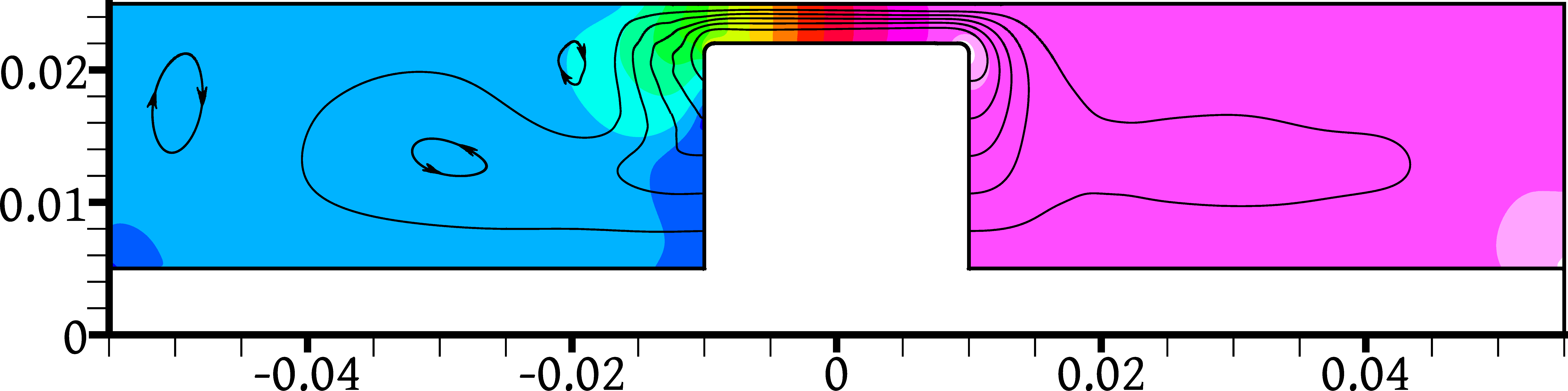}
        \caption{lPTT-100, 32 \si{Hz}; $\delta p$ = 87.5 \si{kPa}}
        \label{sfig: flow lPTT-100 f=32}
    \end{subfigure}

\vspace{0.2cm}    

    \begin{subfigure}[b]{0.49\textwidth}
        \centering
        \includegraphics[width=0.95\linewidth]{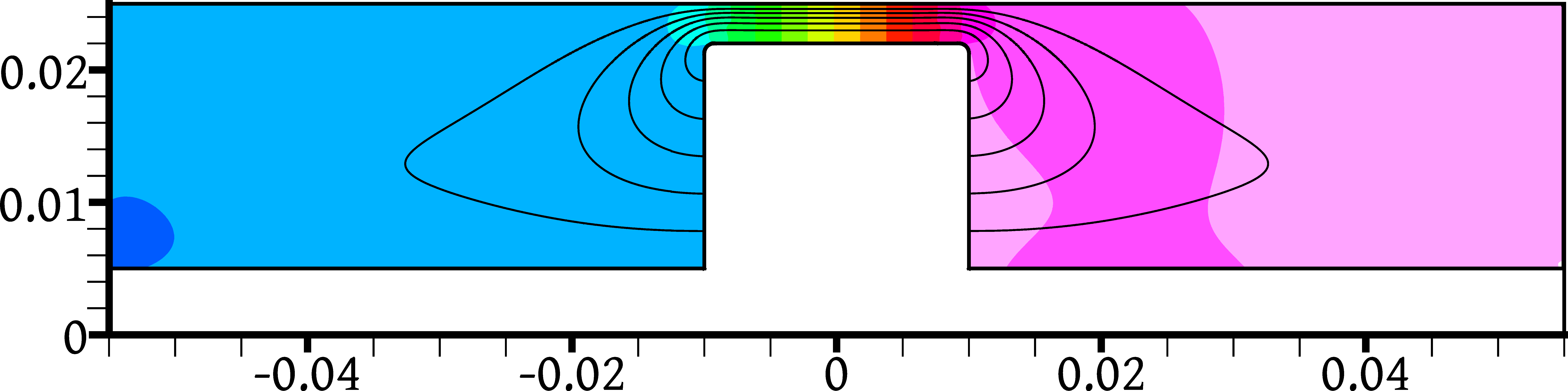}
        \caption{CY-500, 2 \si{Hz}; $\delta p$ = 62 \si{kPa}}
        \label{sfig: flow CY-500 f=2}
    \end{subfigure}
    \begin{subfigure}[b]{0.49\textwidth}
        \centering
        \includegraphics[width=0.95\linewidth]{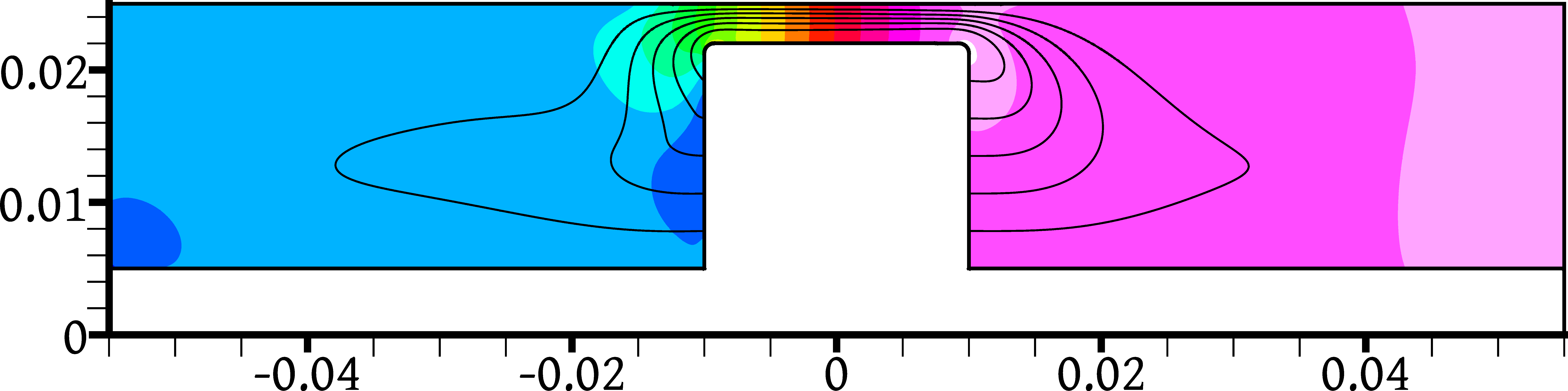}
        \caption{lPTT-500, 2 \si{Hz}; $\delta p$ = 56.25 \si{kPa}}
        \label{sfig: flow lPTT-500 f=2}
    \end{subfigure}
    
\caption{Pressure contours (colour) and instantaneous streamlines (black lines) of the flow inside 
the damper, for different fluids and frequencies. The piston is moving towards the right with 
maximum velocity (the snapshots are taken at time $t = 5T + 3T/4$). The step of the pressure 
contours, $\delta p$, is indicated in each case.}
  \label{fig: flow fields}
\end{figure}

\begin{figure}[tb]
    \centering
    \begin{subfigure}[b]{0.49\textwidth}
        \centering
        \includegraphics[width=0.95\linewidth]{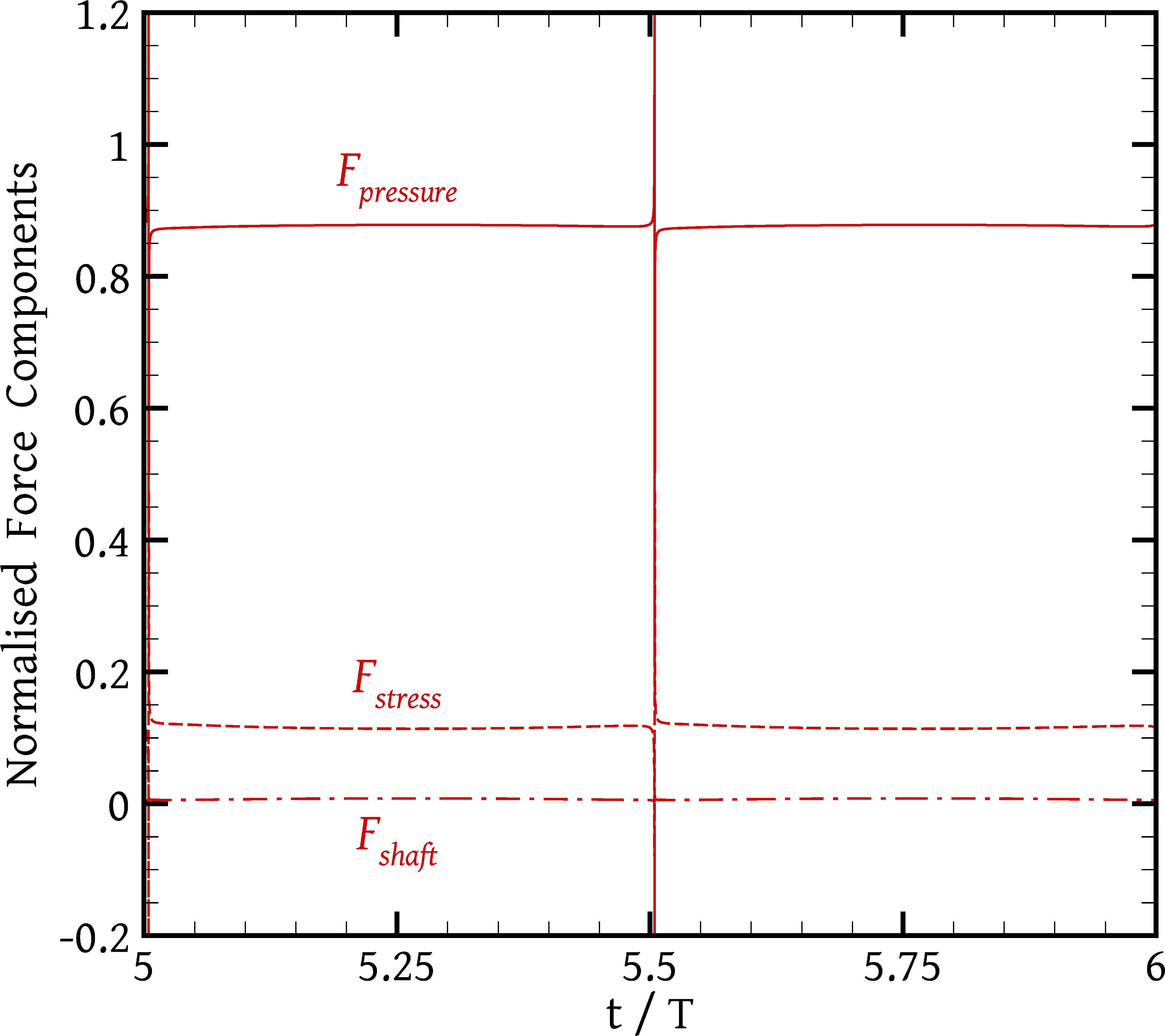}
        \caption{$f$ = 0.5 \si{Hz} ($T$ = 2 \si{s})}
        \label{sfig: force components f=0.5}
    \end{subfigure}
    \begin{subfigure}[b]{0.49\textwidth}
        \centering
        \includegraphics[width=0.95\linewidth]{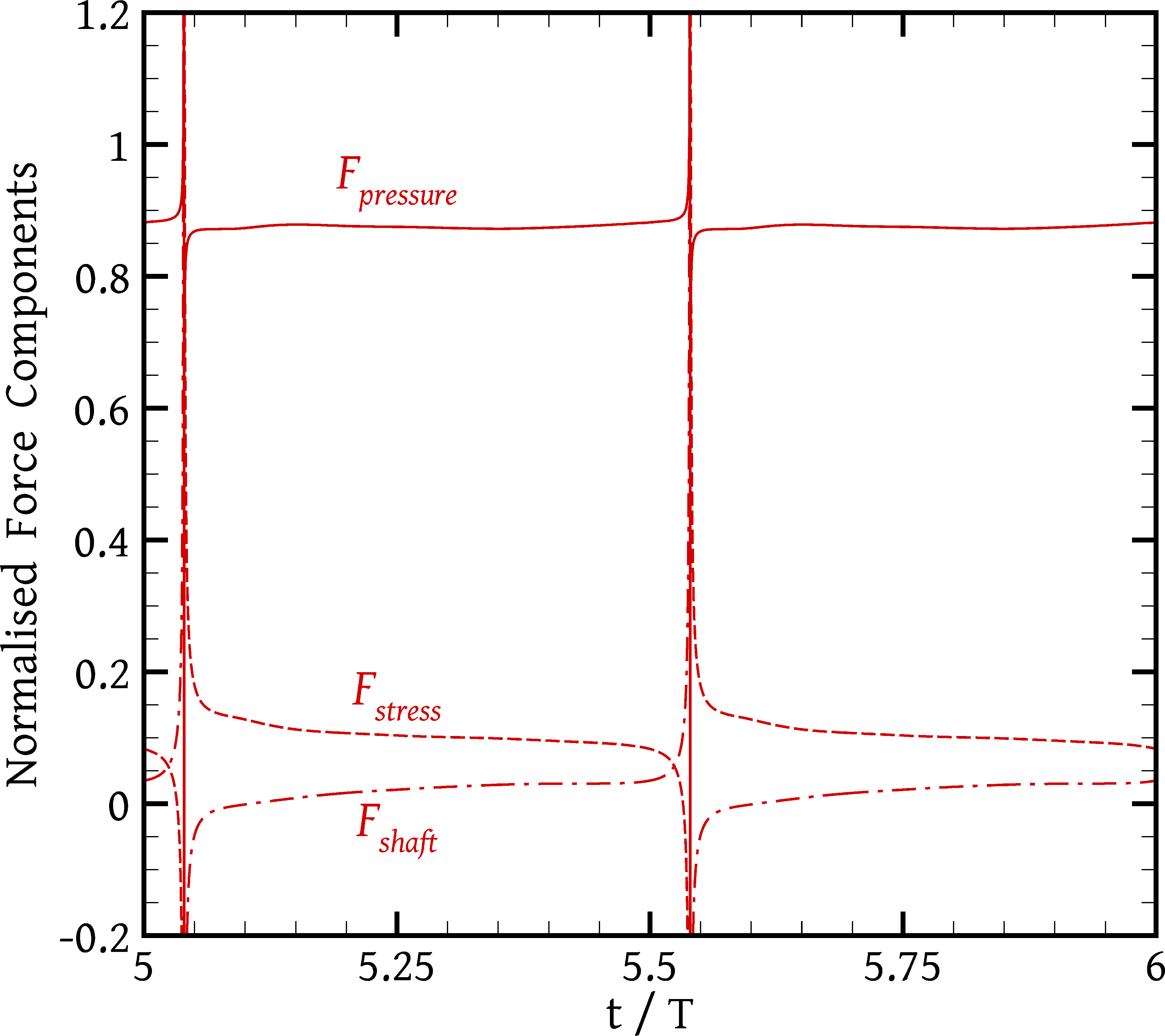}
        \caption{$f$ = 32 \si{Hz} ($T$ = 0.03125 \si{s})}
        \label{sfig: force components f=32}
    \end{subfigure}
    
\caption{Components of the force $F_{fl}$, Eq.\  \eqref{eq: fluid force}, normalised by $F_{fl}(t)$ 
itself, for the lPTT-100 fluid, as a function of time normalised by the oscillation period, over 
the last (6th) period. Continuous line: pressure force on the piston, $\iint_{\mathrm{piston}} -p 
\vf{n} \cdot \vf{e}_x \mathrm{d}s$. Dashed line: force due to viscoelastic stresses on the piston, 
$\iint_{\mathrm{piston}} \vf{n} \cdot \tf{\tau} \cdot \vf{e}_x \mathrm{d}s$. Dash-dot line: force 
due to viscoelastic stresses on the shaft, $\iint_{\mathrm{shaft}} \vf{n} \cdot \tf{\tau} \cdot 
\vf{e}_x \mathrm{d}s$. See the description of Eq.\ \eqref{eq: fluid force} for notation. The 
pressure force on the shaft acts only in the $r$ direction, so it does not contribute to $F_{fl}$.}
  \label{fig: force components}
\end{figure}

Before examining the force $F_{fl}$ it is worthwhile to examine some snapshots of the whole flow 
field in Fig.\ \ref{fig: flow fields}. All snapshots correspond to a time instance when the piston 
moves towards the right with maximum velocity. The general picture that these snapshots present is 
that the rightward piston motion forces oil to flow from the right damper compartment to the left 
compartment through the narrow annular gap, which gives rise to a sharp pressure gradient along the 
gap length in order to overcome the stresses that resist this fluid motion. The resulting large 
pressure difference between the left and right sides of the piston causes a pressure force that 
opposes its motion. The latter is also opposed by the stresses in the gap region, but the pressure 
contribution to the total force $F_{fl}$ is much larger for this damper configuration: as shown in 
Fig.\ \ref{fig: force components}, the pressure force is nearly ten times larger than the force due 
to shear stresses. Figure \ref{fig: force components} also shows that $F_{fl}$ arises mostly on the 
piston surface, while the contribution of the shaft surface to the total force is negligible.

The instantaneous streamlines shown in Fig.\ \ref{fig: flow fields} are drawn equispaced along the 
vertical piston walls in the $r$-direction. The flow rate between any pair of streamlines can be 
calculated from the piston velocity and the area between the points where these streamlines touch 
the vertical piston walls. Because the domain is axisymmetric, this area, and therefore also the 
flow rate, between a pair of successive streamlines that are close to the shaft (small $r$ 
coordinate) is smaller than that between a pair of successive streamlines that lie closer to the 
outer cylinder (large $r$ coordinate). Therefore, it is evident from all snapshots of Fig.\ 
\ref{fig: flow fields} that the flow is mostly restricted to the immediate neighbourhood of the 
piston. Fluid that gets pushed out of the way in front of the piston quickly rises, passes through 
the gap, and travels to the region immediately behind the piston. Fluid that is located farther 
away remains relatively at rest.

In Fig.\ \ref{fig: flow fields}, the left column of snapshots corresponds to the Carreau-Yasuda 
fluids, while the right column contains the corresponding snapshots predicted by the l-PTT fluids. 
The Carreau-Yasuda model predicts a smooth, symmetric flow field in every case examined. This is 
due to the fact that the inertial character of this flow is weak; it is not hard to show that for a 
symmetric domain, such as the present one when the piston lies halfway along the cylinder, the 
momentum equation \eqref{eq: momentum nd} in the limit of $\Rey \rightarrow 0$ (creeping flow), 
with a generalised Newtonian constitutive equation $\tf{\tau} = \eta(\dot{\gamma}) 
\dot{\tf{\gamma}}$, admits symmetric solutions. The vanishing of the time derivative in the momentum 
equation \eqref{eq: momentum nd} as $\Rey \rightarrow 0$ also makes the flow field quasi-steady (not 
dependent on the previous flow history but only on the instantaneous boundary conditions) and 
therefore the choice of time step in the numerical solution procedure has a small impact on the 
accuracy. 

The predictions of the lPTT-100 fluid at the lowest frequency of $f$ = 0.5 \si{Hz} (Fig.\ \ref{sfig: 
flow lPTT-100 f=0.5}) are nearly identical to those of the CY-100 fluid (Fig.\ \ref{sfig: flow 
CY-100 f=0.5}). This suggests that at low frequencies the Carreau-Yasuda generalised Newtonian model 
suffices to describe the behaviour of silicone oil because the Weissenberg number is low and elastic 
effects are not important. However, at higher Weissenberg numbers, which result from higher 
frequency (Fig.\ \ref{sfig: flow lPTT-100 f=32}) or fluid elasticity (Fig.\ \ref{sfig: flow lPTT-500 
f=2}), there are major deviations of the viscoelastic predictions from those of the generalised 
Newtonian model. The viscoelastic flow is unsymmetric and less smooth, and these features become 
more pronounced as the Weissenberg number increases (compare Fig.\ \ref{sfig: flow lPTT-100 f=32} 
with $\Wei \approx 70$ against Fig.\ \ref{sfig: flow lPTT-500 f=2} with $\Wei \approx 25$). Now the 
flow cannot be characterised as quasi-steady, despite the low Reynolds number that suppresses the 
inertial terms in the momentum equation \eqref{eq: momentum nd}, because transient effects are 
important in the constitutive equation \eqref{eq: constitutive lPTT nd} and govern the evolution of 
the stresses. In Figs.\ \ref{sfig: flow lPTT-100 f=32} and \ref{sfig: flow lPTT-500 f=2} one can 
notice that the flow in the upstream chamber (on the right of the piston) resembles more the 
corresponding generalised Newtonian flow than that in the downstream (left) chamber. This can be 
attributed to the fact that the fluid upstream of the piston has not undergone significant 
deformations recently and the stresses have had some time to relax; therefore, the state of the 
fluid is somewhat similar to that of the corresponding Carreau-Yasuda fluid, as can be more clearly 
seen by comparing Figs.\ \ref{sfig: flow lPTT-500 f=2} and \ref{sfig: flow CY-500 f=2}. Once the 
fluid has been pushed through the narrow gap, the large stresses developed there relax gradually in 
the PTT case but immediately in the CY case. Therefore, the PTT and CY flows are more distinctly 
different downstream of the piston (left chamber). 

\begin{figure}[tb]
    \centering
    \begin{subfigure}[b]{0.49\textwidth}
        \centering
        \includegraphics[width=0.95\linewidth]{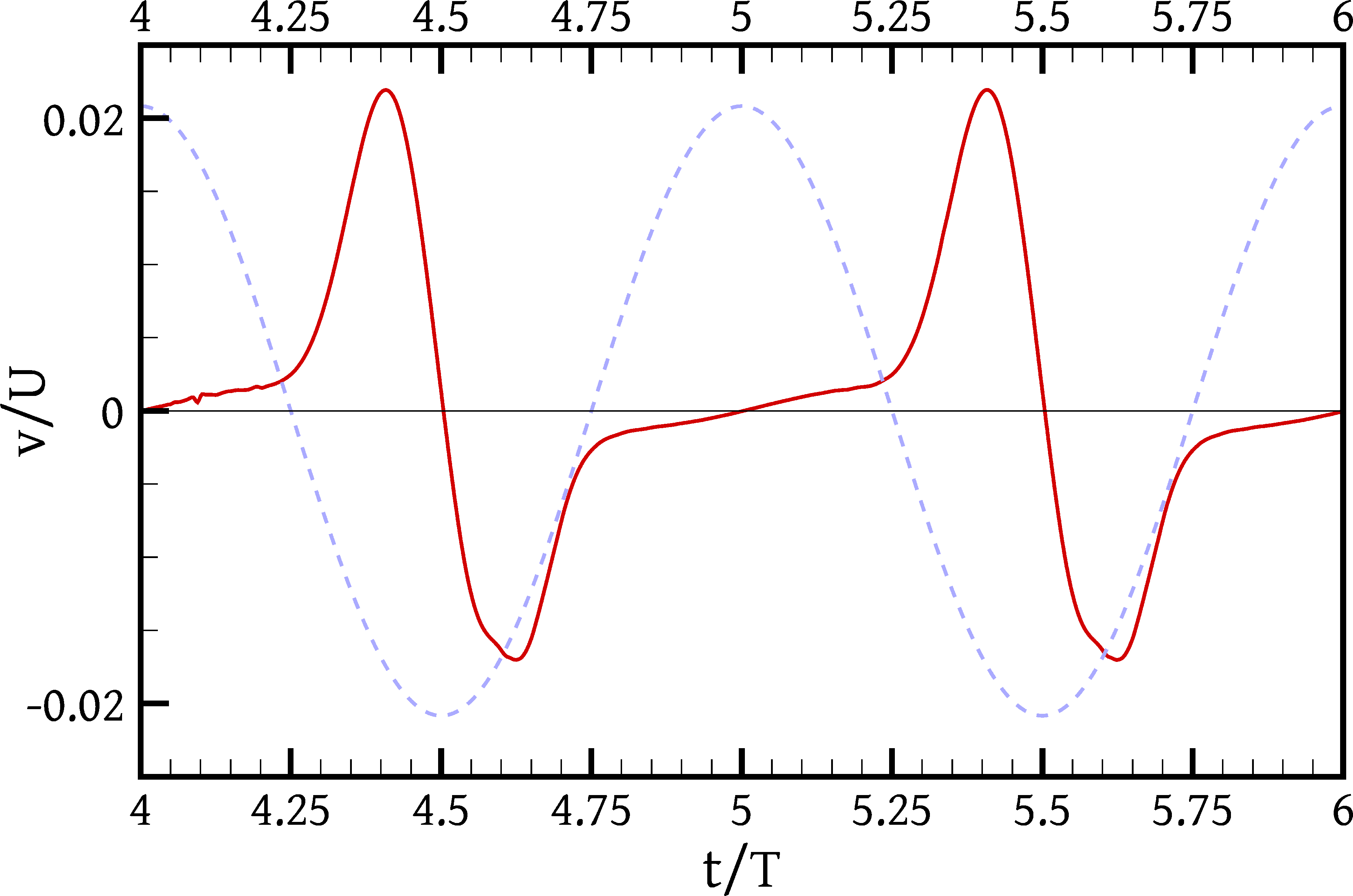}
        \caption{$f$ = 8 \si{Hz}}
        \label{sfig: instability monitor f8}
    \end{subfigure}
    \begin{subfigure}[b]{0.49\textwidth}
        \centering
        \includegraphics[width=0.95\linewidth]{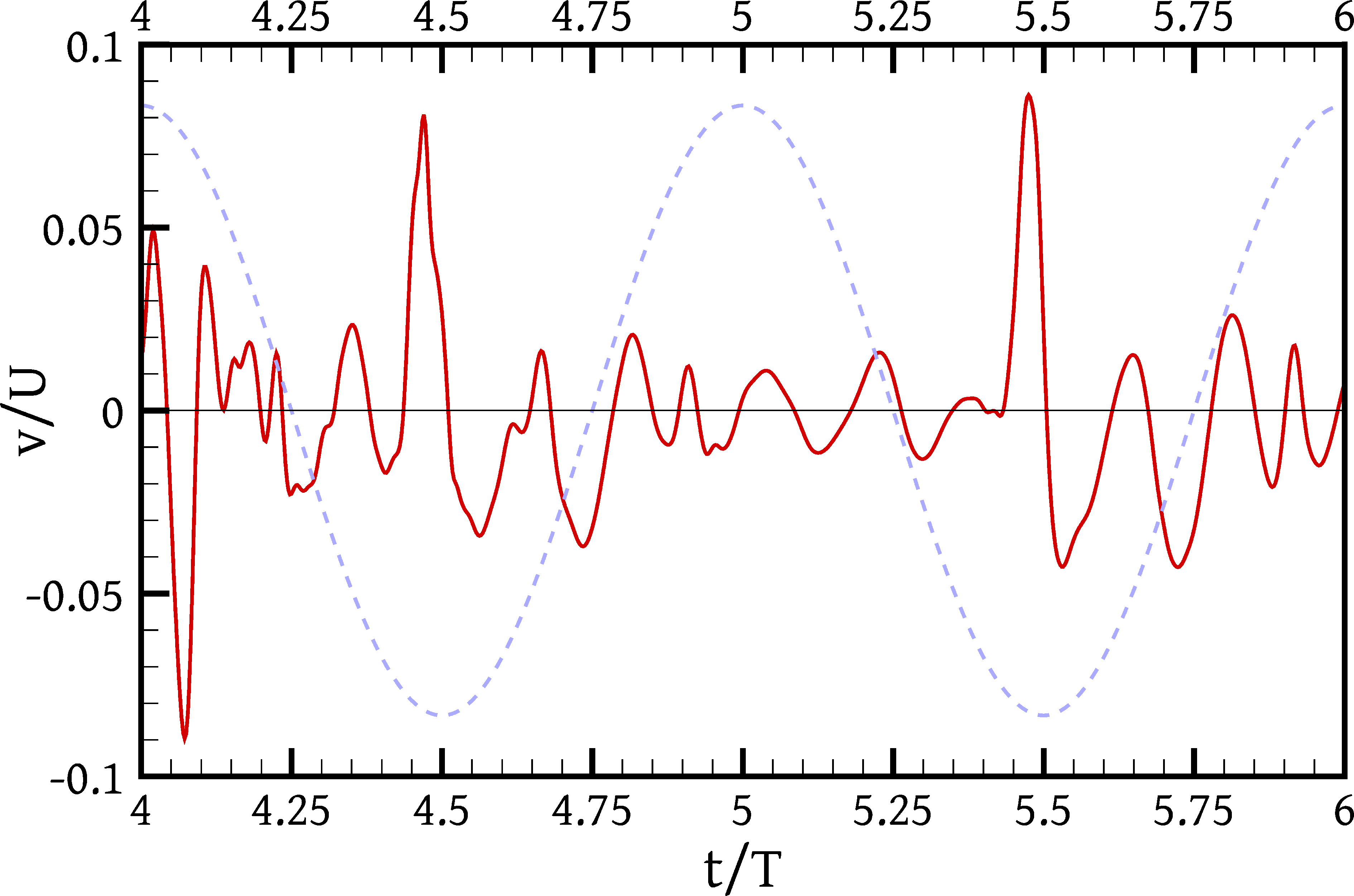}
        \caption{$f$ = 32 \si{Hz}}
        \label{sfig: instability monitor f32}
    \end{subfigure}
    
\caption{History of the radial ($r$-) velocity component $v$ of the lPTT-100 fluid at location $x = 
-35$ \si{mm}, $r = 15$ \si{mm}, for the $f$ = 8 \si{Hz} \subref{sfig: instability monitor f8} and 
the $f$ = 32 \si{Hz} \subref{sfig: instability monitor f32} simulations (drawn in continuous line). 
The velocity $v$ (ordinate) is normalised against the velocity scale $U$ (Table \ref{table: 
operating conditions}), while time (abscissa) is normalised by the oscillation period $T = 1/f$. 
Also drawn, in dashed line, is a cosine wave, which is in phase with the piston displacement.}
  \label{fig: instability monitor}
\end{figure}

In this downstream region, Fig.\ \ref{sfig: flow lPTT-100 f=32}, which corresponds to the highest 
$\Wei$ tested, shows the development of vortices which are of elastic origin since inertia is very 
low. Elasticity in fluids is known to induce flow instabilities, see e.g.\ \cite{Alves_2007, 
Sousa_2009, Sousa_2011, Afonso_2011, Comminal_2016}. Although due to the transiency of the problem, 
with the piston velocity not being constant, it is difficult to precisely describe the vortex 
shedding dynamics, Fig.\ \ref{sfig: instability monitor f32} shows clear evidence of this process in 
the $f$ = 32 \si{Hz} simulation for the lPTT-100 fluid. That figure monitors the radial velocity 
component at a specific location in the left chamber of the damper and shows that it changes sign 
multiple times during a damper oscillation period; the frequency of the vortex shedding does not 
seem to be directly related to the damper oscillation frequency. On the contrary, at the lower 
frequency of $f$ = 8 \si{Hz} Fig.\ \ref{sfig: instability monitor f8} shows that the fluid velocity 
at the same location is completely harmonised with the piston motion (compare with the cosine wave 
drawn in dashed line): the velocity is small when the piston is displaced towards the right, away 
from the given point ($t \in [(k-0.25)T, (k+0.25)T]$ approximately, where $k$ is an integer), 
becomes large and positive when the piston is displaced towards the left and is approaching the 
given point  and is pushing the fluid of the left compartment towards the gap ($t \in 
[(k+0.25)T,(k+0.5)T]$), and becomes negative when the piston is retracting from its extreme left 
position and forcing the fluid of the right compartment to flow into the left compartment ($t \in 
[(k+0.5)T, (k+0.75)T$). In the $f$ = 32 \si{Hz} case (Fig.\ \ref{sfig: instability monitor f32}) the 
kinematics of the vortex shedding are superimposed on this mean fluid motion. The $f$ = 32 \si{Hz} 
lPTT-100 case was the only one among the cases studied where instabilities were observed. A thorough
examination of this phenomenon is beyond the scope of the present paper.

\begin{figure}[tb]
    \centering
    \begin{subfigure}[b]{0.49\textwidth}
        \centering
        \includegraphics[width=0.95\linewidth]{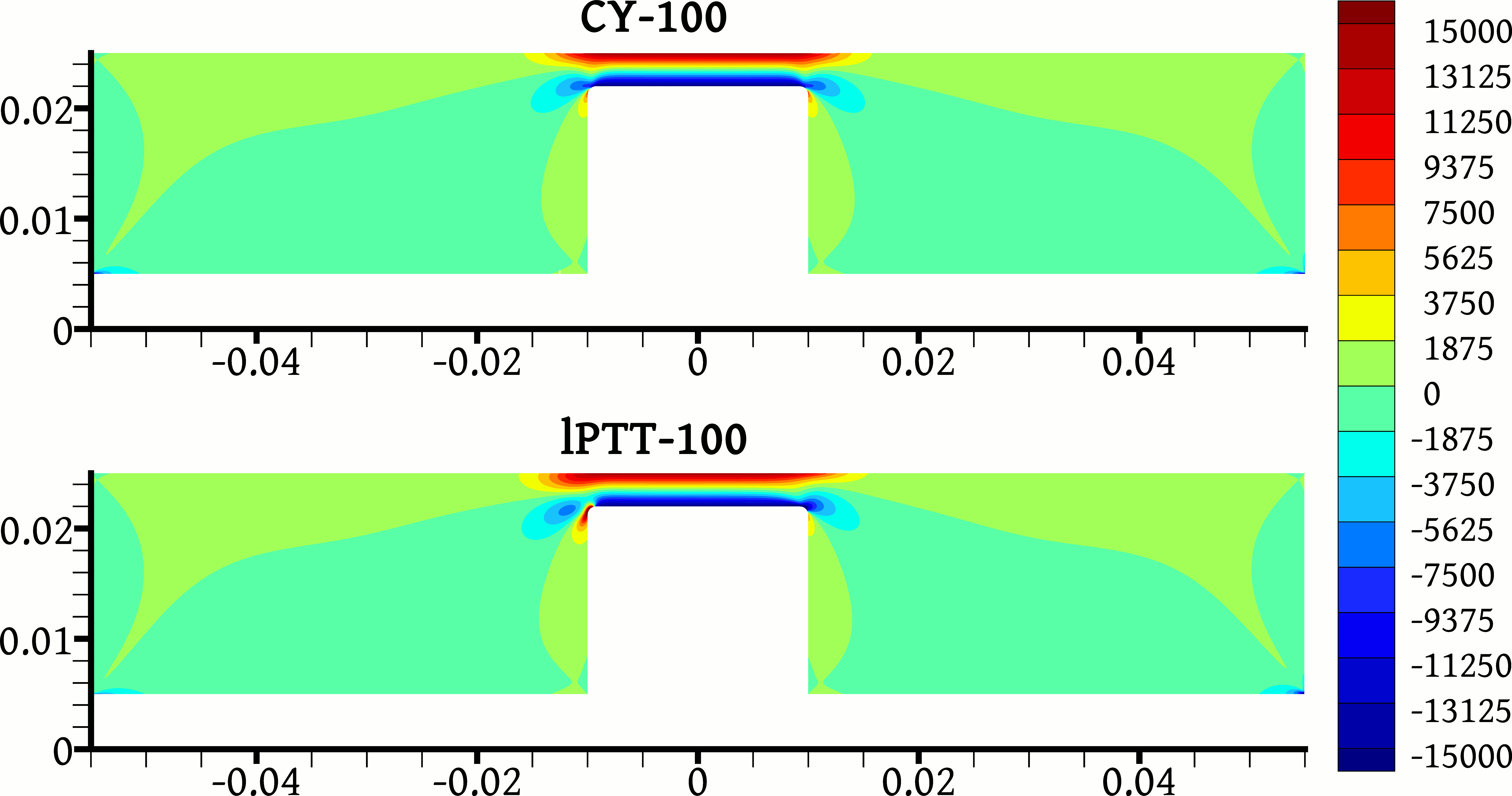}
        \caption{Maximum piston velocity}
        \label{sfig: stress_xy f=0.5 Umax}
    \end{subfigure}
    \begin{subfigure}[b]{0.49\textwidth}
        \centering
        \includegraphics[width=0.95\linewidth]{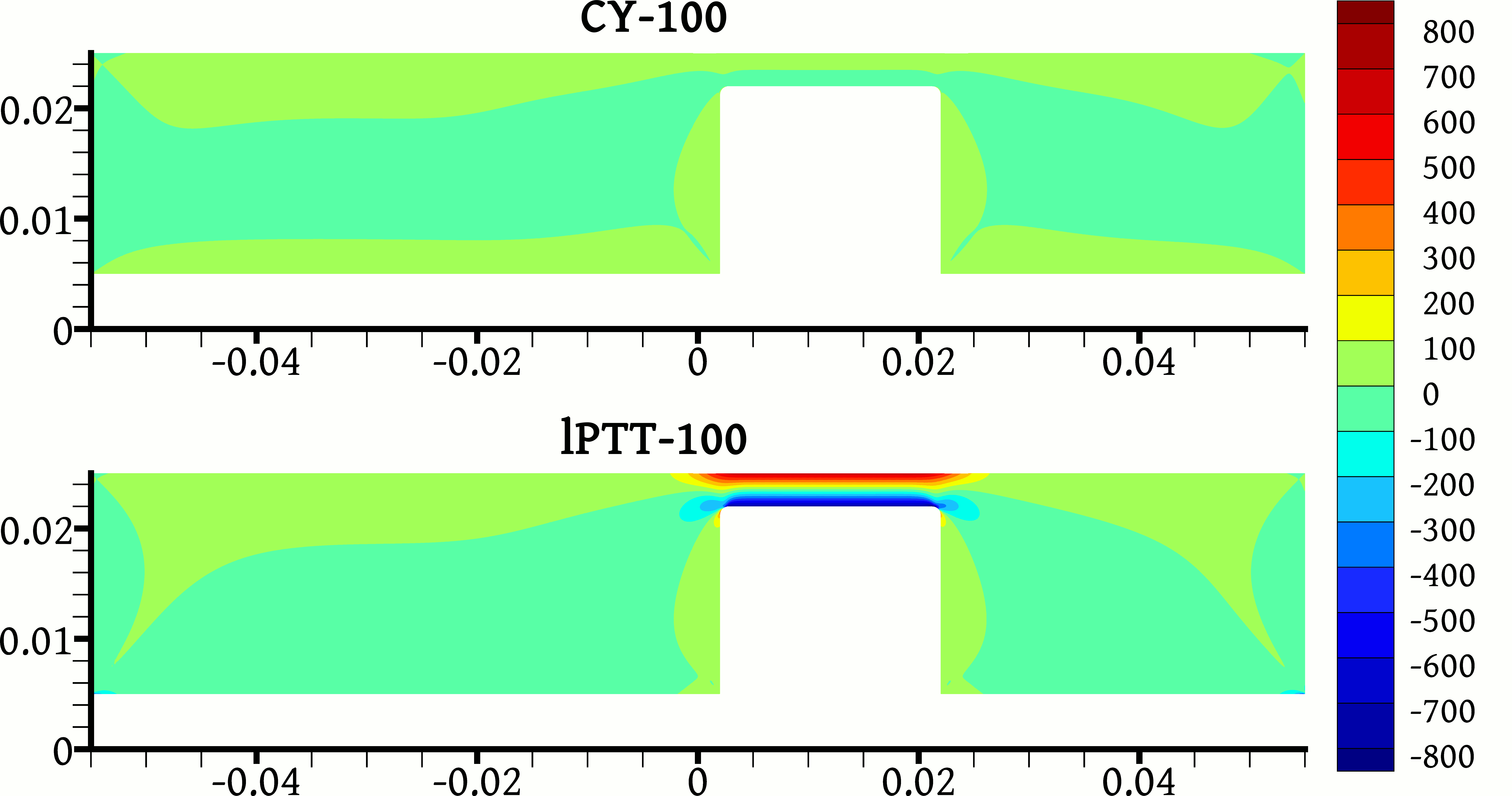}
        \caption{Zero piston velocity}
        \label{sfig: stress_xy f=0.5 Umin}
    \end{subfigure}
    
\caption{Contours of the $\tau_{rx}$ stress component, in \si{Pa}, for the fluids indicated, at a 
frequency of $f$ = 0.5 \si{Hz}. The figures on the left correspond to time $t = 5T + 3T/4$, when 
the piston moves with maximum velocity towards the right; those on the right correspond to time $t 
= 6T$, when the piston momentarily stops moving.}
  \label{fig: stress_xy f=0.5}
\end{figure}

\begin{figure}[tb]
    \centering
    \begin{subfigure}[b]{0.49\textwidth}
        \centering
        \includegraphics[width=0.95\linewidth]{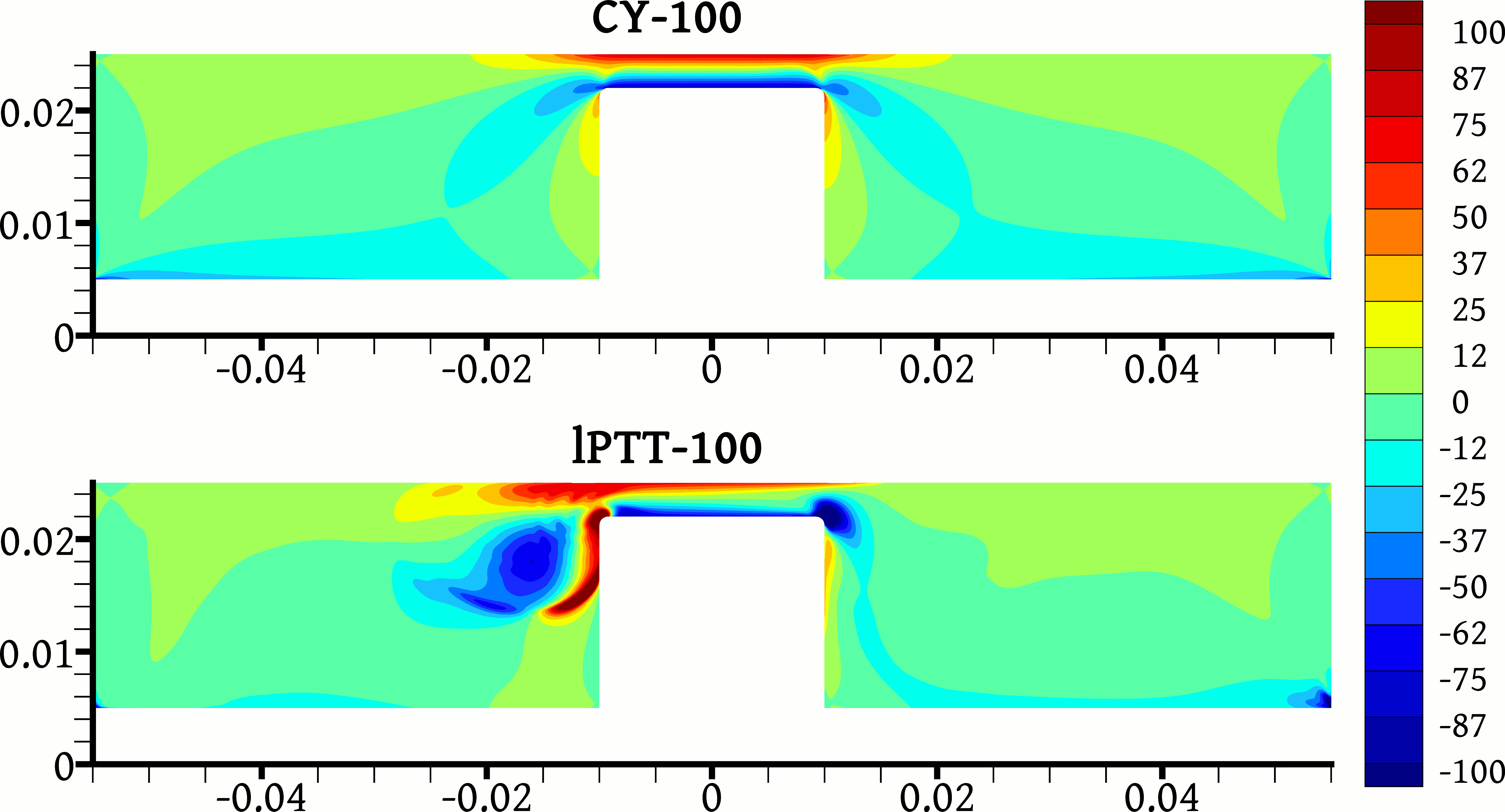}
        \caption{Maximum piston velocity}
        \label{sfig: stress_xy f=32 Umax}
    \end{subfigure}
    \begin{subfigure}[b]{0.49\textwidth}
        \centering
        \includegraphics[width=0.95\linewidth]{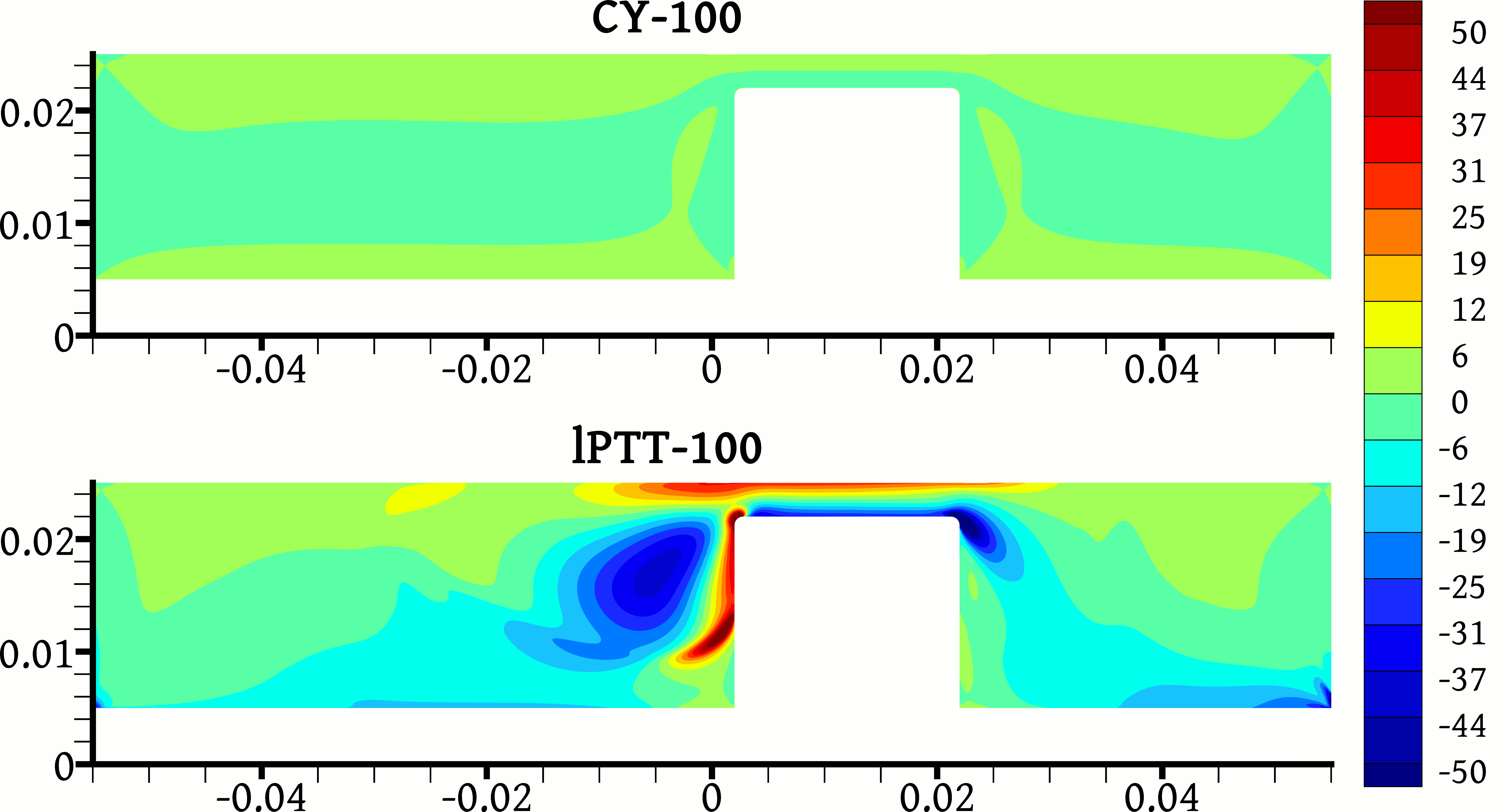}
        \caption{Zero piston velocity}
        \label{sfig: stress_xy f=32 Umin}
    \end{subfigure}
    
\caption{As per Fig.\ \ref{fig: stress_xy f=0.5}, but for a frequency of $f$ = 32 \si{Hz}. Also, 
the $\tau_{rx}$ contours are in \si{kPa} rather than \si{Pa}.}
  \label{fig: stress_xy f=32}
\end{figure}

Figures \ref{fig: stress_xy f=0.5} and \ref{fig: stress_xy f=32} show snapshots of the distribution 
of the stress component $\tau_{rx}$ for frequencies $f$ = 0.5 \si{Hz} and $f$ = 32 \si{Hz}, 
respectively, for the CY-100 and lPTT-100 fluids. The geometry of the present problem is such that 
$\tau_{rx}$ is by far the most crucial stress component for producing $F_{fl}$. Figure \ref{sfig: 
stress_xy f=0.5 Umax} shows that at low frequencies the lPTT-100 and CY-100 models predict very 
similar stress distributions when the piston is moving fast. However, when the piston is coming to 
a 
halt (Fig.\ \ref{sfig: stress_xy f=0.5 Umin}) the CY-100 stresses go to zero because the fluid 
velocity, and therefore also its gradient, vanish since the inertia of the fluid is very low. For 
generalised Newtonian fluids such as the CY-100 the stress tensor depends only on the instantaneous 
velocity gradient which is zero in this case. On the contrary, the PTT stresses depend also on the 
deformation history and at the time instance shown in Fig.\ \ref{sfig: stress_xy f=0.5 Umin} they 
have not had enough time to relax; they are non-zero and continue to push the piston towards the 
left (towards the $x$ = 0 position), despite the near-zero velocity gradient.

These phenomena are much more pronounced at the $f$ = 32 \si{Hz} frequency (Fig.\ \ref{fig: 
stress_xy f=32}). Now, when the piston is stopped (Fig.\ \ref{sfig: stress_xy f=32 Umin}) the PTT 
stresses are quite high, comparable to those when the piston is moving fast (Fig.\ \ref{sfig: 
stress_xy f=32 Umax}). This is because the Deborah number is high (Table \ref{table: operating 
conditions}), i.e.\ the time scale of the oscillatory piston motion is comparable to the relaxation 
time of the fluid and the stresses do not have enough time to relax within an oscillation period. 
On the other hand, the CY stresses again go to zero when the piston stops, a sign that fluid 
inertia is still quite small even at this high frequency. Concerning the stress distributions at 
high piston velocities (Fig.\ \ref{sfig: stress_xy f=32 Umax}), contrary to the $f$ = 0.5 \si{Hz} 
case now they are quite different between the lPTT-100 and CY-100 fluids, with the former exhibiting 
complex patterns.

\subsection{Effects of shear-thinning}
\label{ssec: results; shear-thinning}

Shear-thinning plays the most important role in determining the overall force levels. As mentioned, 
for this flow the most important stress component is $\tau_{rx}$, and consequently the most 
important type of fluid deformation is shear. Therefore, steady shear diagrams such as that of 
Fig.\ \ref{fig: steady shear viscosity} can convey a lot of useful information about this flow. 
This 
is why the Carreau-Yasuda model is relatively successful in its predictions, especially at low 
frequencies: in the concept of a generalised Newtonian fluid it is inherently assumed that a steady 
shear diagram such as Fig.\ \ref{fig: steady shear viscosity} conveys all the information about the 
fluid rheology.

\begin{figure}[tb]
    \centering

    \begin{subfigure}[b]{0.40\textwidth}
        \centering
        \includegraphics[width=0.95\linewidth]{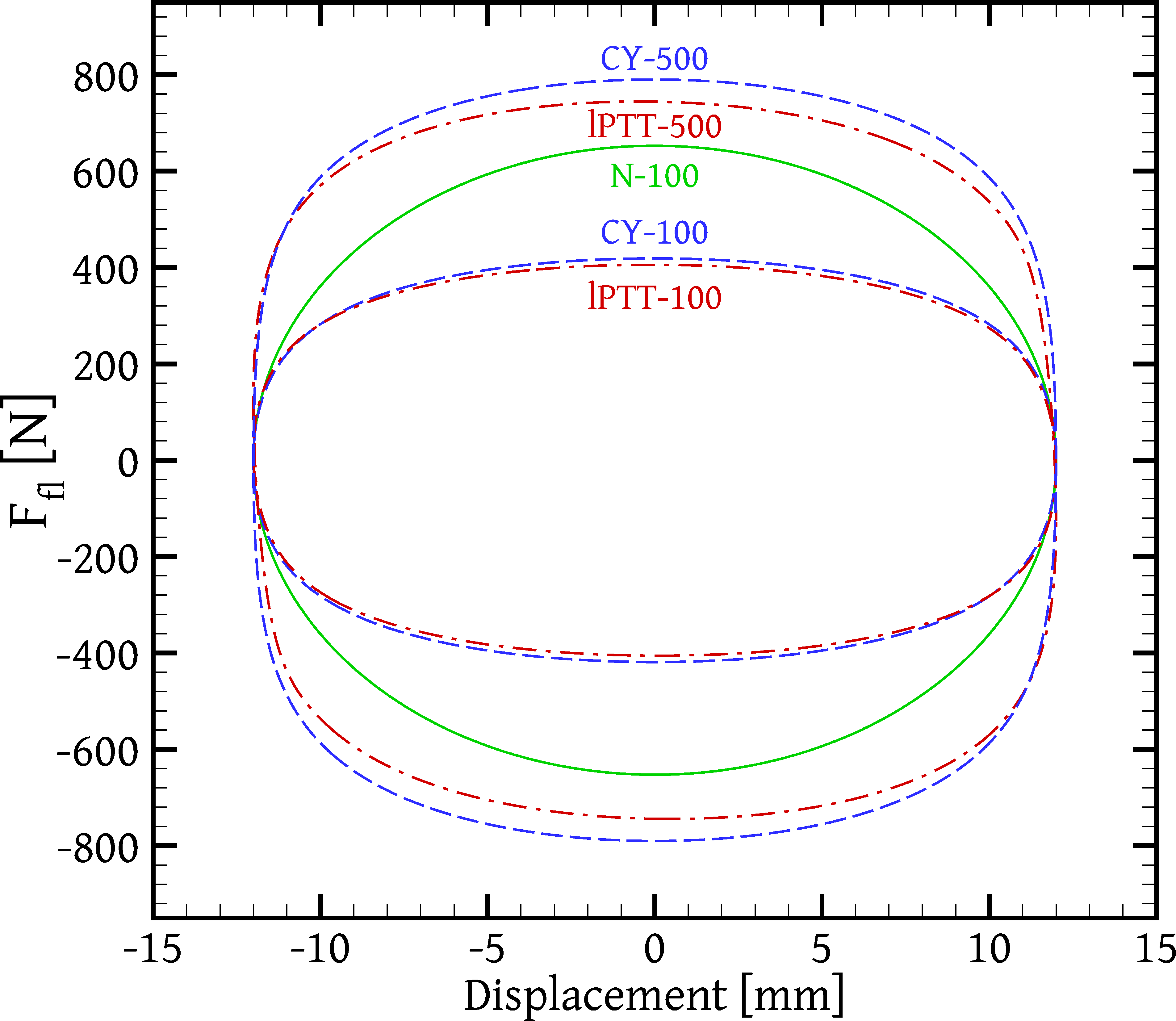}
        \caption{$f$ = 0.5 \si{Hz}}
        \label{sfig: force displacement f=0.5}
    \end{subfigure}
    \begin{subfigure}[b]{0.40\textwidth}
        \centering
        \includegraphics[width=0.95\linewidth]{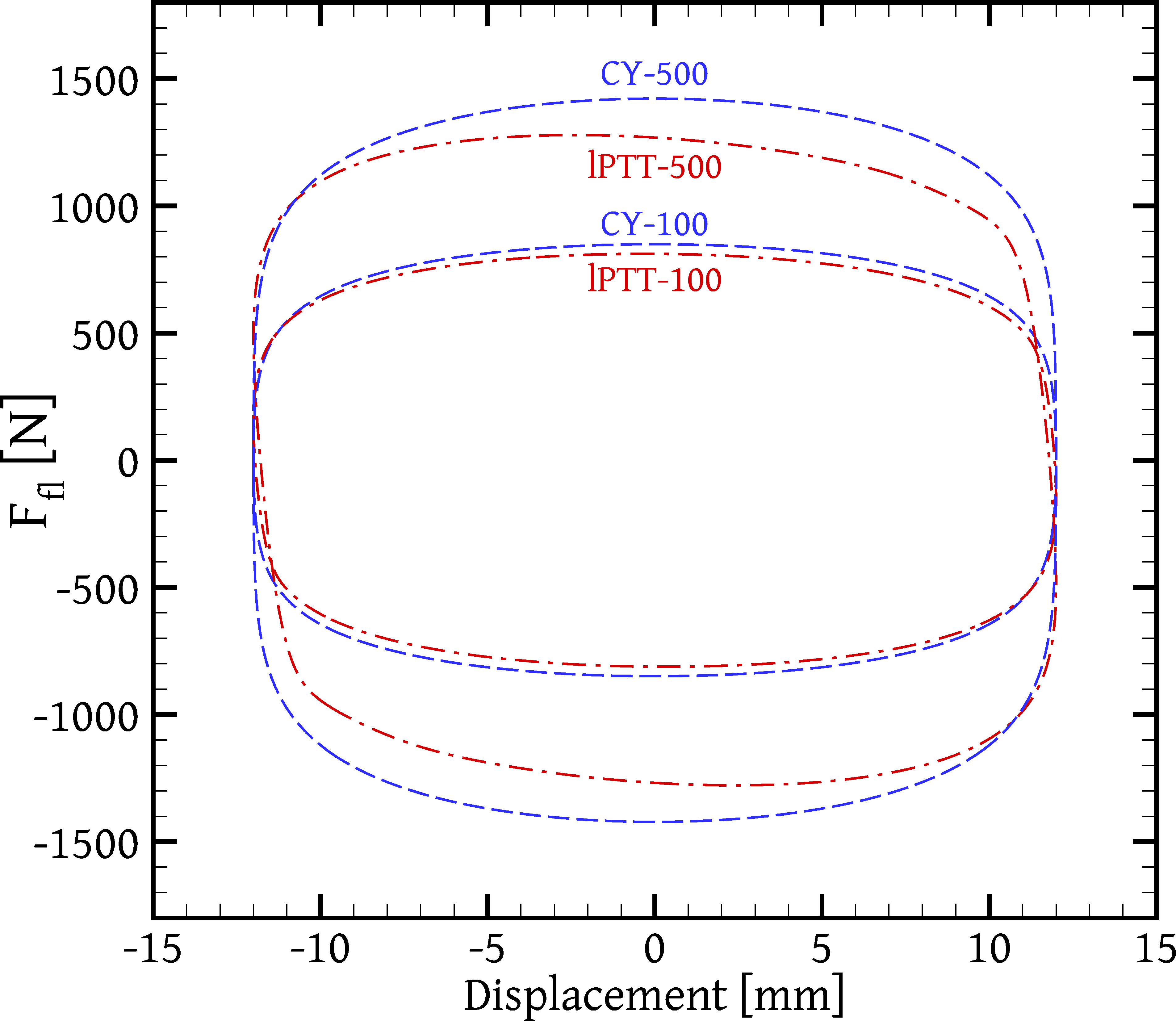}
        \caption{$f$ = 2 \si{Hz}}
        \label{sfig: force displacement f=2}
    \end{subfigure}
    
    \vspace{0.2cm}
    
    \begin{subfigure}[b]{0.40\textwidth}
        \centering
        \includegraphics[width=0.95\linewidth]{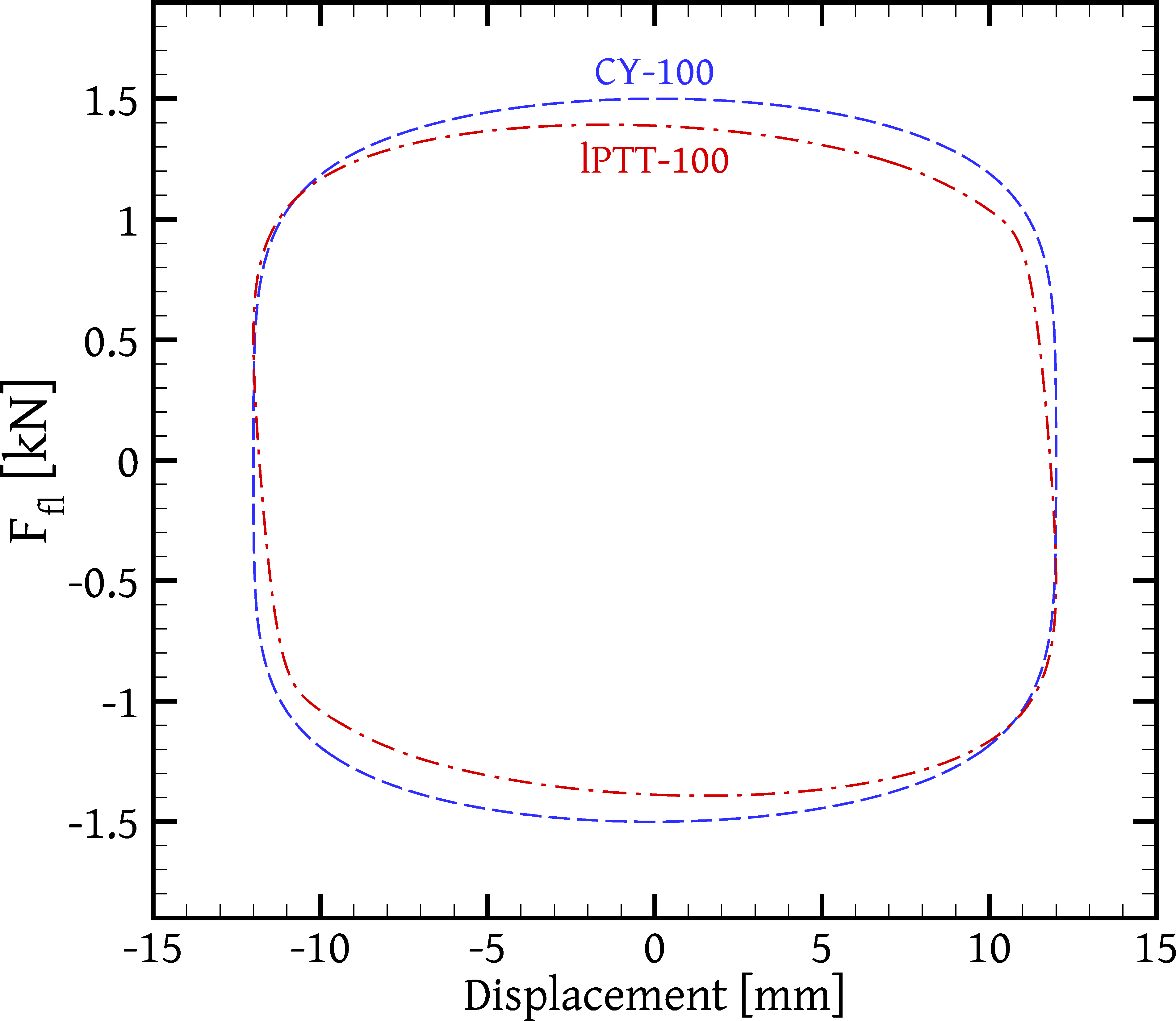}
        \caption{$f$ = 8 \si{Hz}}
        \label{sfig: force displacement f=8}
    \end{subfigure}
    \begin{subfigure}[b]{0.40\textwidth}
        \centering
        \includegraphics[width=0.95\linewidth]{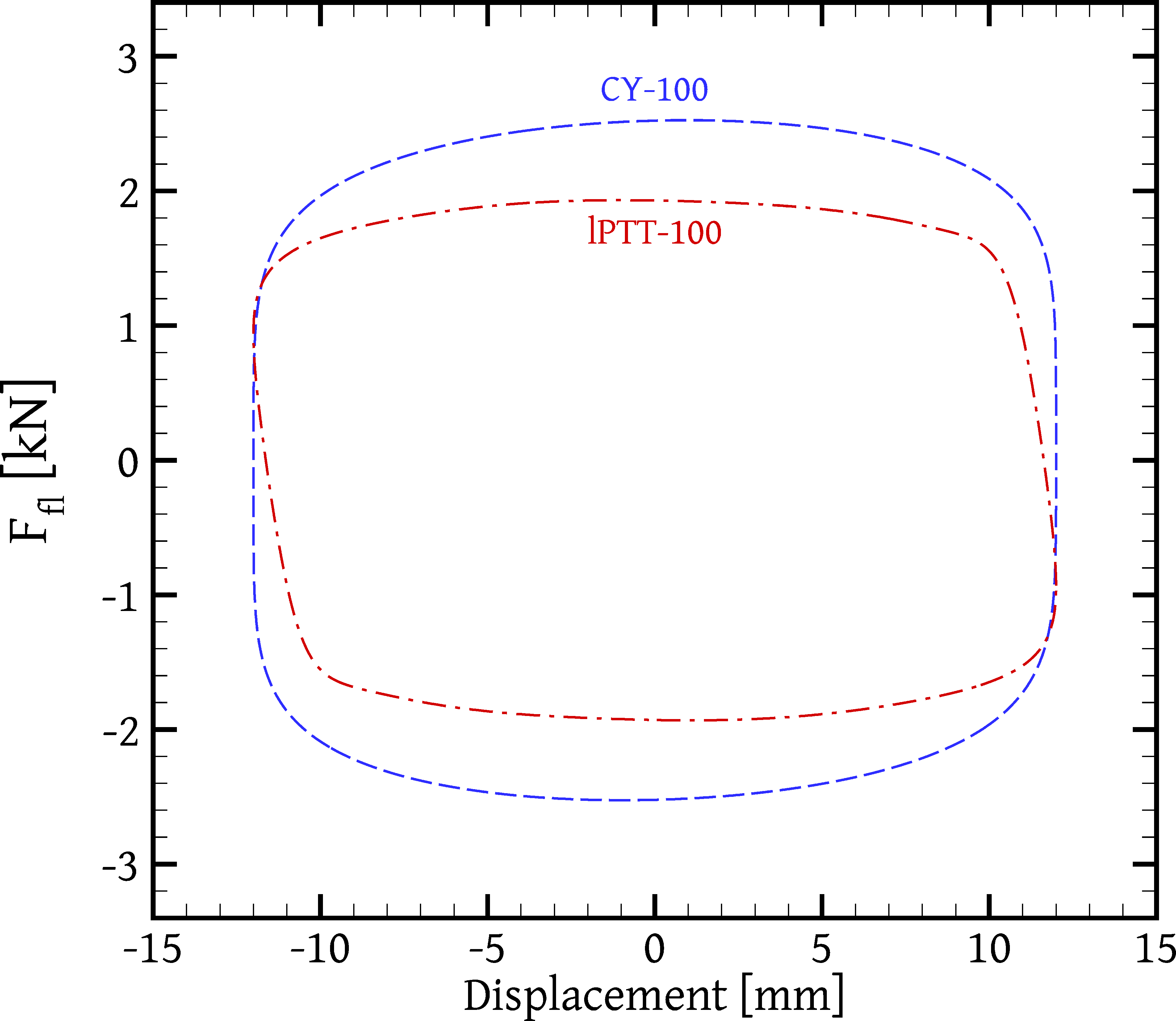}
        \caption{$f$ = 32 \si{Hz}}
        \label{sfig: force displacement f=32}
    \end{subfigure}
    
\caption{Force $F_{fl}$ (Eq.\ \eqref{eq: fluid force}) versus piston displacement plots, for each 
of the oscillation frequencies selected for the simulations. Each plot contains the results of all 
fluids tested at that particular frequency, except for the N-100 fluid which is only plotted in 
\subref{sfig: force displacement f=0.5} for clarity.}
  \label{fig: force displacement}
\end{figure}

The importance of shear-thinning is demonstrated in Fig.\ \ref{sfig: force displacement f=0.5}: 
even at the low frequency of $f$ = 0.5 \si{Hz} the force produced by the CY-100 and lPTT-100 fluids 
when the piston is at $x = 0$ is only 64\% and 62\%, respectively, of that produced by a Newtonian 
fluid of 100 \si{Pa.s} viscosity (N-100). The force reduction due to shear-thinning increases 
overwhelmingly with increasing frequency; at $f$ = 32 \si{Hz} these ratios of CY-100 and lPTT-100 
forces to N-100 force become 6\% and 4.6\%, respectively. The force difference between the CY and 
lPTT fluids will be discussed in Section \ref{ssec: results; elasticity}.

Figure \ref{fig: force displacement} also shows that while the Newtonian force -- displacement 
loops are elliptical in shape, those of both the PTT and CY models are more ``rectangular'', i.e.\ 
the force rises sharply as the piston moves away from an extreme position, and remains relatively 
constant along most of the piston stroke. This is a result of the shear-thinning property of the 
PTT and CY models: as the piston accelerates, the shear rate $\dot{\gamma}$ increases, but the 
viscosity $\eta$ decreases, albeit at a lower pace, (Fig.\ \ref{fig: steady shear viscosity}), 
resulting in a small overall rise of the stress $\tau = \eta \dot{\gamma}$. The exact opposite 
happens during deceleration of the piston. Therefore, the variation of the force is mild along most 
of the piston stroke. This is usually a desirable property, as it maximises the absorbed energy for 
a given force capacity. In this respect shear-thinning fluids such as the CY and PTT behave 
similarly to viscoplastic fluids \cite{Syrakos_2016}.

In terms of the force-velocity diagrams (Figs.\ \ref{fig: ND force - ND velocity 100}, \ref{fig: 
force - ND velocity 100}), shear thinning has the effect of reducing the slope of the curves as 
the velocity increases. On the contrary, for the N-100 fluid the slope of the force curve remains 
constant at all piston velocities (Fig.\ \ref{fig: ND force - ND velocity 100}). Thus, if one wishes 
to fit the power law \eqref{eq: viscous damper} to these force-velocity relationships then the 
exponent $n$ would be $n = 1$ for the Newtonian fluid and $n < 1$ for the CY and PTT fluids.

Another effect of shear-thinning is that it strengthens the role of inertia by weakening the 
viscous / viscoelastic stresses. This will be discussed in Sec.\ \ref{ssec: results; inertia}.

\begin{figure}[tb]
    \centering
    \begin{subfigure}[b]{0.49\textwidth}
        \centering
        \includegraphics[width=0.95\linewidth]{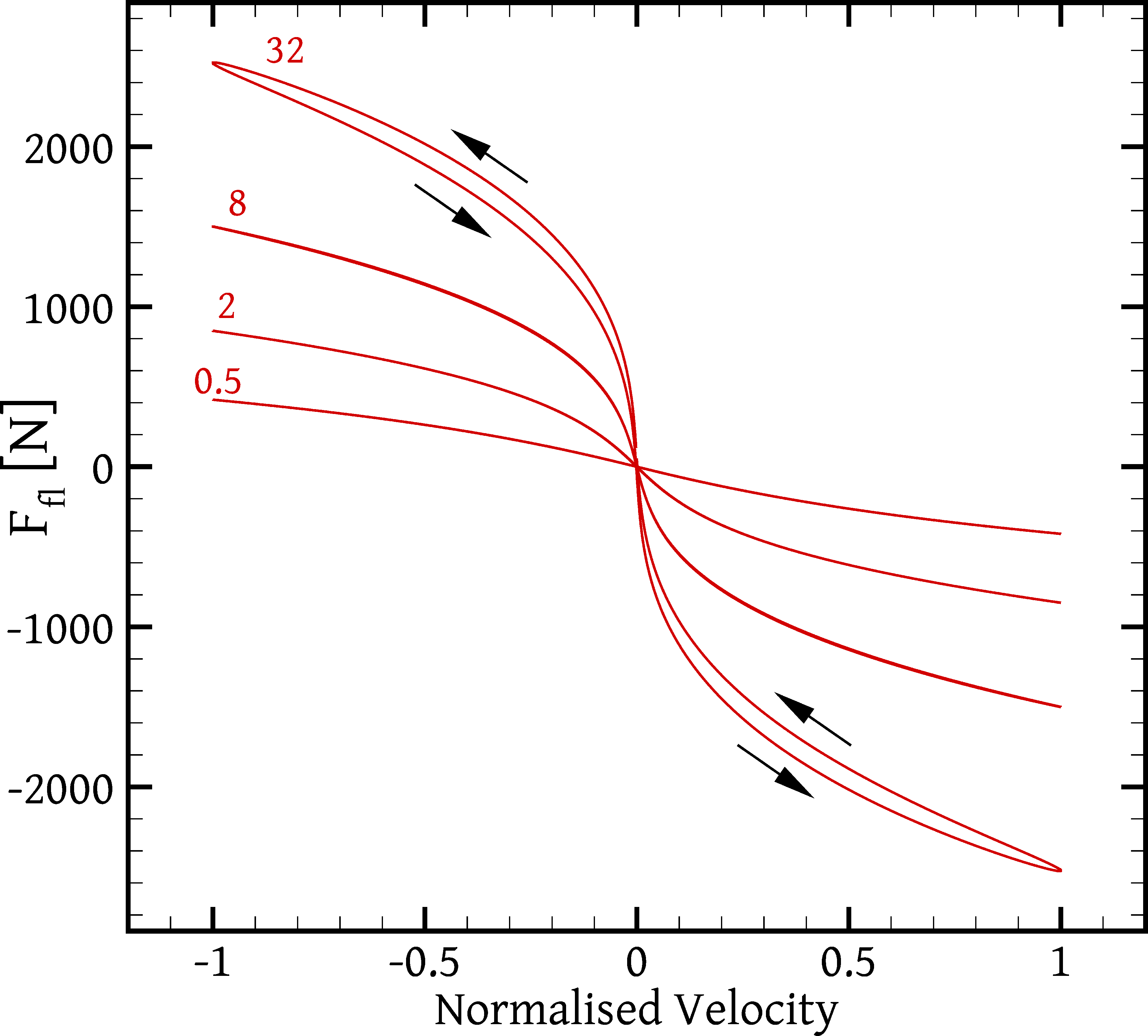}
        \caption{CY-100 fluid}
        \label{sfig: force - ND velocity CY-100}
    \end{subfigure}
    \begin{subfigure}[b]{0.49\textwidth}
        \centering
        \includegraphics[width=0.95\linewidth]{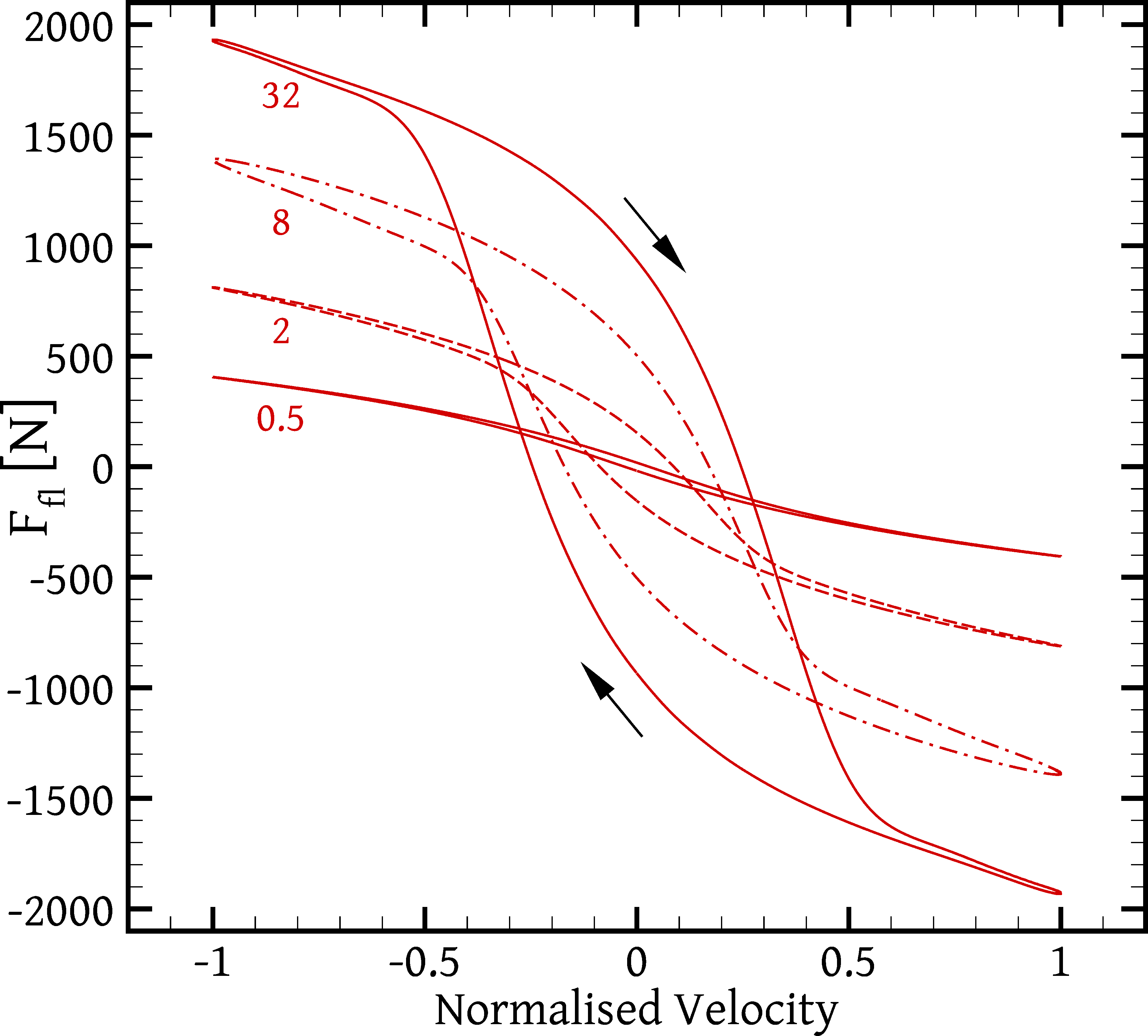}
        \caption{lPTT-100 fluid}
        \label{sfig: force - ND velocity lPTT-100}
    \end{subfigure}
    
 \caption{Diagrams of force $F_{fl}$ (Eq.\ \eqref{eq: fluid force}) versus piston velocity, the 
latter divided by its maximum value $U_p$ (Eq.\ \eqref{eq: piston velocity}, listed in Table 
\ref{table: operating conditions} for each frequency) for all the tested frequencies $f$ = 0.5, 2, 8 
and 32 \si{Hz}; the left diagram \subref{sfig: force - ND velocity CY-100} is for the CY-100 fluid, 
and the right diagram \subref{sfig: force - ND velocity lPTT-100} is for the lPTT-100 fluid.}
 \label{fig: force - ND velocity 100}
\end{figure}

\begin{figure}[tb]
    \centering
    \begin{subfigure}[b]{0.49\textwidth}
        \centering
        \includegraphics[width=0.95\linewidth]{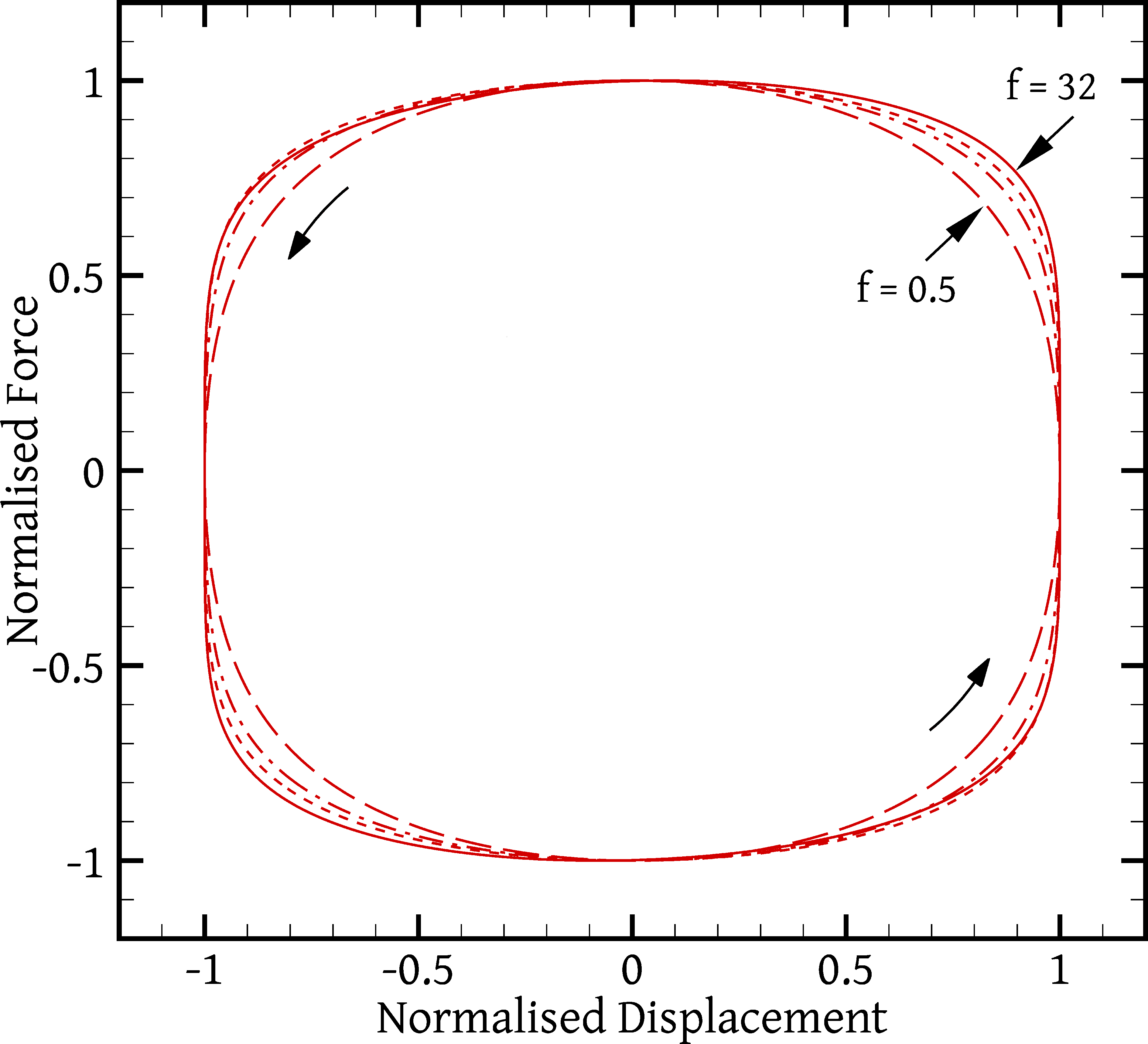}
        \caption{CY-100 fluid}
        \label{sfig: ND force - ND displacement CY-100}
    \end{subfigure}
    \begin{subfigure}[b]{0.49\textwidth}
        \centering
        \includegraphics[width=0.95\linewidth]{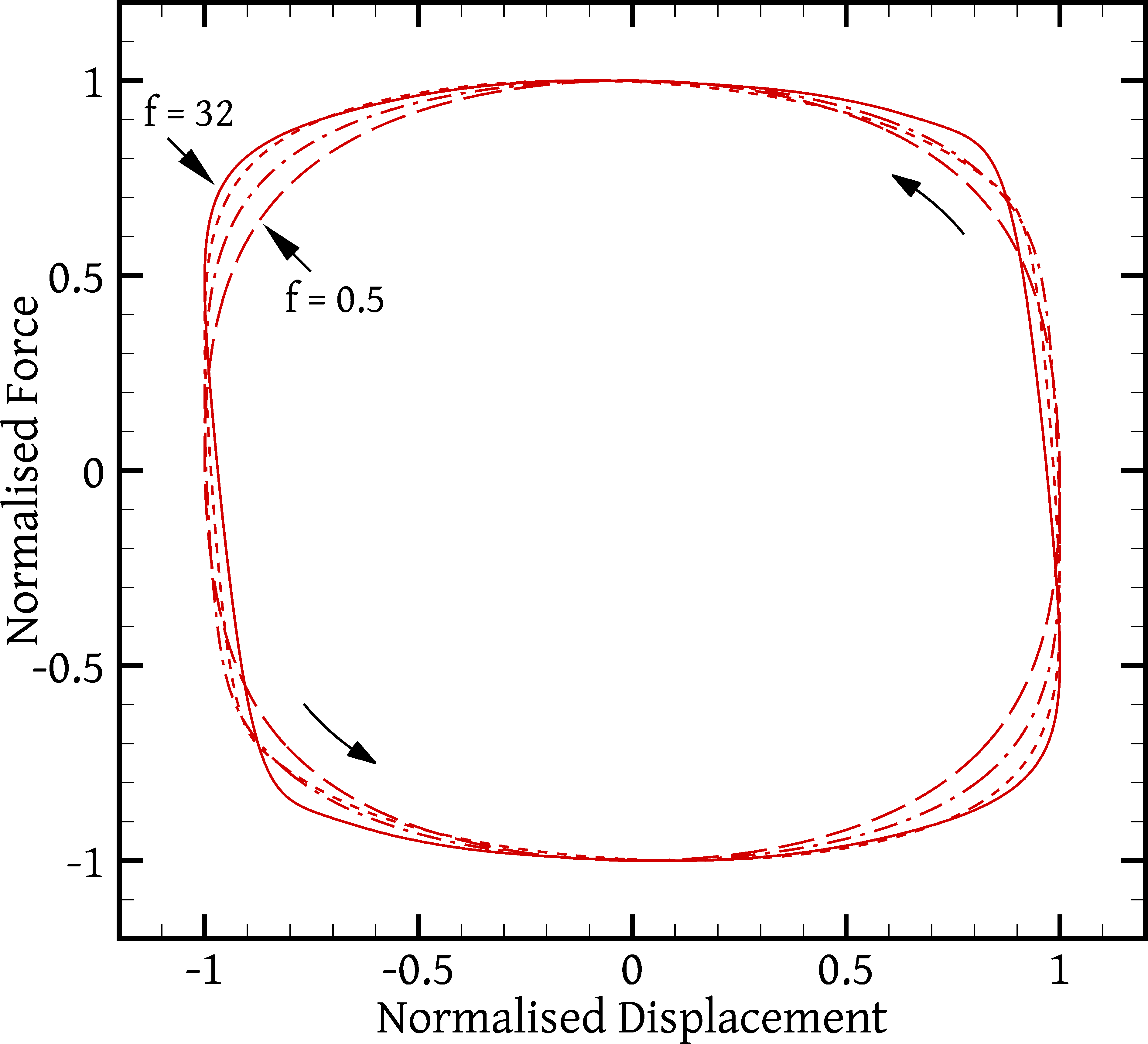}
        \caption{lPTT-100 fluid}
        \label{sfig: ND force - ND displacement lPTT-100}
    \end{subfigure}
    
\caption{Diagrams of force $F_{fl}$ (Eq.\ \eqref{eq: fluid force}) versus displacement during the 
last (sixth) period simulated, normalised by their maximum values in each case, for all the tested 
frequencies $f$ = 0.5 (long dashes), 2 (dash-dot), 8 (short dashes) and 32 (solid) \si{Hz}; the left 
diagram \subref{sfig: ND force - ND displacement CY-100} is for the CY-100 fluid, and the right 
diagram \subref{sfig: ND force - ND displacement lPTT-100} is for the lPTT-100 fluid.}
  \label{fig: ND force - ND displacement}
\end{figure}

\subsection{Effects of inertia}
\label{ssec: results; inertia}

As indicated by the Reynolds numbers listed in Table \ref{table: operating conditions}, fluid 
inertia plays only a minor role in this flow. When the fluid is Newtonian, inertia is not at all 
noticeable due to the high viscosity: force-displacement diagrams are perfectly elliptical (Figs.\ 
\ref{sfig: validation Newtonian}, \ref{sfig: force displacement f=0.5}) and force-velocity 
diagrams are lines rather than loops (Figs.\ \ref{sfig: force velocity 100 f=0.5 nd}, \ref{sfig: 
force velocity 100 f=32 nd}); no hysteresis is observed. On the other hand, some hysteresis, 
although mild, is evident in the corresponding plots of the CY fluids at $f$ = 32 \si{Hz}, and 
inertia is the only mechanism that can be responsible for this. The difference between the 
Newtonian and CY fluids is that the latter exhibit shear-thinning, which means that at high 
velocities the viscous forces acting on CY fluid particles are significantly smaller than those 
acting on the Newtonian fluid particles. Since the force balance is between viscosity, pressure, and 
inertia, smaller viscous forces implies a more significant role for inertia.

How inertia influences the flow can be observed in the force-displacement and force-velocity 
diagrams for the CY-100 fluid at $f$ = 32 \si{Hz}. In Fig.\ \ref{sfig: validation CY-100 f=32}, and 
in Fig.\ \ref{sfig: force velocity 100 f=32 nd} on the CY-100 curve, one observes that 
$|F_{fl}(\mathrm{a})| = |F_{fl}(\mathrm{c})| < |F_{fl}(\mathrm{b})|$. In other words, for the same 
piston displacement and velocity magnitude, the force is greater when the piston is accelerating 
than when it is decelerating. This is because during acceleration the piston has to accelerate also 
the surrounding fluid (added mass effect), and fluid inertia makes this harder; during deceleration, 
on the other hand, the piston also has to decelerate the surrounding fluid which pushes the piston 
forward due to its inertia, thus aiding to overcome the viscous fluid resistance and resulting in a 
smaller overall backwards force. The component of $F_{fl}$ that is related to the fluid inertia is 
depicted schematically in Fig.\ \ref{fig: damper states}. Inertia effects are only noticeable at the 
highest frequency considered: in Fig.\ \ref{sfig: ND force - ND displacement CY-100} one can observe 
that it is only at $f$ = 32 \si{Hz} that the loop is stretched more in the +45 degree direction than 
in the --45 degree direction, and in Fig.\ \ref{sfig: force - ND velocity CY-100} it is only at $f$ 
= 32 \si{Hz} that the force is plotted as a loop rather than a single line. The highly viscous 
nature of the fluids does not allow significant inertial effects in this particular application. 
More pronounced inertial effects were studied in \cite{Syrakos_2016}.

Inertia effects are not particular to the CY fluid, but concern the PTT fluid as well; however, for 
the latter things are complicated by the superposition of elasticity effects which in many respects 
act in an opposite manner to inertia. The discussion is deferred to Sec.\ \ref{ssec: results; 
elasticity}.

\subsection{Effects of elasticity}
\label{ssec: results; elasticity}

Viscoelasticity is responsible for the ``stiffness'' aspect of the damper's behaviour that was 
described in Sec.\ \ref{sec: introduction}. In particular, one can notice in Fig.\ \ref{sfig: 
validation lPTT-500 f=2} that $|F_{fl}(\mathrm{a})| = |F_{fl}(\mathrm{c})| > |F_{fl}(\mathrm{b})|$, 
i.e.\ the fluid reaction force is larger when the piston approaches the extreme positions and 
smaller when it is moving away from them. This can be attributed to fluid elasticity: the piston 
motion deforms the fluid, and the latter, being elastic, tends to recover its original shape 
pushing the piston backwards. In fig.\ \ref{sfig: sketch piston Dp Um} although the piston has 
reversed its direction of motion compared to Fig.\ \ref{sfig: sketch piston Dp Up}, there was not 
enough time to completely undo the fluid deformation caused by the preceding rightward piston motion 
and the elastic component of $F_{fl}$ continues to push the piston towards the left. Thus, when the 
piston is close to the extremities the elastic component of $F_{fl}$ acts in a spring-like manner, 
pushing the piston towards the centre of the damper. This is completely at odds with the effect of 
inertia (Fig.\ \ref{fig: damper states}). In force-displacement and force-velocity diagrams elastic 
effects are manifest in the form of a hysteresis that has the exact opposite sense than that of 
inertia: the force-displacement loops are stretched in the --45 degree direction (Fig.\ \ref{sfig: 
ND force - ND displacement lPTT-100}) rather than in the +45 degree direction (Fig.\ \ref{sfig: ND 
force - ND displacement CY-100}) and the force-velocity loops have a clockwise sense (Fig.\ 
\ref{sfig: force - ND velocity lPTT-100}) rather than a counterclockwise one (Fig.\ \ref{sfig: force 
- ND velocity CY-100}). This opposite manifestation of inertial effects compared to elastic effects 
has been observed also in LAOS experiments \cite{Hashemi_2017}. Of course, at $f$ = 32 \si{Hz} 
inertia effects are also expected to play a role for the lPTT-100 fluid, similarly to the CY-100 
case, but for this particular flow they are somewhat masked by the elastic effects which are 
dominant. Nevertheless, it will be shown below that the weakening of the elastic hysteresis observed 
at normalised velocity magnitudes greater than 0.5 for $f$ = 32 \si{Hz} in Fig.\ \ref{sfig: force - 
ND velocity lPTT-100} compared to $f$ = 8 \si{Hz} is due to a superposition of an opposed inertial 
hysteresis.

It would be nice if the observable effect of elasticity could somehow be quantified instead of 
relying only on qualitative observations. A simple way to do this arises from a decomposition of 
the fluid force $F_{fl}$ into ``viscous'' and ``elastic'' components, which is inspired by an 
analogous procedure applied in LAOS experiments \cite{Cho_2005}. For a purely elastic medium, one 
would expect the force to be a function of only the piston displacement, $F_{fl}(x_p)$, while for a 
purely viscous medium one would expect it to be a function of only the piston velocity, 
$F_{fl}(u_p)$. In practice, due to the mixed character of silicone oil, the force-displacement and 
force-velocity diagrams are hysteretic, meaning that for each piston displacement the force has two 
distinct values, depending on the piston velocity, and similarly for each piston velocity the force 
has two distinct values, depending on the piston displacement. The force therefore depends on both 
the piston displacement and velocity, $F_{fl}(x_p,u_p)$. A simple way to split this force is the 
following:
%^b
\begin{equation} \label{eq: F splitting preliminary}
 F_{fl}(x_p,u_p) \;=\; \frac{F_{fl}(x_p,u_p) + F_{fl}(x_p,-u_p)}{2} \;+\; 
                       \frac{F_{fl}(x_p,u_p) - F_{fl}(x_p,-u_p)}{2}
\end{equation}
%^a
If the periodic state has been reached, then from symmetry considerations it follows that 
$F_{fl}(-x_p,-u_p) = - F_{fl}(x_p,u_p)$ (the forces at instances (a) and (c) of Fig.\ \ref{fig: 
damper states} are of equal magnitude but opposite sense), which also means that $F_{fl}(x_p,-u_p) 
= -F_{fl}(-x_p,u_p)$. Substituting this last equation in Eq.\ \eqref{eq: F splitting preliminary} 
we obtain
%^b
\begin{equation} \label{eq: F splitting}
 F_{fl}(x_p,u_p) \;=\; F^e_{fl}(x_p) \;+\; F^v_{fl}(u_p)
\end{equation}
%^a
where
\begin{equation} \label{eq: F_e}
 F^e_{fl}(x_p) \;\equiv\; \frac{F_{fl}(x_p,u_p) + F_{fl}(x_p,-u_p)}{2}
\end{equation}
\begin{equation} \label{eq: F_v}
 F^v_{fl}(u_p) \;\equiv\; \frac{F_{fl}(x_p,u_p) + F_{fl}(-x_p,u_p)}{2}
\end{equation}
%^a
The reason why the component $F^e_{fl}$ can be considered to be a function of only $x_p$ (for the 
given operating conditions) is that expression \eqref{eq: F_e} is an even function of $u_p$, i.e.\ 
it has the same value at point $(x_p,u_p)$ as it has at the point $(x_p,-u_p)$. Thus, whereas in a 
plot of $F_{fl}$ versus $x_p$, for each piston displacement $x_p$ there are two distinct $F_{fl}$ 
values, one corresponding to the piston moving towards the right with velocity $u_p$ and one 
corresponding to it moving towards the left with velocity $-u_p$, in a plot of $F^e_{fl}$ versus 
$x_p$ these two values coincide, and the plot is a single line rather than a loop (no hysteresis). 
Thus $F^e_{fl}$ is completely determined by the piston displacement $x_p$, as for a force of purely 
elastic origin. The averaging of $F_{fl}(x_p,u_p)$ and $F_{fl}(x_p,-u_p)$ in expression \eqref{eq: 
F_e} can be assumed to cause a cancellation of the viscous effects of the opposite velocities $u_p$ 
and $-u_p$, leaving only the elastic effect of the displacement $x_p$ on the force $F_{fl}$. Hence 
$F^e_{fl}$ will be termed the ``elastic'' component of $F_{fl}$.

In exactly the same way, the ``viscous'' component $F^v_{fl}$, eq.\ \eqref{eq: F_v}, of $F_{fl}$, 
comes from averaging the values $F_{fl}(x_p,u_p)$ and $F_{fl}(-x_p,u_p)$ to eliminate the opposite 
elastic effects of the displacements $x_p$ and $-x_p$, leaving only the viscous effect of the 
piston velocity $u_p$. The expression \eqref{eq: F_v} is an even function of $x_p$ and thus the 
plot of $F^v_{fl}$ versus $u_p$ is a single line, not a loop, as for a force of purely viscous 
origin.

\begin{figure}[tb]
    \centering
    \begin{subfigure}[b]{0.49\textwidth}
        \centering
        \includegraphics[width=0.95\linewidth]{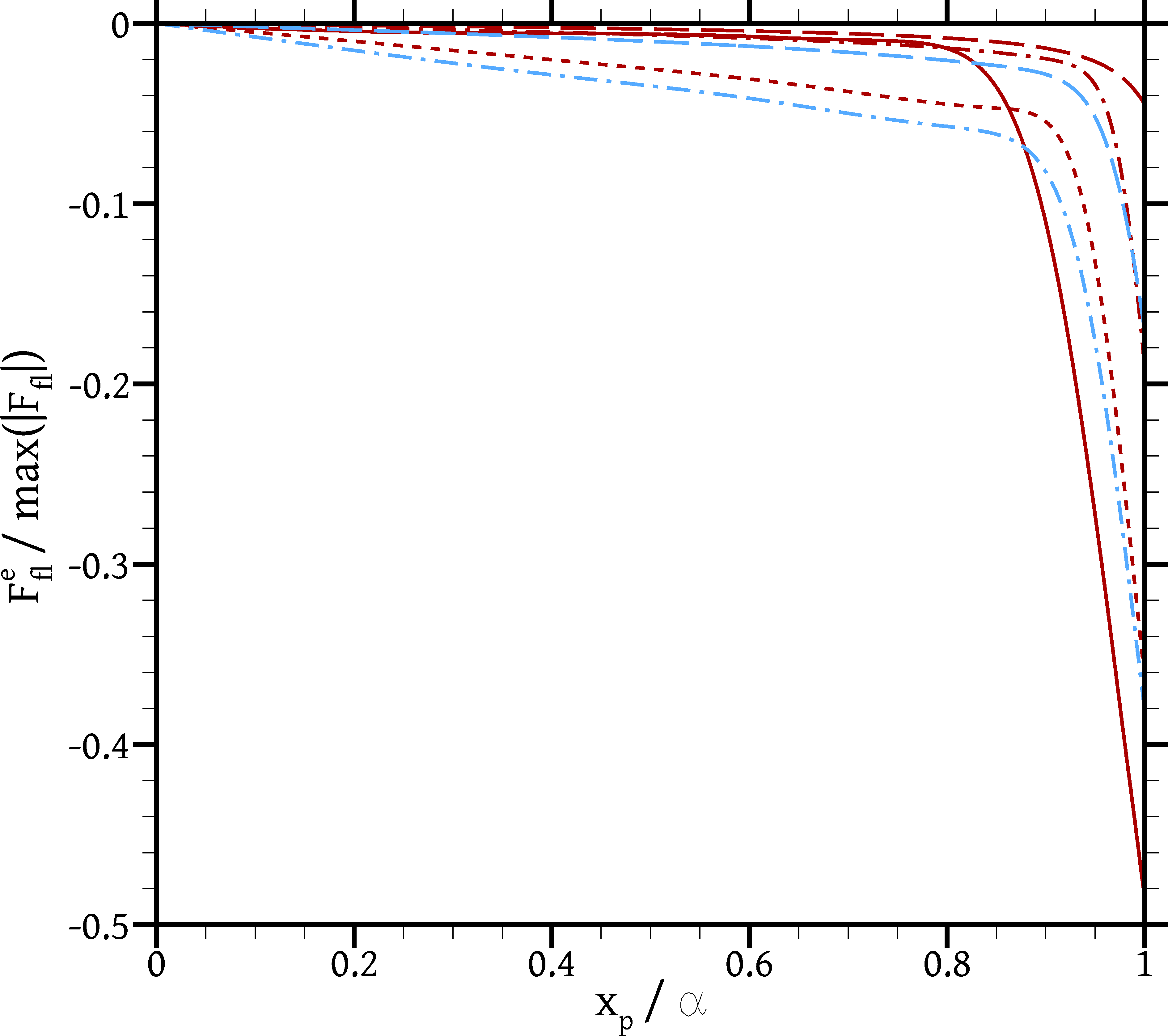}
        \caption{}
        \label{sfig: F elastic}
    \end{subfigure}
    \begin{subfigure}[b]{0.49\textwidth}
        \centering
        \includegraphics[width=0.95\linewidth]{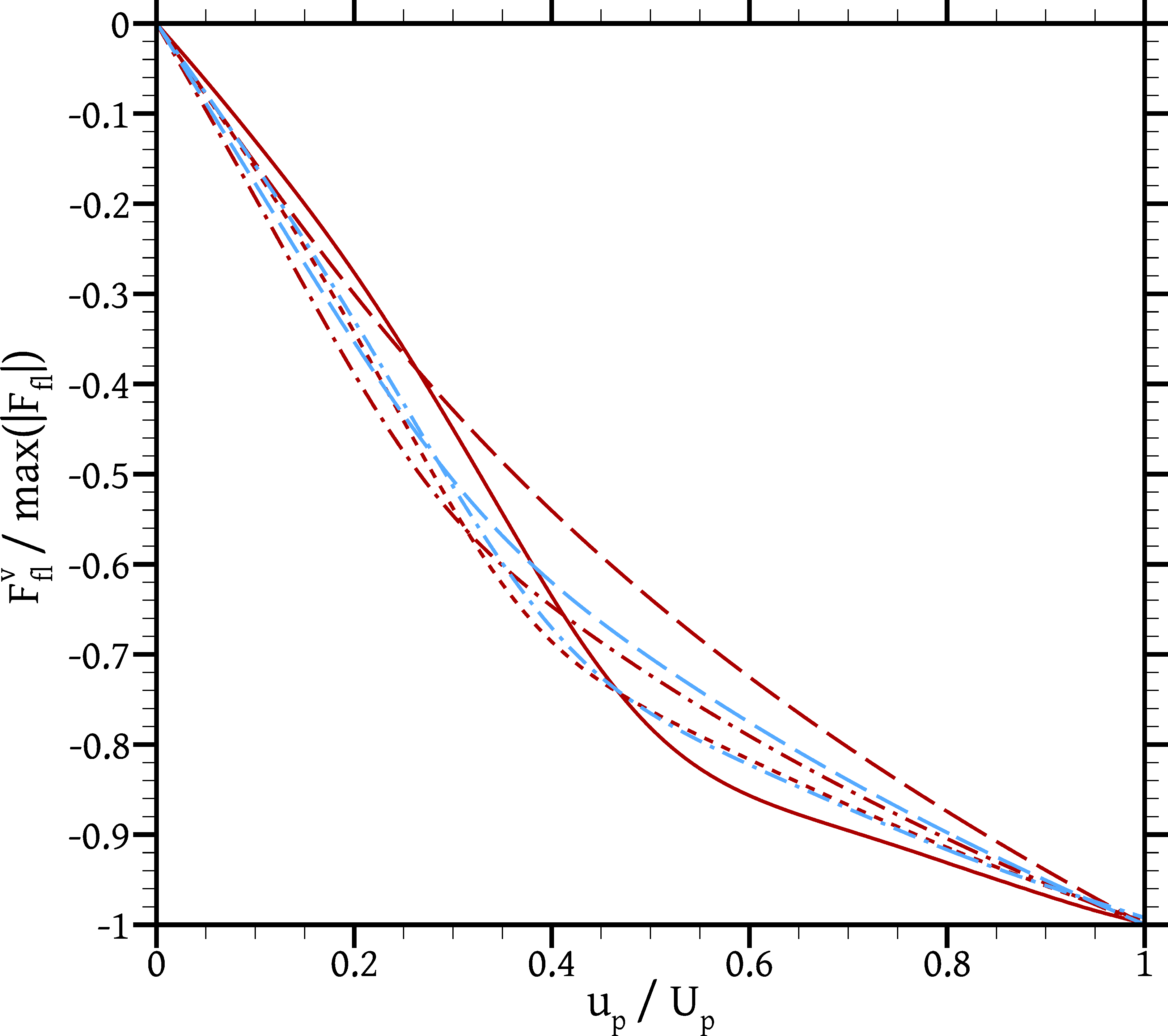}
        \caption{}
        \label{sfig: F viscous}
    \end{subfigure}
    
\caption{\subref{sfig: F elastic} Elastic force component $F^e_{fl}$ (Eq.\ \eqref{eq: F_e}), 
normalised by the maximum magnitude of $F_{fl}$, against piston displacement normalised by the 
oscillation amplitude $\alpha$. \subref{sfig: F viscous} Viscous force component $F^v_{fl}$ (Eq.\ 
\eqref{eq: F_v}), normalised by the maximum magnitude of $F_{fl}$, against normalised piston 
velocity ($u_p / U_p$, Eq.\ \eqref{eq: piston velocity}). The dark red lines correspond to the 
lPTT-100 fluid, and the light blue lines to the lPTT-500 fluid. Line style denotes the oscillation 
frequency: $f$ = 0.5 (long dashes), 2 (dash-dot), 8 (short dashes), and 32 \si{Hz} (continuous).}
  \label{fig: Force decomposition}
\end{figure}

Putting these ideas to practice, normalised diagrams of $F^e_{fl}$ versus displacement and of 
$F^v_{fl}$ versus velocity are plotted in Figs.\ \ref{sfig: F elastic} and \ref{sfig: F viscous}, 
respectively, for all cases. The diagrams are plotted only for positive displacement and velocity, 
due to symmetry ($F^e_{fl}(-x_p) = -F^e_{fl}(x_p)$ and $F^v_{fl}(-u_p) = -F^v_{fl}(u_p)$). It is 
noted that, from Eq.\ \eqref{eq: F_e}, the curve of $F^e_{fl}$ versus displacement is just the 
average between the upper and lower branches of the $F_{fl}$ vs.\ displacement loop, and similarly, 
from Eq.\ \eqref{eq: F_v}, the curve of $F^v_{fl}$ vs.\ velocity is just the average between the 
upper and lower branches of the the $F_{fl}$ vs.\ velocity loop. Figure \ref{sfig: F elastic} shows 
that the elastic component $F^e_{fl}$ is quite small over most of the piston stroke but increases 
sharply as the piston approaches an extreme position. Concerning the viscous component, Fig.\ 
\ref{sfig: F viscous} shows that as the velocity $u_p$ increases the magnitude of the derivative 
$\mathrm{d}F^v_{fl}/\mathrm{d}u_p$ decreases, which is directly related to the shear-thinning 
property of the fluid. In general, this manifestation of shear-thinning becomes more pronounced as 
either the oscillation frequency or the fluid relaxation time increases. 

One possible measure of the elastic character of the fluid force $F_{fl}$ is the ratio of the 
maximum absolute value of $F^e_{fl}$ to the maximum absolute value of $F_{fl}$. The value of this 
ratio is given in Table \ref{table: elasticity magnitudes} for all cases. It can reach quite 
high values -- nearly 0.5 for the lPTT-100, $f$ = 32 \si{Hz} case. From Fig.\ \ref{sfig: F 
elastic}, the maximum absolute value of $F^e_{fl}$ occurs at the extreme displacements $x_p = \pm 
\alpha$, where in fact $F^e_{fl} = F_{fl}$ because $F^v_{fl} = 0$ (this follows from the definition 
\eqref{eq: F_v} combined with the fact that $u_p = 0$ at $x_p = \pm \alpha$ and the aforementioned 
symmetry condition $F_{fl}(x_p,-u_p) = -F_{fl}(-x_p,u_p)$). If the simulation cases are sorted 
according to the values of this ratio then the order is roughly the same as that which results from 
sorting according to the $\Wei$ or $\Deb$ numbers (Table \ref{table: operating conditions}).

\begin{table}[thb]
\caption{Ratios of maximum magnitude of $F^e_{fl}$ (Eq.\ \eqref{eq: F_e}) over maximum magnitude of 
$F_{fl}$ and of $W^e_{fl}$ over $W_{fl}$ (Eqs.\ \eqref{eq: elastic work} and \eqref{eq: viscous 
work}), expressed as percentages, for various cases. The numbers in parentheses correspond to the 
line printed with long dashes in Fig.\ \ref{fig: F inertia}.}
\label{table: elasticity magnitudes}
\begin{center}
\begin{small}   % to make the font smaller
\renewcommand\arraystretch{1.25}   % Adjust row height (default=1)
{%scope of \G
\newcommand{\e}[1]{$\cdot 10^{-#1}$}
\begin{tabular}{ c | c | c | c }
\toprule
 Fluid  &  $f$ [\si{Hz}]  &  $\frac{\max(F^e_{fl})}{\max(F_{fl})}$ [\%]  &  $W^e_{fl}/ W_{fl}$ [\%] 
\\ \midrule
\multirow{4}{*}{lPTT-100} &     0.5    &    4.5           &    0.33
\\
                       &        2      &   19.0           &    0.67
\\
                       &        8      &   36.3           &    2.12
\\
                       &       32      &   48.4 (52.2)    &    2.08 (3.18)
\\ \midrule
\multirow{2}{*}{lPTT-500} &     0.5    &   17.0           &    0.91
\\
                       &        2      &   38.1           &    2.76
\\
\bottomrule
\end{tabular}
} %end scope of \G
\end{small}
\end{center}
\end{table}

The large values of the ratio $\max(F^e_{fl})/\max(F_{fl})$ may give an exaggerated picture about 
the overall effect of elasticity, since $F^e_{fl}$ has very small values over most of the piston 
stroke and only increases close to $x = \pm \alpha$ (Fig.\ \ref{sfig: F elastic}). A better picture 
can be obtained by comparing the integrals
%^b
\begin{equation} \label{eq: viscous work}
 W_{fl} \;\equiv\; \oint F_{fl} \; \mathrm{d} x_p 
\end{equation}
%^a
and
%^b
\begin{equation} \label{eq: elastic work}
 W^e_{fl} \;\equiv\; 2 \int_{x_p = 0}^{x_p = \alpha} F^e_{fl} \; \mathrm{d} x_p
\end{equation}
%^a
The integral $W_{fl}$ \eqref{eq: viscous work} is the work done by the fluid force $F_{fl}$ during 
an oscillation cycle, which is the energy dissipated into heat. Since the elastic component 
$F^e_{fl}$ exhibits no hysteresis when plotted against $x_p$ (its plots in Fig.\ \ref{sfig: F 
elastic} are single lines and not loops), its net work during a full oscillation is zero, $\oint 
F^e_{fl} \, \mathrm{d} x_p = 0$. Hence, all of the energy dissipation occurs through the 
viscous component $F^v_{fl}$:
%^b
\begin{equation*}
 \oint F_{fl} \; \mathrm{d} x_p \;=\; \oint \left( F^v_{fl} + F^e_{fl} \right) \, \mathrm{d} x_p
 \;=\; \oint F^v_{fl} \, \mathrm{d} x_p  \;+\;  \oint F^e_{fl} \, \mathrm{d} x_p
 \;=\; \oint F^v_{fl} \, \mathrm{d} x_p  \;+\;  0
\end{equation*}
%^a
However, $F^e_{fl}$ causes storage and release of energy within an oscillation cycle. As the piston 
approaches an extreme position, $F^e_{fl}$ opposes this motion and converts some of the supplied 
energy into potential energy in the form of elastic energy of the fluid. This energy is then 
released as the piston retracts from the extreme position, when $F^e_{fl}$ is pushing it along. The 
energy storage during an oscillation cycle, omitting the equal energy release, is given by the 
integral $W^e_{fl}$, Eq.\ \eqref{eq: elastic work}. Thus the ratio $W^e_{fl} / W_{fl}$, given in 
Table \ref{table: elasticity magnitudes} for the various cases, is an indicator of the importance 
of elastic effects. It can be seen from the low values of this ratio in Table \ref{table: 
elasticity magnitudes} that the action of the damper is mostly dissipative, with the largest energy 
storage recorded at about 3\%.

\begin{figure}[tb]
  \centering
  \includegraphics[scale=1.00]{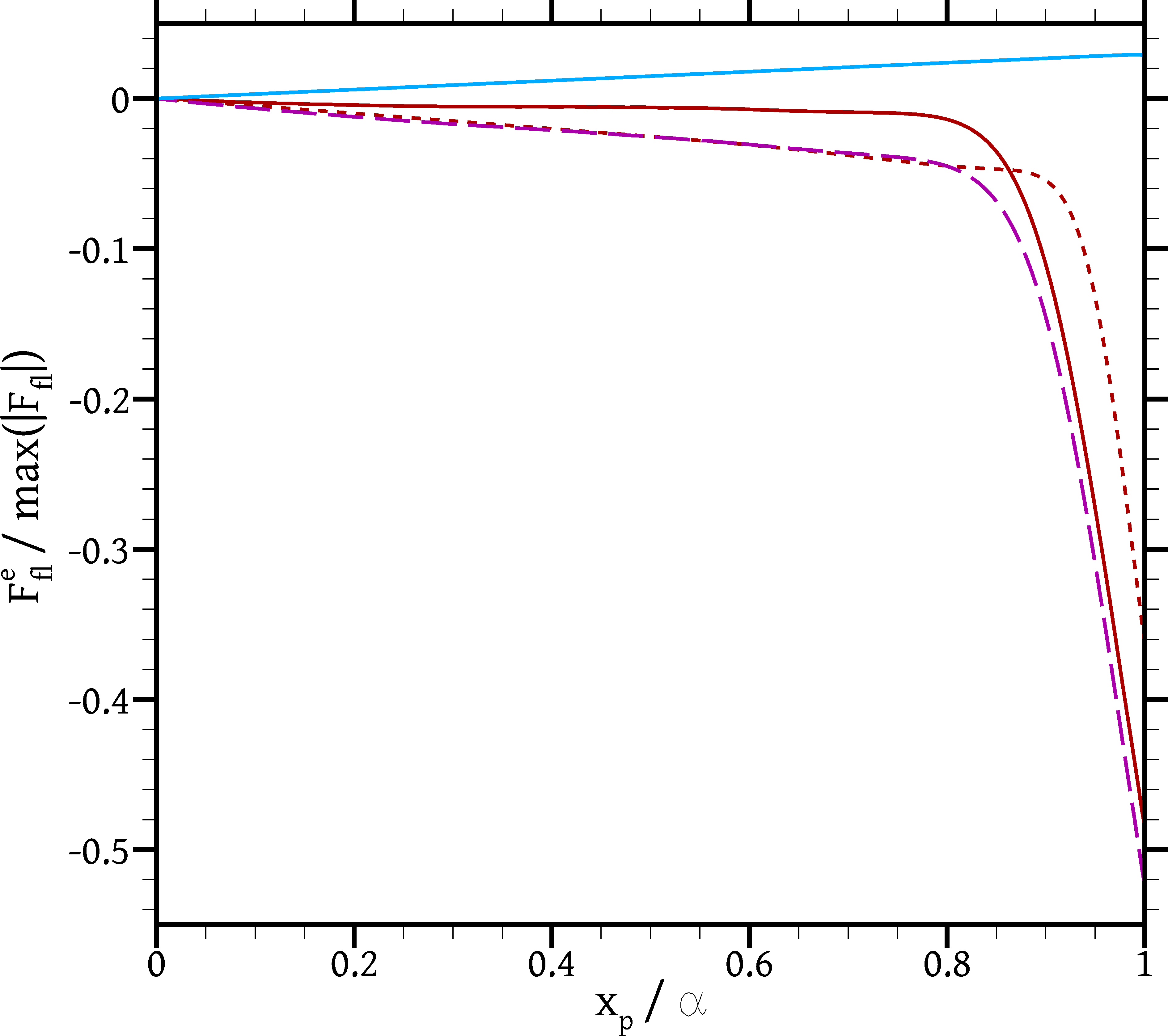}
  \caption{The force component $F^e_{fl}$, Eq.\ \eqref{eq: F_e}, normalised by the maximum value of 
$F_{fl}$, against normalised piston displacement $x_p / \alpha$, for the following cases: lPTT-100 
at $f$ = 32 \si{Hz} (continuous dark red line), lPTT-100 at $f$ = 8 \si{Hz} (short dashes, dark 
red), and CY-100 at $f$ = 32 \si{Hz} (continuous light blue line). The purple line of long dashes 
corresponds again to the lPTT-100, $f$ = 32 \si{Hz} case where, however, from its $F^e_{fl}$ force 
the corresponding component $F^e_{fl}$ of the CY-100, $f$ = 32 \si{Hz} case has been subtracted 
(normalisation is again by the maximum $F_{fl}$ value of the lPTT-100, $f$ = 32 \si{Hz} case).}
  \label{fig: F inertia}
\end{figure}

Interestingly, the ratio $W^e_{fl} / W_{fl}$ for the lPTT-500, $f$ = 2 \si{Hz} case is 
significantly larger than for the lPTT-100, $f$ = 32 \si{Hz} case for which the $\Wei$ and $\Deb$ 
numbers are highest. This is connected to the fact that $F^e_{fl}$ for the latter case is very small 
up to $x_p \approx 0.8\alpha$ (Fig.\ \ref{sfig: F elastic}), and one cannot help but wonder whether 
this is due to inertia effects masking the elastic effects. This can be checked by plotting 
$F^e_{fl}$ for the corresponding CY-100, $f$ = 32 \si{Hz} case for which there is no elasticity. 
This plot, in Fig.\ \ref{fig: F inertia}, shows that the ``elastic'' component $F^e_{fl}$ is in fact 
non-zero, and that its direction (the sign of $F^e_{fl}$ in Fig.\ \ref{fig: F inertia}) is exactly 
the opposite of what would be expected of an elastic force: as the piston moves towards $x_p = 
\alpha$, $F^e_{fl}$ keeps pushing it forward instead of opposing the motion. Obviously, $F^e_{fl}$ 
is in this case the inertial component of the total force $F_{fl}$ (Fig.\ \ref{fig: damper states}). 
One notices also in Fig.\ \ref{fig: F inertia} that in the CY-100 case $F^e_{fl}$ increases linearly 
with the piston displacement $x_p$. This is not surprising, and shows that the inertial component of 
the force is proportional to the piston acceleration (which is in phase with the displacement, 
comparing Eqs.\ \eqref{eq: shaft position} and \eqref{eq: shaft acceleration}); the fluid 
acceleration, and hence the required force, follows that of the piston. In the lPTT-100 case (for 
$f$ = 32 \si{Hz}, which is the only case when inertia is noticeable) $F^e_{fl}$ is the net sum of 
the elastic and inertial components, which oppose each other. Both components store energy to 
release it later (elastic potential energy in the case of the elastic component and fluid kinetic 
energy in the case of the inertial component) each with a net work of zero during a full 
oscillation, but as one is storing the other is releasing and vice versa. If we want to isolate only 
the elastic part of the force, we can assume that the inertial part is the same as for the CY-100 
fluid at the same frequency (which has no elastic part) and subtract it from the whole, i.e.\ the 
purely elastic component of $F^e_{fl}$ for a lPTT fluid can be approximated as 
$F^e_{fl}(\mathrm{lPTT}) - F^e_{fl}(\mathrm{CY})$. This is plotted in Fig.\ \ref{fig: F inertia} 
(line with long dashes) and it may be seen to no longer be negligible for displacements $x_p/\alpha 
< 0.8$ compared, for example, to the lPTT-100, $f$ = 8 \si{Hz} case. Also, in Table \ref{table: 
elasticity magnitudes} the corresponding metrics are printed in parentheses, and the $W^e_{fl} / 
W_{fl}$ ratio can be seen to now be the highest among all cases, in correspondence with that case's 
$\Wei$ and $\Deb$ numbers' ranking.

\begin{figure}[tb]
  \centering
  \includegraphics[scale=0.90]{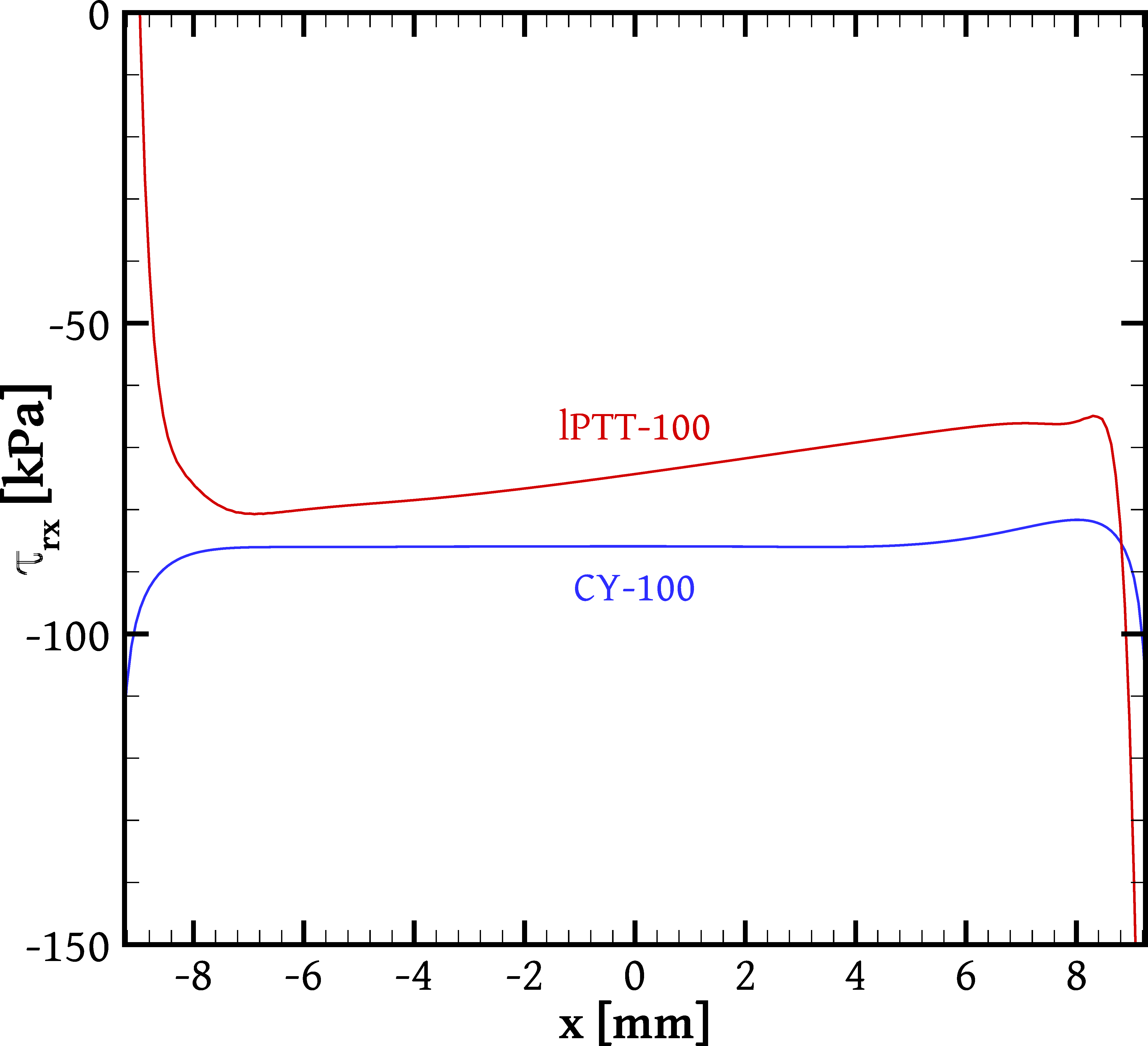}
  \caption{Distributions of the $\tau_{rx}$ stress component along the horizontal surface of the 
piston for the CY-100 (blue) and lPTT-100 (red) fluids, at a time instance when the piston is 
moving with maximum velocity towards the right; the oscillation frequency is $f$ = 32 \si{Hz}.}
  \label{fig: stress on piston}
\end{figure}

Another phenomenon that must be attributed to the elasticity of the fluid is the systematically 
lower $F_{fl}$ values observed for the PTT fluids compared to the corresponding CY fluids, at the 
same frequencies. Figure \ref{fig: force displacement} and Table \ref{table: operating conditions} 
show that the higher the Deborah number is, the lower $F_{fl}$ is for a PTT fluid compared to the 
CY fluid of the same nominal viscosity, at the same frequency. Since the CY parameters have been 
fitted so that the CY fluids' behaviour exactly matches that of the corresponding PTT fluids in 
steady shear (Fig.\ \ref{fig: steady shear viscosity}), this observation indicates that for the PTT 
fluids shear-thinning is not the only factor that causes a force reduction compared to Newtonian 
fluids, but there is an additional factor that cannot be any other than elasticity. To investigate 
this, in Fig.\ \ref{fig: stress on piston} we plot the distribution of $\tau_{rx}$ on the 
horizontal surface of the piston, i.e.\ along its length excluding the rounded corners, at maximum 
piston velocity towards the right, for both the CY-100 and lPTT-100 fluids at the $f$ = 32 \si{Hz} 
frequency for which the force difference between the two fluids is the largest. The stresses are 
negative, pushing the piston towards the left. As mentioned, this $\tau_{rx}$ distribution is 
mostly the cause of $F_{fl}$ both directly and indirectly through the pressure difference that it 
generates across the piston. Figure \ref{fig: stress on piston} shows that whereas for the CY fluid 
$\tau_{rx}$ is constant over most of the piston length, for the PTT fluid this stress component 
increases in magnitude as one moves downstream (i.e.\ towards the left). At the left end of the 
piston the PTT stress magnitude has not yet caught up with the CY stress levels; overall, 
$\tau_{rx}$ has larger magnitude for the CY fluid, resulting in a larger force $F_{fl}$.

\begin{figure}[tb]
    \centering
    \begin{subfigure}[b]{0.49\textwidth}
        \centering
        \includegraphics[width=0.95\linewidth]{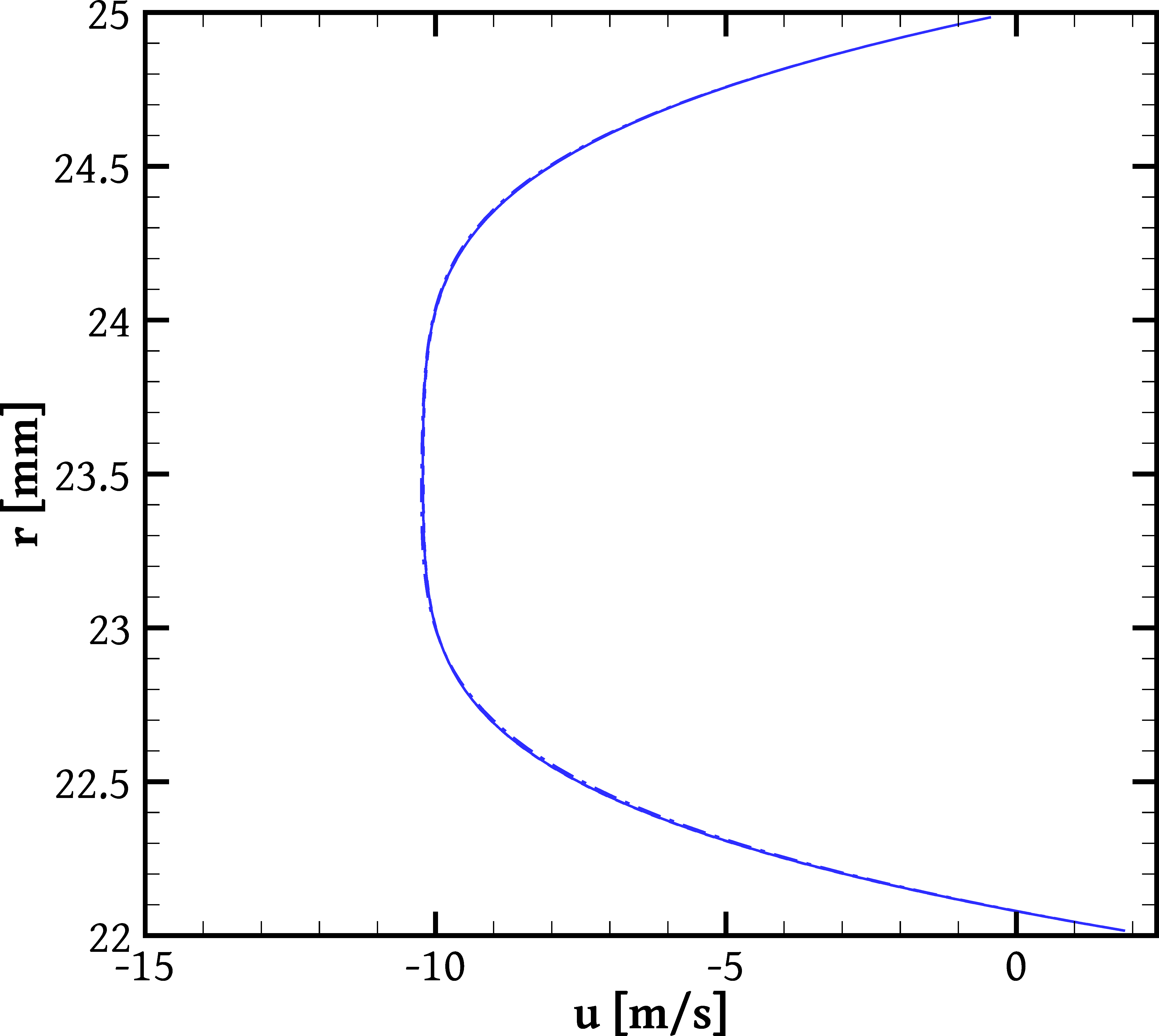}
        \caption{CY-100 fluid}
        \label{sfig: velocity profiles CY-100}
    \end{subfigure}
    \begin{subfigure}[b]{0.49\textwidth}
        \centering
        \includegraphics[width=0.95\linewidth]{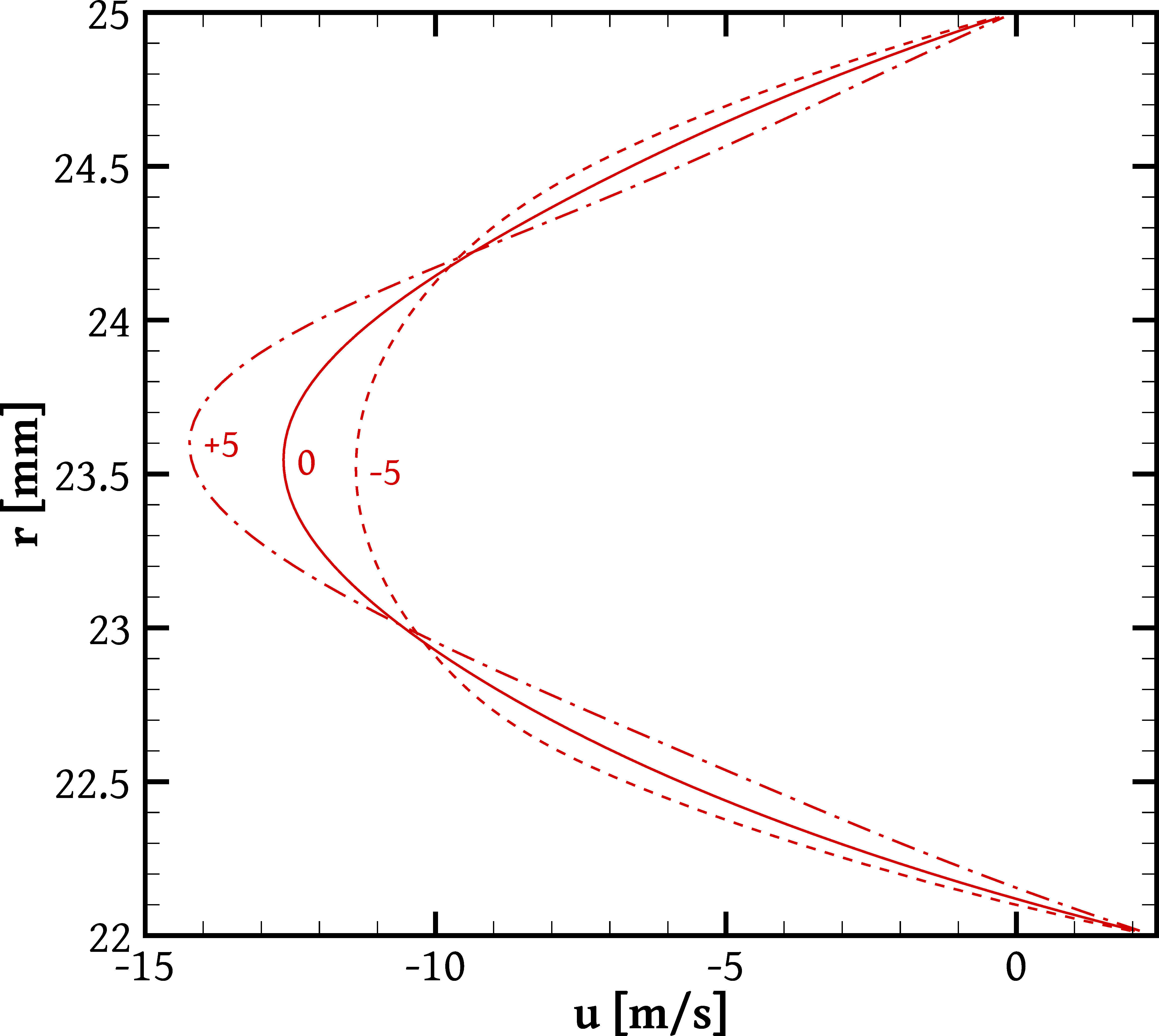}
        \caption{lPTT-100 fluid}
        \label{sfig: velocity profiles lPTT-100}
    \end{subfigure}
    
\caption{Profiles of the $x$-component of fluid velocity across the gap, at a time instance when 
the piston is moving towards the right with maximum velocity, with its midpoint at $x = 0$, for the 
CY-100 \subref{sfig: velocity profiles CY-100} and lPTT-100 \subref{sfig: velocity profiles 
lPTT-100} fluids. The frequency is $f = 32$ \si{Hz}. Three profiles are plotted in each case: at 
halfway along the piston ($x = 0$, solid line), at $5$ \si{mm} upstream ($x = +5$ \si{mm}, dash-dot 
line), and at $5$ \si{mm} downstream ($x = -5$ \si{mm}, dashed line). The profiles nearly coincide 
in the CY-100 case \subref{sfig: velocity profiles CY-100}.}
  \label{fig: velocity profiles}
\end{figure}

These observations concerning $\tau_{rx}$ can be explained by looking at velocity profiles across 
the gap, from the piston surface to the outer casing, such as those drawn in Fig.\ \ref{fig: 
velocity profiles}. Figure \ref{sfig: velocity profiles CY-100} shows that once the CY fluid enters 
the gap it very quickly acquires its fully developed velocity profile, a profile which is typical 
of a shear-thinning generalized Newtonian fluid, exhibiting flatness away from the walls and a 
steep gradient close to the walls. In contrast, Fig.\ \ref{sfig: velocity profiles lPTT-100} shows 
that the flow of the PTT fluid that enters the gap evolves much more slowly and does not reach a 
fully developed profile before the fluid exits at the other side of the gap. The PTT profile is 
initially more ``pointed'', near the gap entrance, and evolves towards a flatter profile, more 
reminiscent of the CY profile, as the fluid moves downstream. A more pointed profile entails a less 
steep velocity gradient at the piston wall, hence the lower $\tau_{rx}$ magnitudes for the PTT 
fluid 
compared to the CY fluid. Similar observations have been reported for contraction flows of 
viscoelastic fluids (both shear-thinning and non-shear-thinning (Boger) fluids), for which both 
experimental and numerical studies \cite{Saramito_1994, Alves_2003b, Oliveira_2007, Alves_2007, 
Sousa_2009, Sousa_2011} have shown that the fluid velocity along the channel centreline is maximum 
at the contraction entrance and drops towards a limit value as the fluid moves downstream.

This different behaviour between the two fluids reflects the different importance of convective 
terms in their respective governing equations. For the CY fluid, the convective terms of the 
momentum equation \eqref{eq: momentum nd} are not significant due to the low value of $\Rey$, and 
therefore the flow is elliptic in character and adapts almost instantaneously to its local 
surroundings, thus resulting in a very short development length in the gap. For the PTT fluid, the 
convective terms of the momentum equation are insignificant as well, but now the constitutive 
equation \eqref{eq: constitutive lPTT nd} also includes convection terms which are quite important 
and make a fluid particle's state heavily dependent on its flow history and not only on its 
surroundings. This results in a large development length that essentially stretches throughout the 
gap. To get a better feel of this, one can compare the average time that a fluid particle spends 
travelling along the gap against the relaxation time of the fluid. At maximum piston velocity, the 
mean relative velocity between the fluid and the piston is $U$, listed in Table \ref{table: 
operating conditions} for the various cases. Dividing the gap length $L_p$ (Table \ref{table: 
damper dimensions}) by this velocity gives the average time that it takes for a fluid particle to 
cross the gap (at maximum piston velocity), $T_p = L_p / U$ say. Dividing this time with the 
relaxation time of the fluid, we get the following results: for the lPTT-100 fluid, $T_p / \lambda$ 
= 12.5 ($f$ = 0.5 \si{Hz}), 3.1 ($f$ = 2 \si{Hz}), 0.8 ($f$ = 8 \si{Hz}) and 0.2 ($f$ = 32 
\si{Hz}); 
for the lPTT-500 fluid, $T_p / \lambda$ = 2.1 ($f$ = 0.5 \si{Hz}) and 0.5 ($f$ = 2 \si{Hz}). 
Therefore, in the higher $\Deb$ number cases considered, the particle travel time is smaller than 
the relaxation time of the fluid, and a fluid particle does not have enough time to reach a 
``steady 
state'' until it reaches the exit of the gap.

The conclusion is then that viscoelastic flow can entail large development lengths in the gap, and 
simplified one-dimensional analyses such as those mentioned in Sec.\ \ref{sec: introduction} may 
lead to significant underestimation of the force.

\subsection{Applied force}
\label{ssec: results; Fap}

\begin{figure}[tb]
    \centering
    \begin{subfigure}[b]{0.35\textwidth}
        \centering
        \includegraphics[width=0.95\linewidth]{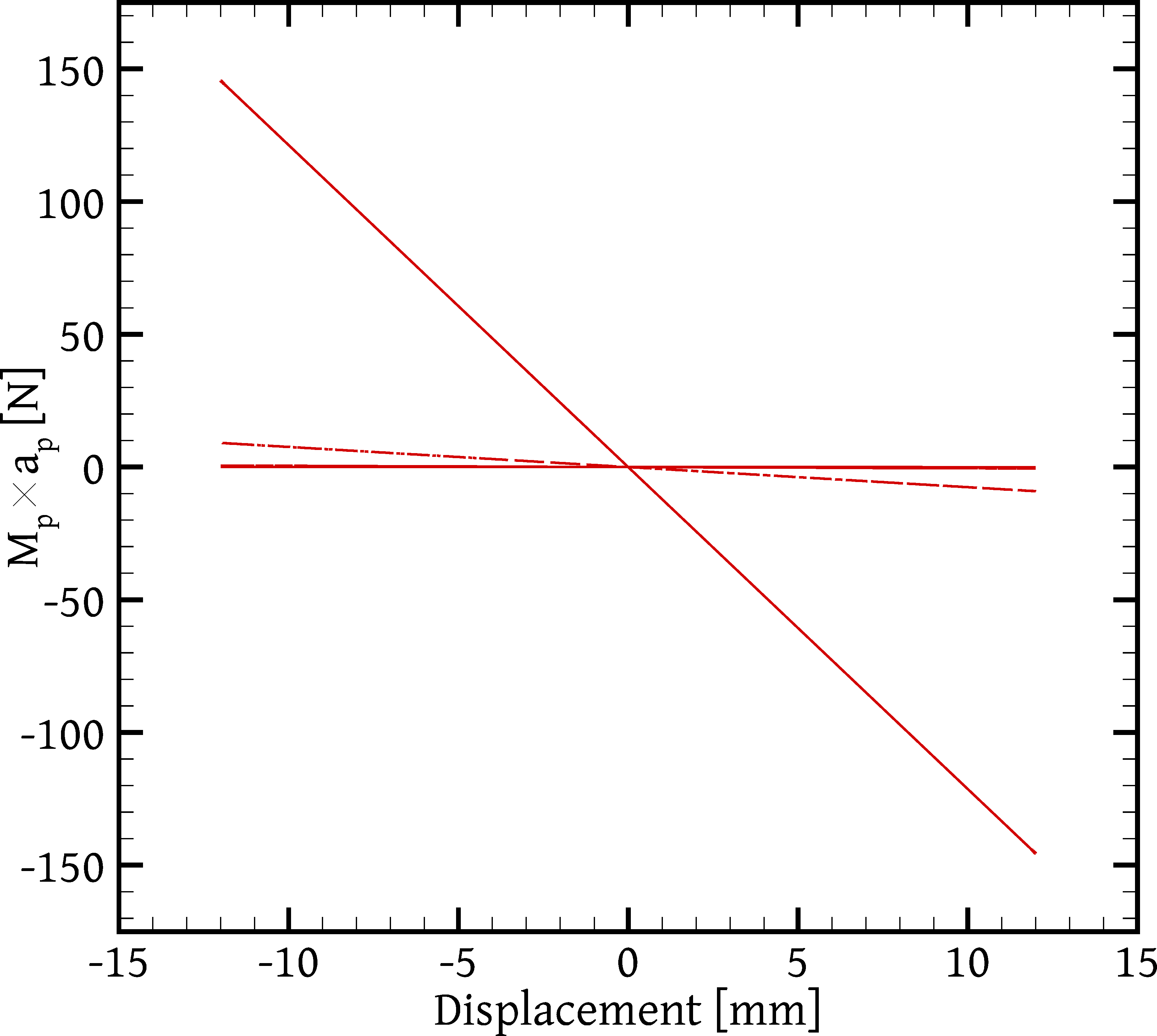}
        \caption{force-displacement}
        \label{sfig: piston inertia vs displacement}
    \end{subfigure}
    \begin{subfigure}[b]{0.35\textwidth}
        \centering
        \includegraphics[width=0.95\linewidth]{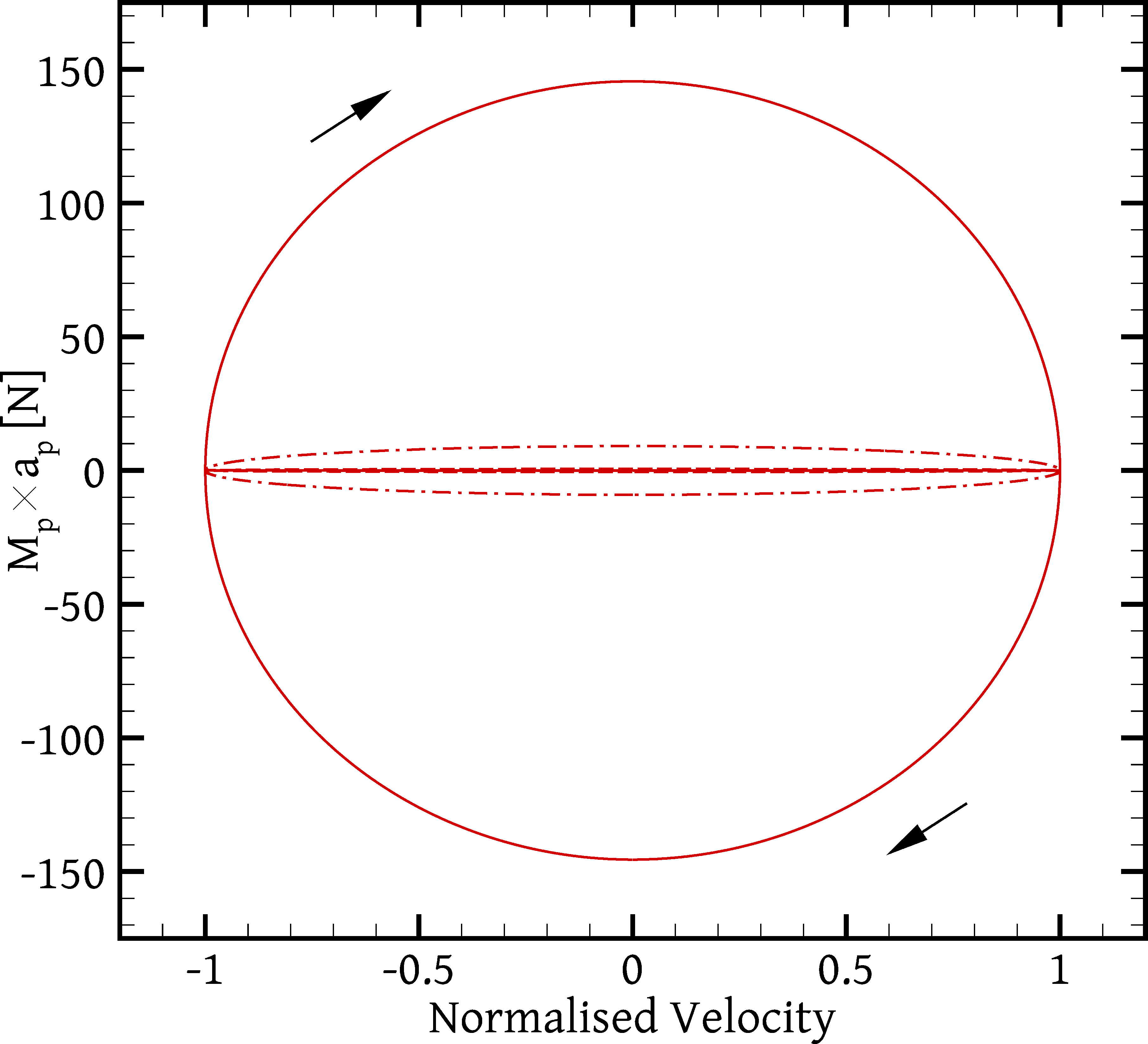}
        \caption{force-velocity}
        \label{sfig: piston inertia vs velocity}
    \end{subfigure}
    
\caption{Mass of the shaft-piston assemblage $M_p$ times its acceleration $a_p$, versus piston 
displacement \subref{sfig: piston inertia vs displacement} or velocity \subref{sfig: piston inertia 
vs velocity} at various oscillation frequencies: $f$ = 0.5 \si{Hz} (solid line), 2 \si{Hz} (dashed 
line), 8 \si{Hz} (dash-dot line) and 32 \si{Hz} (solid line, again). At $f$ = 0.5 and 2 \si{Hz}, 
$M_p a_p$ is very small and the corresponding lines nearly coincide and are horizontal.}
  \label{fig: piston inertia}
\end{figure}

Since the damper geometry and kinematics are known, the shaft-piston mass and acceleration in Eq.\ 
\eqref{eq: force balance} can be calculated and the equation solved for the applied force $F_{ap}$. 
Neglecting the friction force $F_{fr}$ leads to
%^b
\begin{equation} \label{eq: F_ap}
 F_{ap} \;=\; M_p a_p \;-\; F_{fl}
\end{equation}
%^a
The fluid reaction force $F_{fl}$ has been computed and the results were presented in the preceding 
sections. The mass $M_p$ of the piston-shaft assemblage can be estimated at about 0.3 \si{kg} from 
the geometrical dimensions and assuming a material density of 8000 \si{kg/m^3} (steel). From Eq.\ 
\eqref{eq: shaft acceleration} it is seen that the extra component of $F_{ap}$ required to overcome 
the inertia of the piston, $M_p a_p$, is in phase with the piston displacement \eqref{eq: shaft 
position}. The plot of $M_p a_p$ versus piston displacement is a straight line of slope $-M_p 
\omega^2$ (Fig.\ \ref{sfig: piston inertia vs displacement}). Since the plot collapses to a single 
line rather than a loop, the net work of this force component is zero, causing no energy 
dissipation. By inverting the plots of $F_{fl}$ of Fig.\ \ref{fig: force displacement} and 
superimposing the corresponding plots of $M_p a_p$, Fig.\ \ref{sfig: piston inertia vs 
displacement}, we obtain the plot of $F_{ap}$, Eq.\ \eqref{eq: F_ap}, shown in Fig.\ \ref{sfig: 
force displacement with piston inertia}.

\begin{figure}[tb]
    \centering
    \begin{subfigure}[b]{0.49\textwidth}
        \centering
        \includegraphics[width=0.95\linewidth]{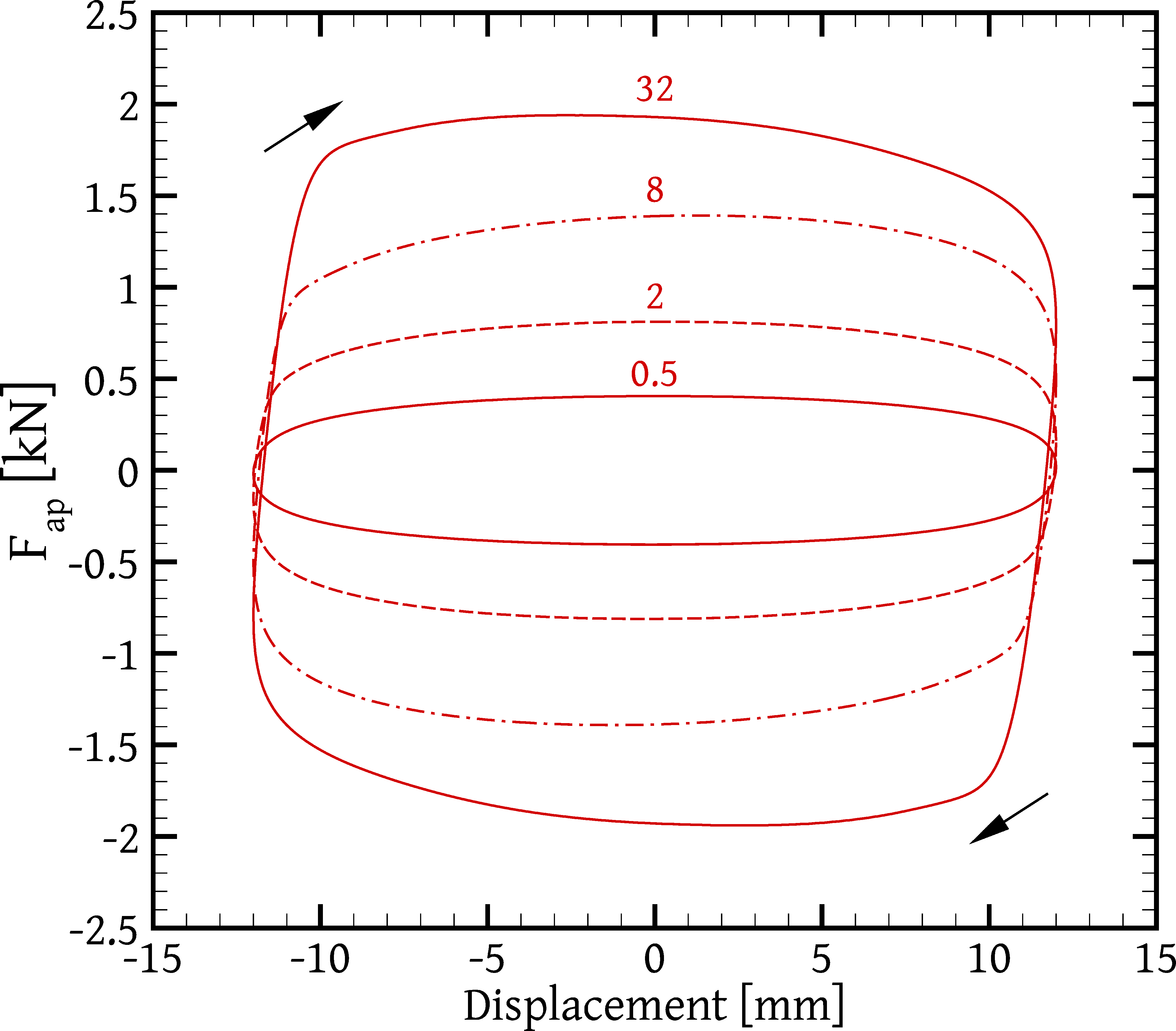}
        \caption{force-displacement}
        \label{sfig: force displacement with piston inertia}
    \end{subfigure}
    \begin{subfigure}[b]{0.49\textwidth}
        \centering
        \includegraphics[width=0.95\linewidth]{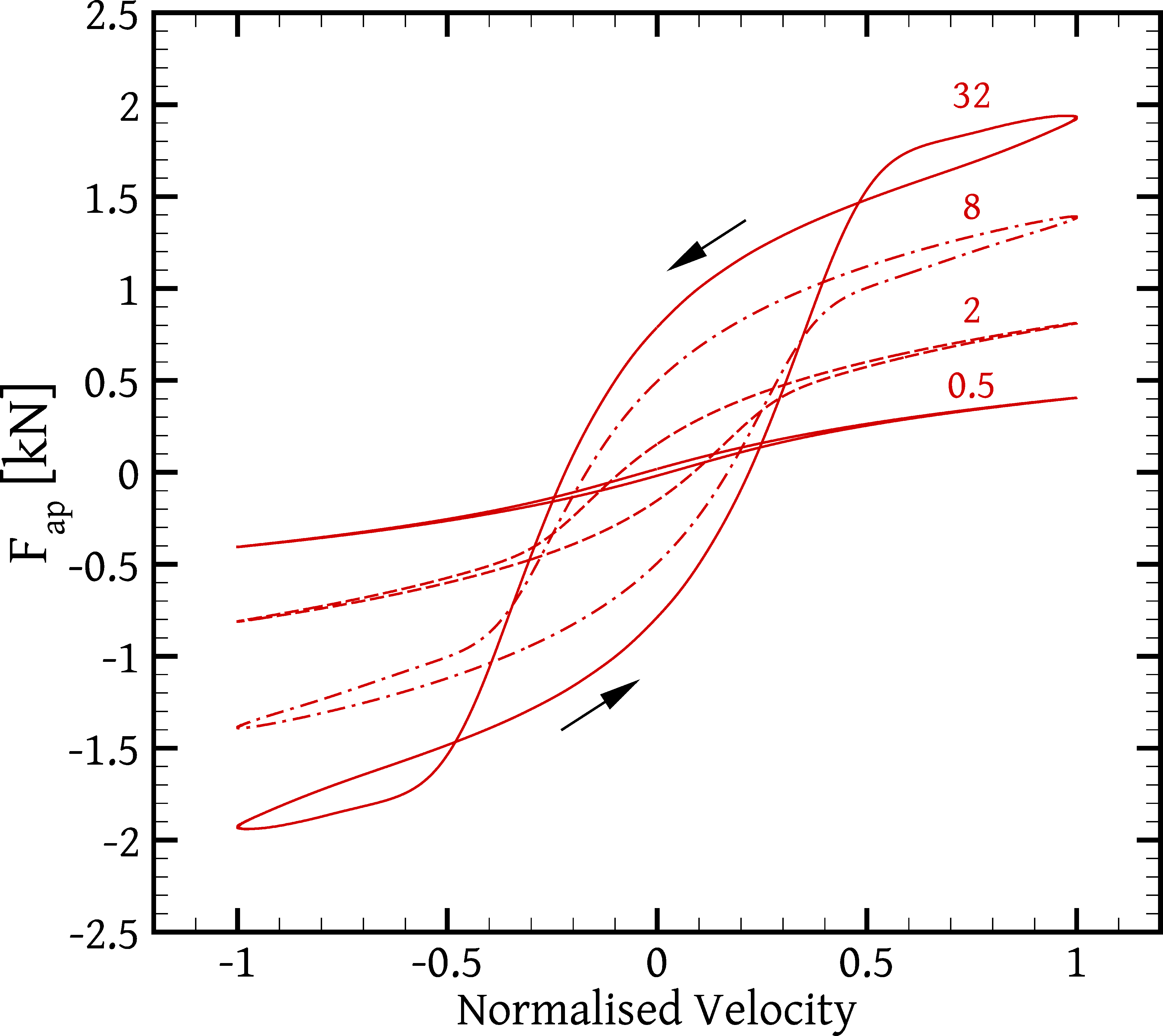}
        \caption{force-velocity}
        \label{sfig: force velocity with piston inertia}
    \end{subfigure}
    
\caption{Diagrams of applied force $F_{ap}$, Eq.\ \eqref{eq: F_ap}, versus piston displacement 
\subref{sfig: force displacement with piston inertia} and versus normalised piston velocity 
\subref{sfig: force velocity with piston inertia} for the lPTT-100 fluids, for all the tested 
frequencies $f$ = 0.5, 2, 8 and 32 \si{Hz}.}
  \label{fig: forces with piston inertia}
\end{figure}

Similarly, we can draw $F_{ap}$ versus piston velocity in Fig.\ \ref{sfig: force velocity with 
piston inertia} by inverting the plot of $F_{fl}$ in Fig.\ \ref{sfig: force - ND velocity lPTT-100} 
(to account for the minus sign in Eq.\ \eqref{eq: F_ap}) and superimposing the plot of $M_p a_p$ 
shown in Fig.\ \ref{sfig: piston inertia vs velocity}. Evidently, at $f$ = 32 \si{Hz} the 
superposition of the inertial component causes a hysteresis inversion, from elastic-type to 
inertia-type, when the magnitude of the normalised velocity is greater than 0.5 (note that due to 
the inversion of Fig.\ \ref{sfig: force - ND velocity lPTT-100} the sense of the hystereses has 
also been inverted: elastic hysteresis is now counterclockwise and inertial hysteresis is 
clockwise). At low velocities (when the piston is near the extremities) the hysteresis remains 
elastic. At lower frequencies the $M_p a_p$ term is much smaller due to the acceleration $a_p$ being 
proportional to $\omega^2$ (Eq.\ \eqref{eq: shaft acceleration}) and hence to the square of the 
frequency, and the piston inertia is not felt much.

The plots of $F_{ap}$ shown in Fig.\ \ref{fig: forces with piston inertia} are in qualitative 
agreement with experimental data for viscous dampers found in the literature \cite{Hou_2008, 
Yun_2008, Jiao_2016}, and even for magnetorheological dampers \cite{Snyder_2001, Yao_2002, 
Wang_2007, Wang_2011, Parlak_2012} (these employ elastoviscoplastic fluids, but plasticity has 
similar effects to shear-thinning as far as the damper force is concerned). In these studies there 
is a prominent elastic hysteresis at low velocities in force-velocity diagrams, which is not 
predicted by simulations with the Carreau-Yasuda model. Hence we only plotted $F_{ap}$ as predicted 
by the PTT model, which accounts for all fluid properties, viscosity, elasticity, and inertia.

\subsection{Using a silicone oil of lower viscosity}
\label{ssec: results; low viscosity oil}

For completeness, we briefly investigate the consequences of using a silicone oil whose viscosity 
lies significantly below the high range on which this investigation has focused. Lower viscosity 
is related to lower relaxation times, and thus one may expect that elastic effects will become 
negligible. However, these effects do not depend solely on the relaxation time but also on the 
geometrical / kinematic parameters of the problem, as expressed in the definitions of the 
Weissenberg and Deborah numbers. In order to achieve a similar response to that of a high-viscosity 
oil damper from a damper employing an oil of lower viscosity, the gap width of the latter would 
have to be reduced as otherwise the force would be too weak due to the low viscosity. The reduced 
gap width will partially counterbalance the reduction in $\Wei$ due to the lower relaxation time 
($\Deb$ will be unaffected). 

\begin{figure}[tb]
  \centering
  \includegraphics[scale=0.80]{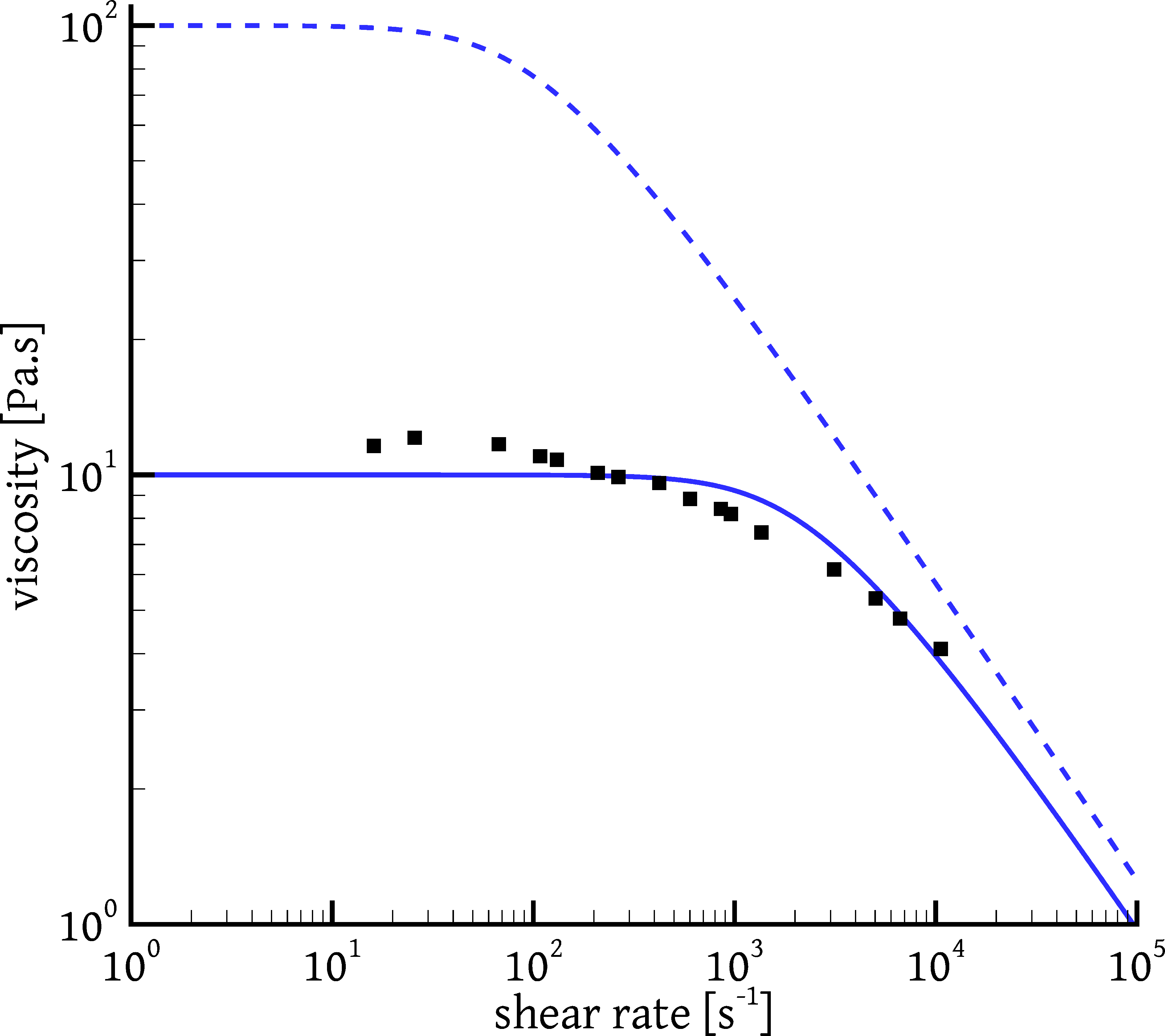}
  \caption{Variation of steady shear viscosity with shear rate for the lPTT-10 (continuous line) 
and lPTT-100 (dashed line) fluids, along with experimental measurements for a 12.5 \si{Pa.s} 
silicone oil from \cite{Currie_1950} (${\scriptstyle\blacksquare}$).}
  \label{fig: steady shear viscosity lPTT-10}
\end{figure}

To investigate this, we performed a couple of simulations, at $f$ = 2 and 32 \si{Hz}, respectively, 
on the intermediate grid 2 (Table \ref{table: grids}), using a fluid labelled ``lPTT-10'' with 
parameters $\eta_0$ = 10 \si{Pa.s}, $\lambda$ = 0.7 \si{ms} and $\epsilon$ = 0.10. The relaxation 
time was chosen on the following basis, since detailed characterisations such as those available 
for the high-viscosity oils were not found: the viscosity $\eta_0$ = 10 \si{Pa.s} is still high 
enough such that the polymer chains are entangled \cite{Longin_1998} in which case both the 
reptation time of the polymer chains, which can be used as an approximation to the fluid relaxation 
time \cite{Likhtman_2002}, and the zero-shear viscosity are proportional to the molecular weight 
raised to the power 3.4 \cite{Wool_1993}. In \cite{Longin_1998} detailed relaxation spectra are 
given for three silicone oils of zero-shear viscosities $\eta_0$ of 60, 100, and 500 \si{Pa.s} from 
which Eq.\ \eqref{eq: average relaxation time} gives average relaxation times $\lambda$ of 4.2, 10.5 
and 40.4 \si{ms}, respectively. These values do not deviate substantially from the theoretical 
proportionality between viscosity and relaxation time, so that it is realistic to select a value of
$\lambda$ = 0.7 \si{ms} for the lPTT-10 fluid by determining the proportionality constant from the 
closest experimental data pair of \{$\eta_0$ = 60 \si{Pa.s}, $\lambda$ = 4.2 \si{ms}\}. The PTT 
parameter $\epsilon = 0.10$ was set such that the steady shear viscosity of lPTT-10 compares well 
against available experimental data (Fig.\ \ref{fig: steady shear viscosity lPTT-10}). We note that 
comparison of the lPTT-10 curve in Fig.\ \ref{fig: steady shear viscosity lPTT-10} against the 
experimental data shows that for low viscosity silicone oils even the linear PTT model predicts an
excessive rate of shear thinning.

To counteract the reduction of viscosity, the gap width was reduced to half (1.5 \si{mm}) by 
enlarging the piston radius to 23.5 \si{mm} (the rest of the damper dimensions are unaltered, as in 
Table \ref{table: damper dimensions}). This modification was based on the following simple 
reasoning: given that the radius difference between the piston and the cylinder is small, the 
cross-sectional area of the gap, $A_g$ say, can be assumed to be proportional to the gap width $h$. 
Then, from continuity, the fluid velocity in the gap is inversely proportional to this area and 
therefore to $h$. The shear stress $\tau_{rx}$ on the piston surface along the gap is then assumed 
to be proportional to $\eta_0 U / (h/2) \approx \eta_0 U_f / (h/2)$ which is proportional to 
$h^{-2}$. Next we consider the force balance on the fluid in the gap, neglecting its inertia: the 
viscous force, $\tau_{rx}$ times the area of the piston and cylinder surrounding the gap, is 
balanced by the pressure force $\Delta p \times A_g$ where $\Delta p$ is the pressure difference 
between the left and right damper compartments. Therefore, $\Delta p$ is proportional to $\tau_{rx} 
/ A_g$ which is proportional to $h^{-3}$. The pressure component dominates $F_{fl}$ (Fig.\ \ref{fig: 
force components}) and so we also expect $F_{fl} \sim h^{-3}$. Thus by reducing the gap width to 
half we would expect an 8-fold increase in the force if we used the same fluid, so that using a 
fluid with 10 times smaller viscosity, such as the lPTT-10 compared to the lPTT-100, we may expect 
similar levels of force. This analysis depends on several simplifications, one of which is to 
neglect the effect of shear-thinning. The consequences will be seen below.

The narrowing of the gap and the associated rise of the fluid velocity therein result in large 
characteristic shear rates $\dot{\gamma}_c$ of 1658 \si{s^{-1}} at $f$ = 2 \si{Hz} and 26532 
\si{s^{-1}} at 32 \si{Hz}. This compensates somewhat for the low value of $\lambda$ as far as $\Wei 
\equiv \lambda U / H = \lambda \dot{\gamma}_c$ is concerned, resulting in $\Wei$ = 1.16 and 18.6 
for $f$ = 2 and 32 \si{Hz}, respectively, which is not vastly smaller than the corresponding values 
for the lPTT-100 fluid. The $\Deb$ number, on the other hand, is quite small: 0.0014 ($f$ = 2 
\si{Hz}) and 0.0224 ($f$ = 32 \si{Hz}). Concerning shear-thinning, the ratio 
$\eta(\dot{\gamma}_c)/\eta_0$ is 0.840 ($f$ = 2 \si{Hz}) and 0.222 ($f$ = 32 \si{Hz}), 
significantly larger than the corresponding values for the lPTT-100 fluid. This reflects the fact 
that shear-thinning is less pronounced and its onset occurs at higher shear rates for lower 
viscosity silicone oils compared to higher viscosity ones (Figs.\ \ref{fig: steady shear viscosity} 
and \ref{fig: steady shear viscosity lPTT-10}). Finally, concerning inertia, we have $\Rey_c$ = 
0.11 ($f$ = 2 \si{Hz}) and 6.71 ($f$ = 32 \si{Hz}) which is significantly larger than for lPTT-100 
and thus a more noticeable role is expected of inertia\footnote{Note, however, that the lPTT-10 
geometry differs slightly from that of the lPTT-100 and lPTT-500 simulations, and comparing 
$\Rey_c$, $\Deb$ and $\Wei$ values between different geometries is not a completely valid way of 
assessing the relative importance of the effects that these numbers quantify (see our previous 
publication \cite{Syrakos_2016}).}.

\begin{figure}[tb]
    \centering
    \begin{subfigure}[b]{0.32\textwidth}
        \centering
        \includegraphics[width=0.95\linewidth]{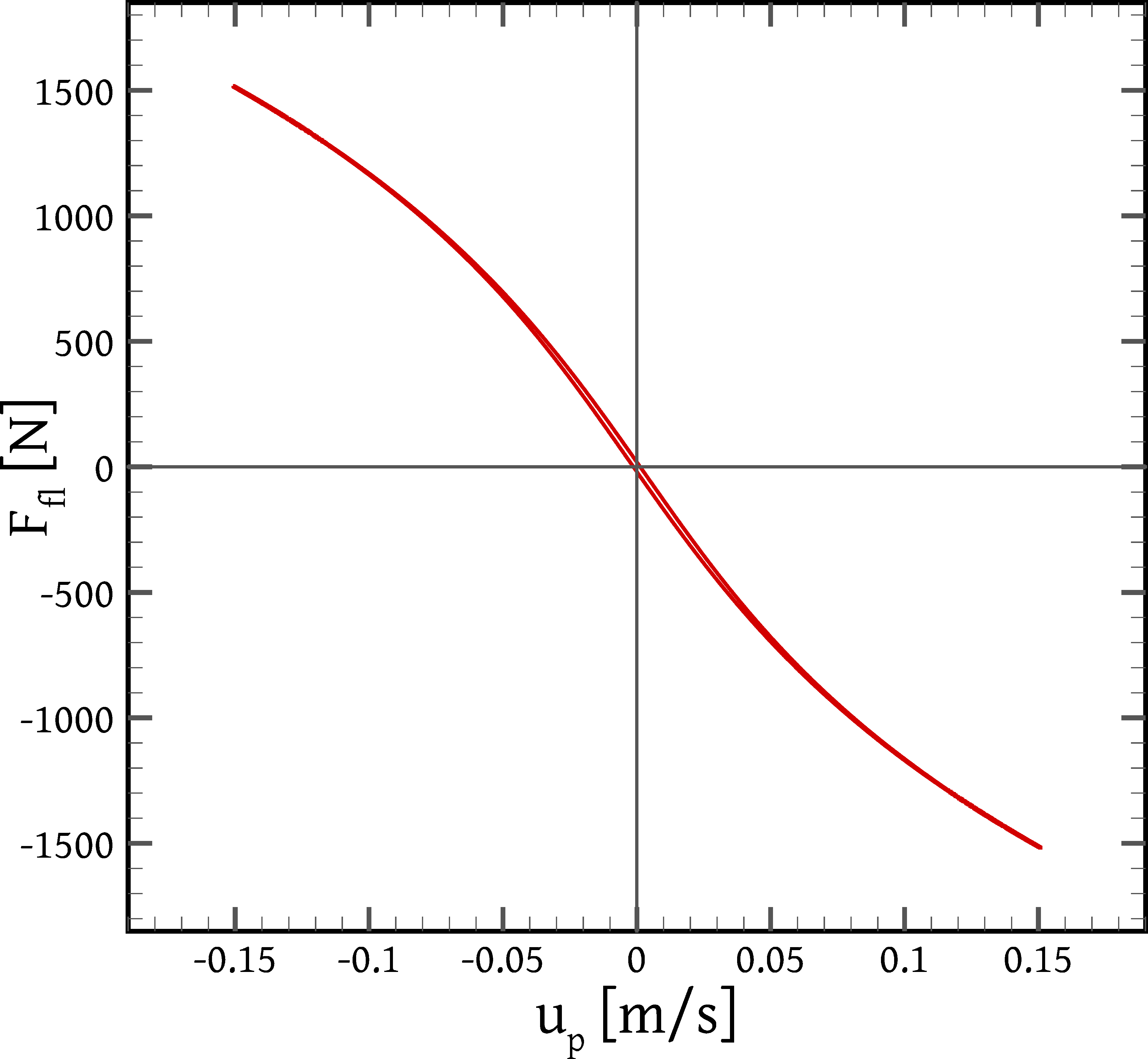}
        \caption{$F_{fl}$ vs.\ $u_p$, $f$ = 2 \si{Hz}}
        \label{sfig: lPTT10 f=2 force velocity}
    \end{subfigure}
    \begin{subfigure}[b]{0.32\textwidth}
        \centering
        \includegraphics[width=0.95\linewidth]{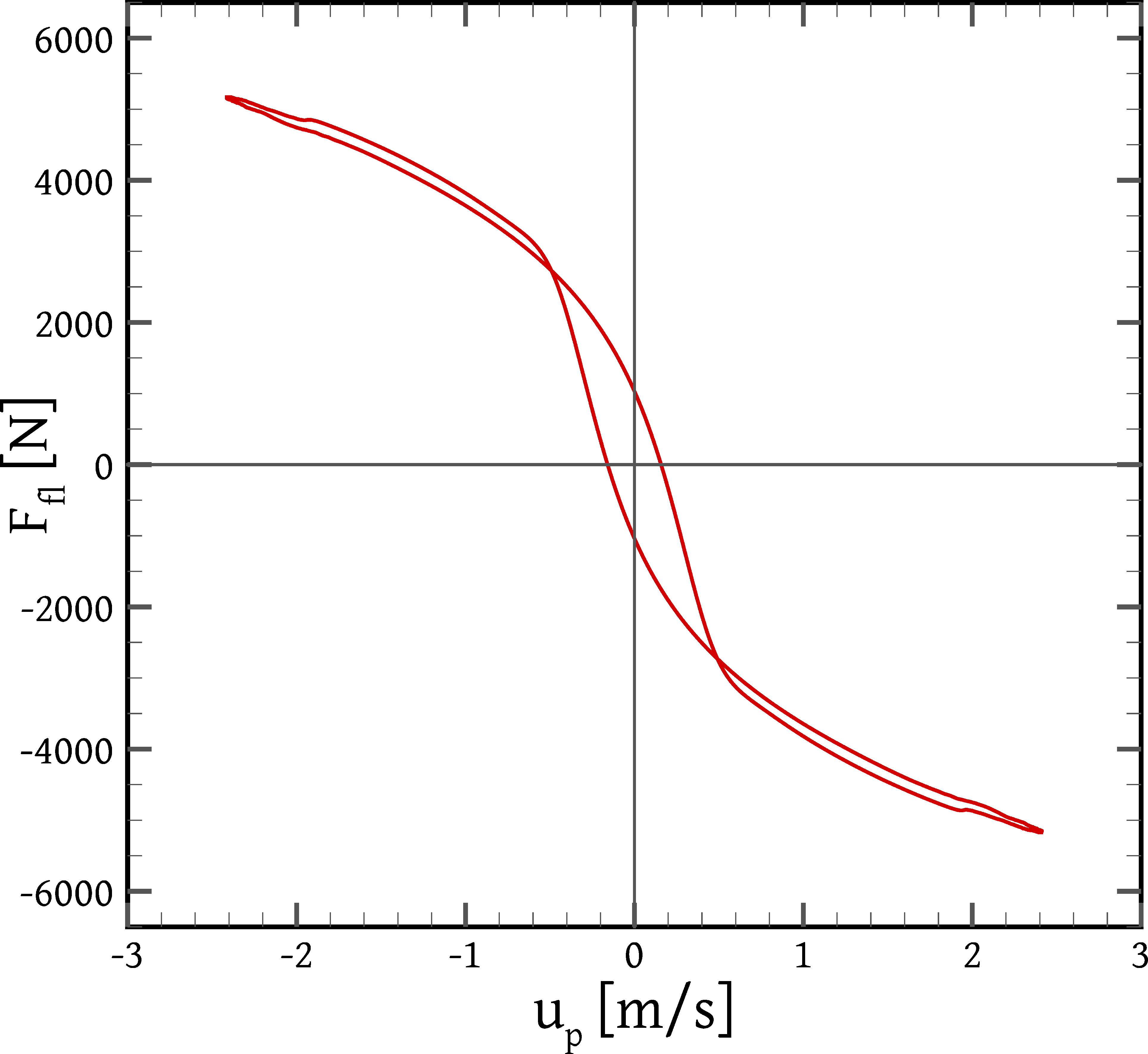}
        \caption{$F_{fl}$ vs.\ $u_p$, $f$ = 32 \si{Hz}}
        \label{sfig: lPTT10 f=32 force velocity}
    \end{subfigure}
    \begin{subfigure}[b]{0.32\textwidth}
        \centering
        \includegraphics[width=0.95\linewidth]{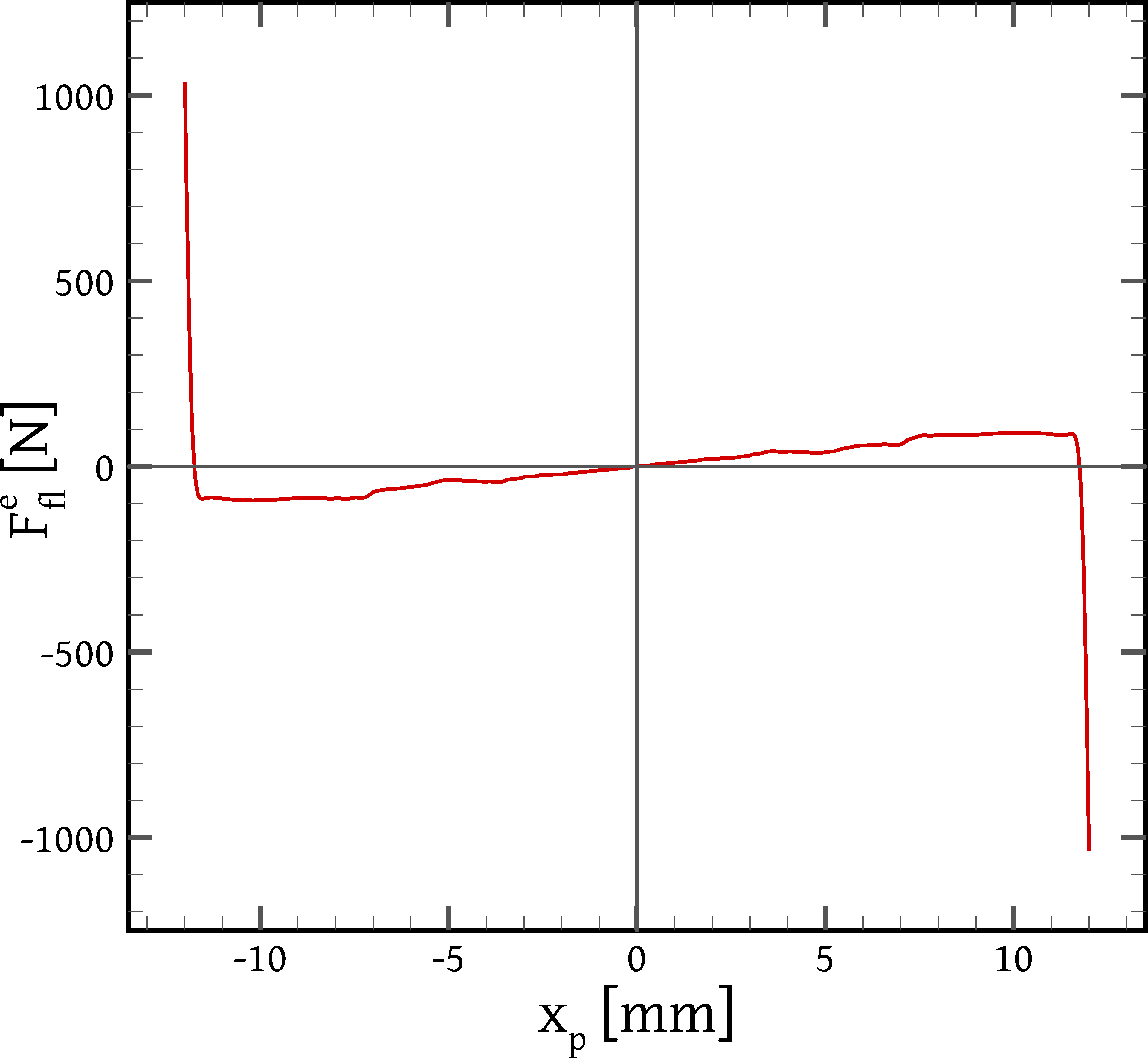}
        \caption{$F^e_{fl}$ vs.\ $x_p$, $f$ = 32 \si{Hz}}
        \label{sfig: lPTT10 f=32 F_e displacement}
    \end{subfigure}
    
\caption{Plots \subref{sfig: lPTT10 f=2 force velocity} and \subref{sfig: lPTT10 f=32 force 
velocity}: $F_{fl}$ versus piston velocity for the lPTT-10 fluid, at oscillation frequencies of $f$ 
= 2 \si{Hz} \subref{sfig: lPTT10 f=2 force velocity} and 32 \si{Hz} \subref{sfig: lPTT10 f=32 force 
velocity}. Plot \subref{sfig: lPTT10 f=32 F_e displacement}: $F^e_{fl}$ versus piston displacement 
for the lPTT-10 fluid at $f$ = 32 \si{Hz}.}
  \label{fig: lPTT10 force diagrams}
\end{figure}

Figure \ref{sfig: lPTT10 f=2 force velocity} shows that at the low frequency of $f$ = 2 \si{Hz} 
elastic effects are hardly noticeable\footnote{For the $f$ = 2 \si{Hz} simulation we reduced the 
time step to $\Delta t = T / 32000$ to ensure that it is quite smaller than the relaxation time; 
this results in $\lambda / \Delta t \approx 45$.}; the hysteretic loop almost collapses to a single 
curve. Shear-thinning appears to be weak as well, with the slope of the curve not changing 
significantly at higher velocities. On the other hand, Fig.\ \ref{sfig: lPTT10 f=32 force velocity} 
shows a clear elastic hysteretic loop at low velocities, i.e.\ when the piston is close to the 
extremities. At higher velocities the hysteresis changes sense and becomes inertial. As explained in 
Appendix \ref{appendix: LAOS flow}, hysteresis reversals can originate also from nonlinearities in 
the elastic behaviour, but in the present case we believe that they are caused by fluid inertia; 
Fig.\ \ref{sfig: lPTT10 f=32 F_e displacement} shows that the ``elastic'' component $F^e_{fl}$ 
pushes the piston towards the extremities over nearly all of the stroke, and only very close to the 
extremities does it act in a spring-like manner pushing it away from the extremities. $F^e_{fl}$ 
varies almost linearly with $x_p$ in the inertial region, as we saw also for the CY-100 fluid in 
Fig.\ \ref{fig: F inertia}. Also, the relatively high value of $\Rey_c$ = 6.71 suggests that 
inertial effects should be expected. Concerning shear-thinning, comparison of the slope variations 
between Figs.\ \ref{sfig: lPTT10 f=2 force velocity} and \ref{sfig: lPTT10 f=32 force velocity} 
shows that it is significantly more pronounced in the $f$ = 32 \si{Hz} case, as expected.

\begin{figure}[tb]
  \centering
  \includegraphics[scale=0.75]{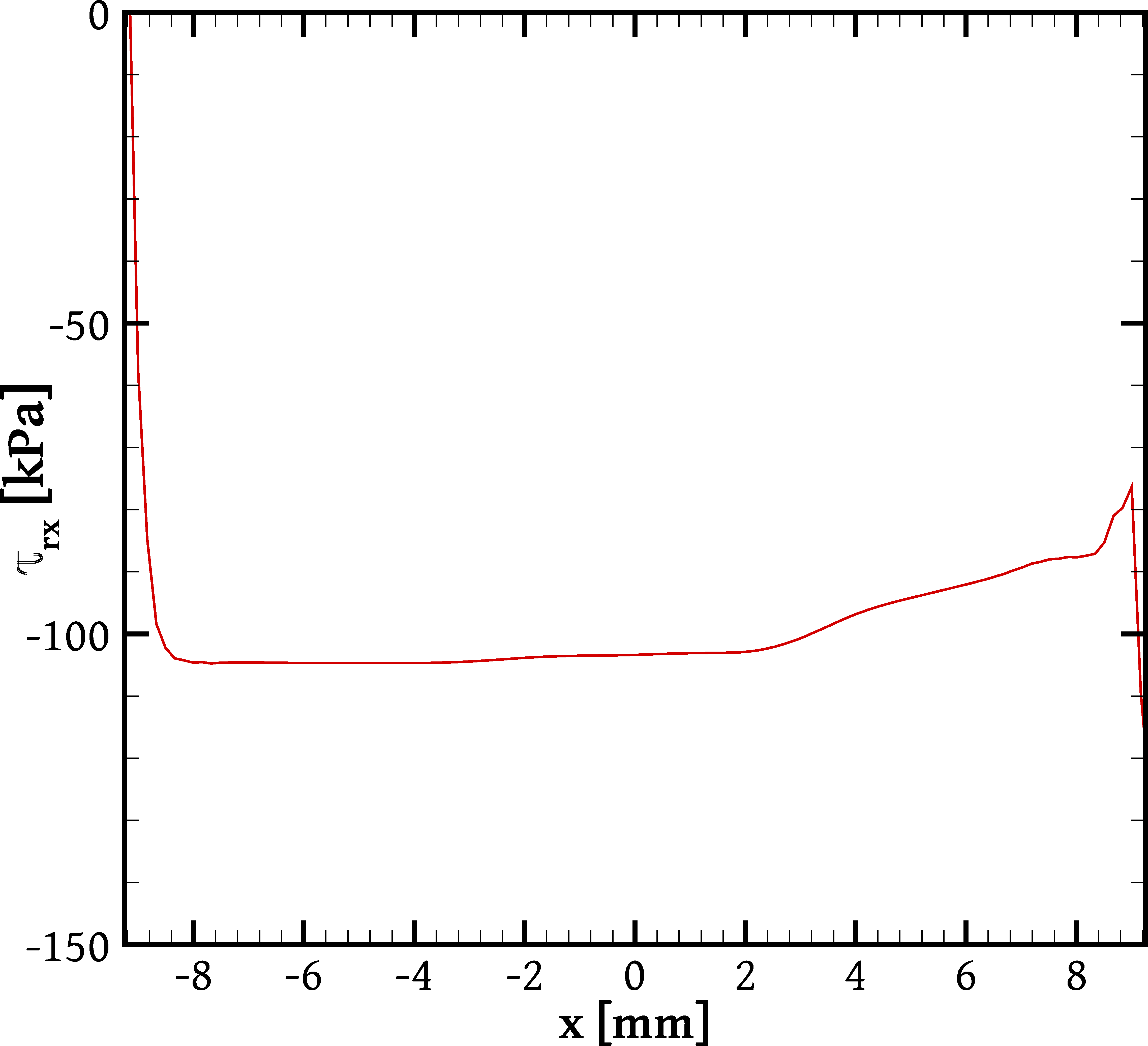}
  \caption{Distributions of the $\tau_{rx}$ stress component along the horizontal surface of the 
piston for the lPTT-10 fluid oscillating at $f$ = 32 \si{Hz}, at a time instance when the piston is 
moving with maximum velocity towards the right.}
  \label{fig: stress on piston lPTT10 f32}
\end{figure}

Figure \ref{fig: stress on piston lPTT10 f32} shows that in the $f$ = 32 \si{Hz} case there does 
exist a substantial flow development length inside the gap, however it is smaller than that of the 
lPTT-100 fluid (Fig.\ \ref{fig: stress on piston}). A comparison between Figs.\ \ref{fig: stress on 
piston lPTT10 f32} and \ref{fig: stress on piston} also shows that the stress in the lPTT-10 case 
is nearly 1.5 times that of the lPTT-100 case, which can be attributed to the weaker shear-thinning 
property of the lPTT-10 fluid. In fact a comparison between Figs.\ \ref{sfig: lPTT10 f=32 force 
velocity} and \ref{sfig: force - ND velocity lPTT-100} shows that the lPTT-10 force is more than 
2.5 times that of the lPTT-100 fluid at the same frequency. This is because the narrowing of the 
gap brings about a larger rise in the pressure component of $F_{fl}$ (which is the dominant one, 
$O(h^{-3})$) than in the stress component ($O(h^{-2})$), as discussed above. In particular, Fig.\ 
\ref{fig: force components} shows that for lPTT-100 the stress component is more than 10\% of the 
total force, but for the lPTT-10 fluid the results showed that it is only 6\%. Interestingly, even 
for the $f$ = 2 \si{Hz} case the lPTT-10 force is still nearly twice that for lPTT-100 (Figs.\ 
\ref{sfig: lPTT10 f=2 force velocity} and \ref{sfig: force - ND velocity lPTT-100}). We conclude 
that the difference in shear-thinning character between high- and low-viscosity oils is responsible 
for higher forces in the latter, under conditions that would produce similar forces had the 
viscosity been constant. The lPTT-100 and lPTT-10 fluids will produce similar forces at frequencies 
low enough for shear rates to drop below the onset of shear-thinning for the lPTT-100 fluid. 
Overall then, utilization of high-viscosity silicone oils can be beneficial in applications where 
it is desirable to moderate the variation of the damper force with oscillation frequency.

\section{Conclusions}
\label{sec: conclusions}

The present study investigated the operation of a silicone oil fluid damper of a common design 
under a forced sinusoidal piston displacement through numerical simulations. In order to relate the 
damper behaviour to the fluid rheology, three different rheological models were used to describe 
the behaviour of the fluid: the Newtonian, Carreau-Yasuda, and Phan-Thien \& Tanner models. Their 
parameters were selected such as to describe the behaviour of high-viscosity silicone oil.

Due to the high viscosity, the Newtonian model predicts creeping, quasi-steady flow without any 
hysteresis. Its force-displacement plots are perfectly elliptical, and its force-velocity plots are 
straight lines.

The Carreau-Yasuda model defines a shear-thinning behaviour that predicts much lower force levels 
than in the Newtonian case. It also makes the force-displacement loops more rectangular, i.e.\ it 
predicts relatively constant force over most of the piston stroke, and makes the force roughly 
proportional to the piston velocity raised to a power less than one. Furthermore, at high 
frequencies, where shear-thinning is intense enough to lower viscous stresses to the point that 
fluid inertia is not negligible any more, the Carreau-Yasuda model predicts some inertial 
hysteresis, i.e.\ that the force is greater when the piston is accelerating than when it is 
decelerating, at a given piston displacement or velocity.

Finally, the PTT model in addition to describing a shear-thinning behaviour also describes the 
elastic behaviour of the fluid. This is the most complete model among those considered and in 
addition to the features predicted also by the CY model it predicts an elastic hysteresis where the 
force has an elastic component that pushes the piston away from the extreme positions, like in a 
spring. Furthermore, a PTT fluid that exhibits the same shear-thinning as a CY fluid was seen to 
produce lower force levels than the latter, especially the higher the frequency or fluid 
elasticity. This was shown to arise from elasticity, which delays the development of a fully 
developed velocity profile in the piston-cylinder gap.

Among these three rheological models, the one whose predictions qualitatively match better the 
experimental results available in the literature (e.g.\ \cite{Hou_2008, Yun_2008, Jiao_2016}) is 
the PTT viscoelastic model. The Newtonian model is completely inaccurate, while the CY model fails 
to predict the stiffness characteristics of the damper behaviour and cannot account for a further 
force reduction due to elasticity. Thus it is shown that fluid elasticity suffices to induce 
stiffness in the damper behaviour even if fluid compressibility is not accounted for. The PTT 
predictions resemble also the experimental results concerning magnetorheological dampers 
\cite{Snyder_2001, Yao_2002, Wang_2007, Wang_2011, Parlak_2012}, which is explained by the fact 
that shear-thinning is in some respects similar to plasticity. For more accurate modelling of MR 
dampers an elastoviscoplastic constitutive equation should be used instead.

The present results also reveal that, since the viscoelastic flow in the critical region of the gap 
varies along the gap length, the accuracy obtainable by simplified studies that assume 
one-dimensional flow in that region is limited. Full two-dimensional (axisymmetric) simulations are 
necessary in order to obtain an accurate understanding of the flow.

To the best of our knowledge this is the first numerical study of the flow in a damper which models 
the fluid through a viscoelastic constitutive equation, and it has shown that numerical simulation 
can be a valuable aid in damper research if the fluid physics is modelled accurately. Future 
studies can incorporate temperature changes due to viscous dissipation, temperature-dependent fluid 
rheology, fluid compressibility, and plasticity (for MR dampers).

\section*{Acknowledgements}
This research was funded by the LIMMAT Foundation under the Project ``MuSiComPS''.

% APPENDICES
% -----------------------------------------------------------------------------
% The Appendices part is started with the command \appendix;
% appendix sections are then done as normal sections
% \appendix

% \section{}
% \label{}

%\appendix
%\appendixpage
\begin{appendices}
\renewcommand\theequation{\thesection.\arabic{equation}}
\setcounter{equation}{0}

\section{Steady shear flow of PTT fluids}
\label{appendix: steady shear flow}

In steady shear flow the velocity vector is given by $\vf{u} = ( \dot{\gamma} \, x_2, 0, 0)$, where 
$\dot{\gamma}$ is the imposed shear rate, while the pressure gradient is zero (as in Couette flow 
between parallel plates, where the fluid flows in the $x_1$ direction while the plate normal 
vectors are aligned with the $x_2$ direction). The flow is steady and one-dimensional so that the 
only (possibly) non-zero partial derivatives of the flow variables are those with respect to the 
$x_2$ direction. Under these conditions the first component of the momentum equation \eqref{eq: 
momentum} implies that the stress component $\tau_{12}$ is constant ($\partial\tau_{12}/\partial x_2 
= 0$). The (2,2) component of the constitutive equation \eqref{eq: constitutive lPTT} gives 
$\tau_{22} = 0$, while the components (1,1) and (1,2)  reduce to the following:
%^b
\begin{align}
\label{eq: steady shear PTT 11}
 Y\left(\mathrm{tr}(\tf{\tau})\right) \, \tau_{11} \;&=\; 2 \, \lambda \, \tau_{12} \, \dot{\gamma}
\\
\label{eq: steady shear PTT 12}
 Y\left(\mathrm{tr}(\tf{\tau})\right) \, \tau_{12} \;&=\; \eta_0 \, \dot{\gamma}
\end{align}
%^a
The explicit form of the function $Y$ has not yet been substituted, so that the above equations 
hold for both the linear and the exponential PTT models. Next, Eq.\ \eqref{eq: steady shear PTT 11} 
is divided by Eq.\ \eqref{eq: steady shear PTT 12} to get
%^b
\begin{equation}
 \tau_{11} \;=\; \frac{2\lambda}{\eta_0} \tau_{12}^2
\end{equation}
%^a
Noting that $\mathrm{tr}(\tf{\tau}) = \tau_{11} + \tau_{22} + \tau_{33} = \tau_{11}$ we can 
substitute the above equation into \eqref{eq: steady shear PTT 12} to obtain an equation with only 
one unknown, $\tau_{12}$:
%^b
\begin{equation} \label{eq: steady shear tau_12 eqn}
 Y\left(\frac{2\lambda}{\eta} \tau_{12}^2\right) \, \tau_{12} \;=\; \eta_0 \, \dot{\gamma}
\end{equation}
%^a
In the linear PTT case ($Y(x) = 1 + (\epsilon\lambda/\eta_0)x$) the above equation becomes
%^b
\begin{equation} \label{eq: steady shear tau_12 eqn lPTT}
 \tau_{12} \;+\; \frac{2 \, \epsilon \, \lambda^2}{\eta_0^2} \tau_{12}^3 \;=\; \eta_0 \, 
\dot{\gamma}
\end{equation}
%^a
which is a cubic polynomial equation whose analytic solution can be found in \cite{Azaiez_1996}. In 
the exponential PTT case ($Y(x) = \exp((\epsilon\lambda/\eta_0)x)$) Eq.\ \eqref{eq: steady shear 
tau_12 eqn} becomes
%^b
\begin{equation} \label{eq: steady shear tau_12 eqn ePTT}
 \exp\left({\frac{2\,\epsilon\,\lambda^2}{\eta_0^2}\tau_{12}^2}\right) \tau_{12} \;=\; \eta_0 \, 
\dot{\gamma}
\end{equation}
%^a
This seems more unwieldy than \eqref{eq: steady shear tau_12 eqn lPTT} but it turns out that it 
also has an analytic solution (although it does not seem to be reported in the literature). For 
simplicity, let $2\epsilon\lambda^2/\eta_0^2 \equiv a > 0$, $\tau_{12} \equiv x$ and $\eta_0 
\dot{\gamma} \equiv b > 0$ so that the equation becomes
%^b
\begin{equation*}
 e^{ax^2} x \;=\; b \quad\Rightarrow\quad  e^{2ax^2} x^2 \;=\; b^2 
                    \quad\Rightarrow\quad  2a x^2 e^{2ax^2} \;=\; 2 a b^2 
                    \quad\Rightarrow\quad  2a x^2 \;=\; W(2 a b^2)
\end{equation*}
%^a
where $W$ is the Lambert W function \cite{Corless_1996}, i.e.\ the inverse function of $f(x) = 
x\mathrm{e}^x$: $x\mathrm{e}^x = y \Leftrightarrow x = W(y)$. Solving for $x \equiv \tau_{12}$ we 
obtain\footnote{The Lambert $W$ function is double-valued on the interval $(-1/e,0)$, but here 
$2ab^2 > 0$ and there is only one branch to follow.}
%^b
\begin{equation} \label{eq: steady shear ePTT tau_12}
 \tau_{12}(\dot{\gamma}) \;=\;
 \sqrt{ \frac{\eta_0^2 \, W(4\epsilon\lambda^2\dot{\gamma}^2)}{4\epsilon\lambda^2} }
\end{equation}

Having obtained the shear stress $\tau_{12}$ from \eqref{eq: steady shear tau_12 eqn lPTT} or 
\eqref{eq: steady shear ePTT tau_12}, the steady shear viscosity $\eta(\dot{\gamma}) = 
\tau_{12}(\dot{\gamma}) / \dot{\gamma}$ is plotted in Fig.\ \ref{fig: steady shear viscosity} for 
the selected models. It may be seen from Eq.\ \eqref{eq: steady shear tau_12 eqn lPTT} that in the 
limit of very small shear rates $\dot{\gamma}$ the stress $\tau_{12}$ also becomes very small and 
since in this case $\tau_{12}^3 \ll \tau_{12}$, Eq.\ \eqref{eq: steady shear tau_12 eqn lPTT} 
asymptotically reduces to $\tau_{12} = \eta_0 \dot{\gamma}$. Thus $\eta_0$ is the zero-shear 
viscosity, $\eta_0 = \lim \eta(\dot{\gamma})$ as $\dot{\gamma} \rightarrow 0$. The same can be said 
of the exponential PTT model. In fact, expanding the exponential in Eq.\ \eqref{eq: steady shear 
tau_12 eqn ePTT} in a Maclaurin series with respect to $\tau_{12}$ and discarding all terms of this 
series except the first two, we obtain Eq.\ \eqref{eq: steady shear tau_12 eqn lPTT}. The discarded 
terms contain higher powers of $\tau_{12}$ which will be negligible if $\dot{\gamma}$ is small 
enough (which implies small $\tau_{12}$). Thus at small shear rates the linear and exponential PTT 
models become equivalent.

Finally, we note that in both Eq.\ \eqref{eq: steady shear tau_12 eqn lPTT} and \eqref{eq: steady 
shear tau_12 eqn ePTT} (or \eqref{eq: steady shear ePTT tau_12}) the individual values of $\lambda$ 
and $\epsilon$ do not make a difference, but it is the product $\epsilon \lambda^2$ that determines 
the solution. Thus even if $\lambda$ and $\epsilon$ of two different PTT fluids differ by orders of 
magnitude, if the product $\epsilon \lambda^2$ is the same for both fluids then their 
$\tau_{12}(\dot{\gamma})$ curves will coincide in a diagram such as that of Fig.\ \ref{fig: steady 
shear viscosity}.

\section{LAOS flow of PTT fluids}
\label{appendix: LAOS flow}

\begin{figure}[tb]
    \centering
    \begin{subfigure}[b]{0.49\textwidth}
        \centering
        \includegraphics[width=0.90\linewidth]{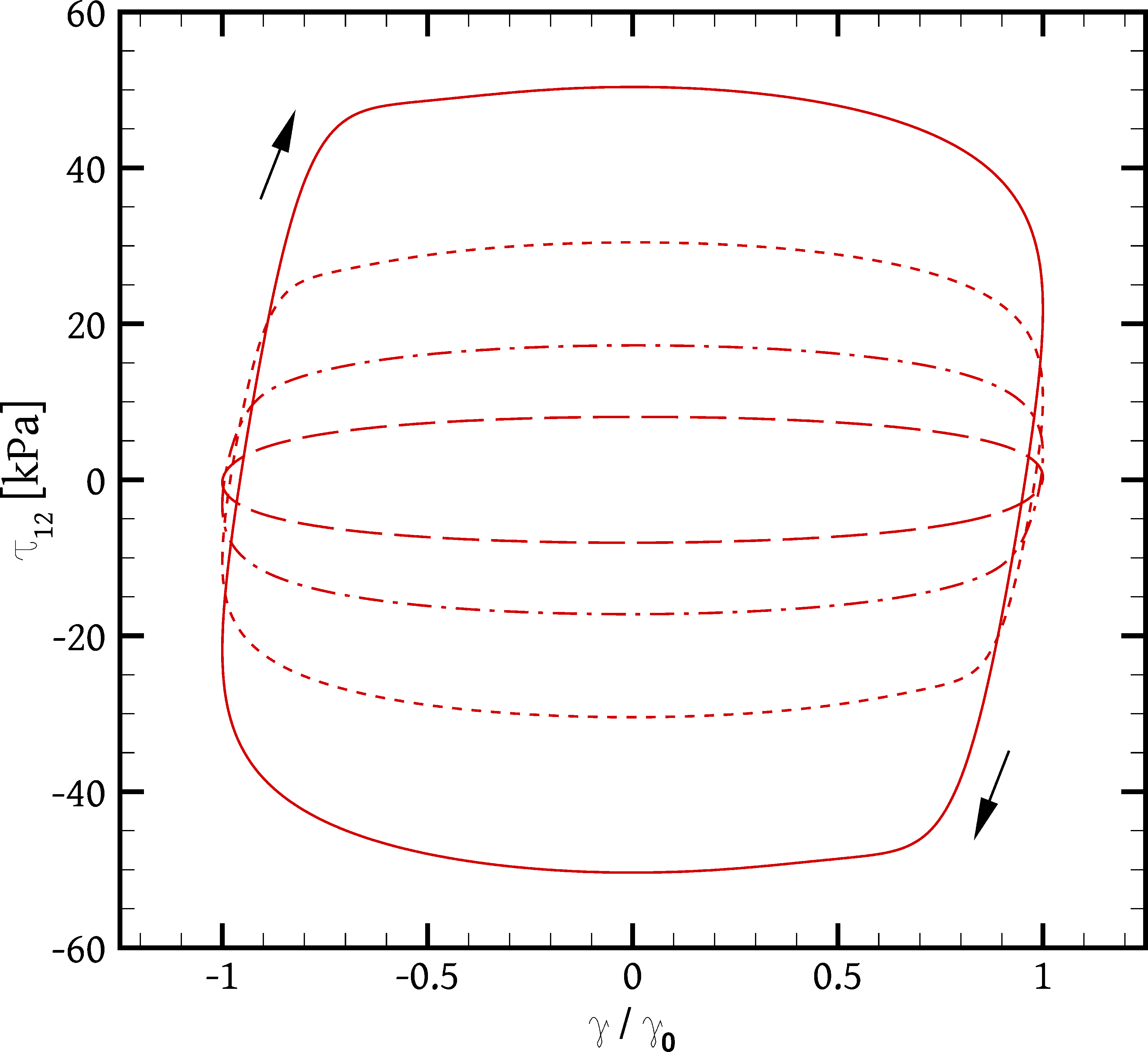}
        \caption{}
        \label{sfig: laos gamma}
    \end{subfigure}
    \begin{subfigure}[b]{0.49\textwidth}
        \centering
        \includegraphics[width=0.90\linewidth]{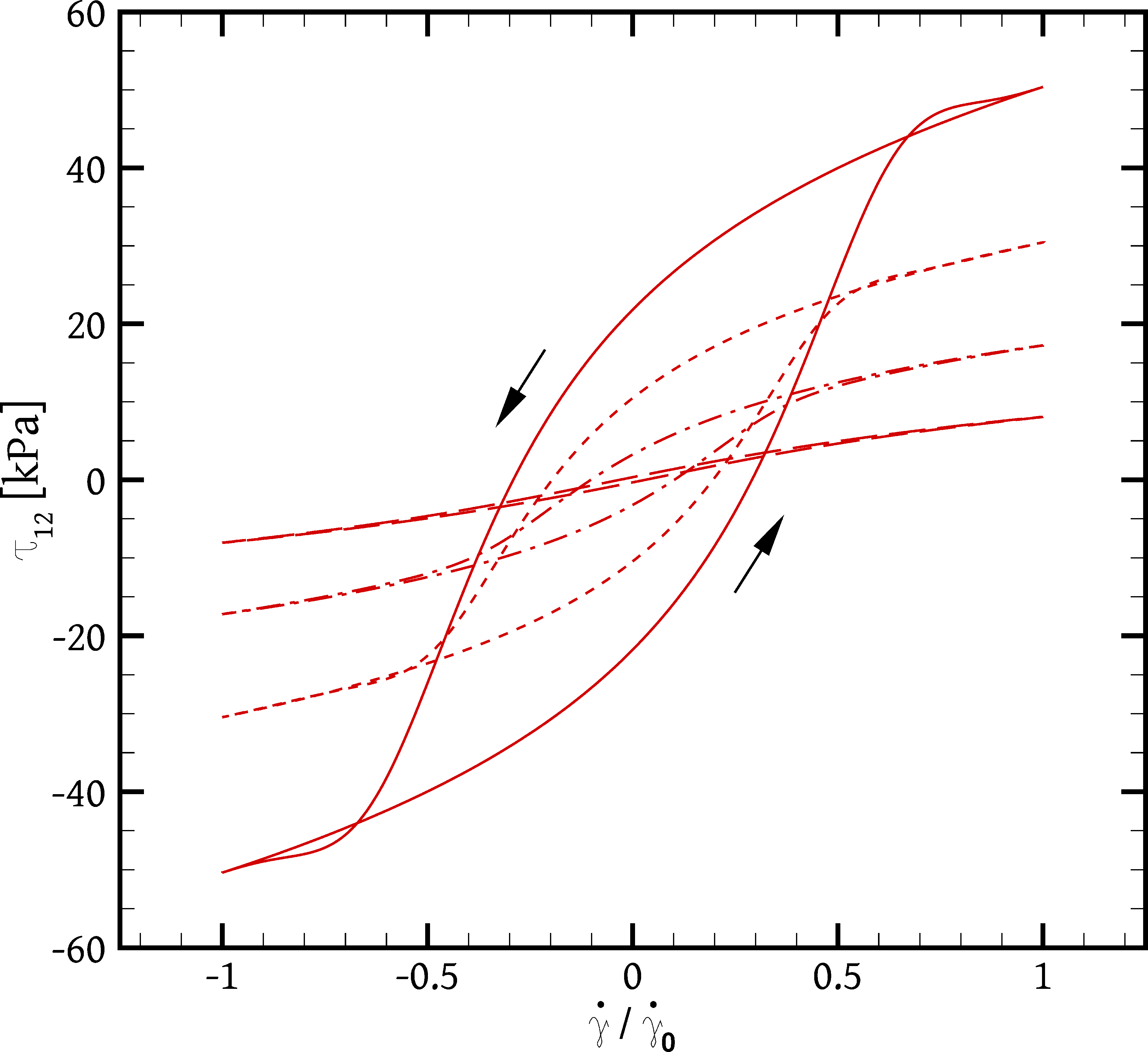}
        \caption{}
        \label{sfig: laos gamma dot}
    \end{subfigure}
    
\caption{Diagrams of shear stress versus normalised shear \subref{sfig: laos gamma} or shear rate 
\subref{sfig: laos gamma dot} for simulations of LAOS flow of the lPTT-100 fluid. The oscillation 
frequencies are $f$ = 0.5 (long dashes), 2 (dash-dot), 8 (short dashes) and 32 (continuous line) 
\si{Hz}, and the corresponding shear rate amplitudes $\dot{\gamma}_0$ are equal to the respective 
$\dot{\gamma}_c$ values of Table \ref{table: operating conditions}.}
  \label{fig: laos}
\end{figure}

For unsteady flow of a PTT fluid between parallel plates, driven by the motion of the top plate, 
under the assumption that the distance between the plates is so small that fluid inertia is 
negligible and the velocity varies linearly as $\vf{u} = (\dot{\gamma}(t) x_2,0,0)$, the governing 
equations contain an additional time derivative compared to the steady-state Eqs.\ \eqref{eq: 
steady shear PTT 11} -- \eqref{eq: steady shear PTT 12}:
%^b
\begin{align}
\label{eq: transient shear PTT 11}
 \lambda \pd{\tau_{11}}{t} \;+\;
 Y\left(\mathrm{tr}(\tf{\tau})\right) \, \tau_{11} \;&=\; 2 \, \lambda \, \tau_{12} \, \dot{\gamma}
\\
\label{eq: transient shear PTT 12}
 \lambda \pd{\tau_{12}}{t} \;+\;
 Y\left(\mathrm{tr}(\tf{\tau})\right) \, \tau_{12} \;&=\; \eta_0 \, \dot{\gamma}
\end{align}
%^a
A particular such transient flow which is commonly used in rheometry and shares some similarities 
with the flow in a damper that is the topic of this paper, is Large Amplitude Oscillatory Flow 
(LAOS) \cite{Hyun_2011}, where the moving plate oscillates so that $\dot{\gamma}(t) = 
\dot{\gamma}_0 \cos(\omega t)$ where $\dot{\gamma}_0$ is the maximum shear rate during an 
oscillation and $\omega$ is the angular frequency. In Fig.\ \ref{fig: laos} we plot 
Lissajous-Bowditch curves, i.e.\ curves of shear stress versus strain (Fig.\ \ref{sfig: laos 
gamma}, where the strain is given by $\gamma(t) = \int_0^t \dot{\gamma} \mathrm{d}t = \gamma_0 
\sin(\omega t)$, $\gamma_0 = \dot{\gamma}_0 / \omega$) or strain rate (Fig.\ \ref{sfig: laos gamma 
dot}) for the lPTT-100 fluid for $\omega = 2\pi f$, $f$ = 0.5, 2, 8 and 32 \si{Hz} and 
$\dot{\gamma}_0$ being equal to the corresponding $\dot{\gamma}_c$ values of Table \ref{table: 
operating conditions}. These plots are very similar to the force-displacement and force velocity 
diagrams of the damper simulations, such as Figs.\ \ref{fig: force displacement} and \ref{sfig: 
force - ND velocity lPTT-100} (only that they are travelled in opposite senses because the force 
plots record the force on the moving wall, while the LAOS plots record the stress on the stationary 
wall). These LAOS plots reveal fluid properties such as elasticity and shear-thinning in the same 
way that the force diagrams of the damper simulations do. Nevertheless, there are significant 
differences between the two types of flow, as noted in the text. The plots of Fig.\ \ref{fig: laos} 
do not account for inertia, since it has been omitted from the governing equations; to account for 
it, one would have to solve, instead of the system of ordinary differential equations \eqref{eq: 
transient shear PTT 11} -- \eqref{eq: transient shear PTT 12}, a system of one-dimensional partial 
differential equations that includes the momentum equation, where stress and inertia vary in the 
direction perpendicular to the plates. It is interesting to notice in Fig.\ \ref{sfig: laos gamma 
dot} that in the $f$ = 32 \si{Hz} case as the shear rate approaches its maximum magnitude there 
appear a pair of secondary loops where the hysteresis is reversed, which is something that inertia 
can also cause (e.g.\ Figs.\ \ref{sfig: force - ND velocity CY-100}, \ref{sfig: force velocity with 
piston inertia}) but in Fig.\ \ref{sfig: laos gamma dot} it is caused by strong nonlinearities in 
the elastic behaviour of the fluid \cite{Ewoldt_2010}.

\end{appendices}

% \end{linenumbers}

%% References
%%
%% Following citation commands can be used in the body text:
%% Usage of \cite is as follows:
%%   \cite{key}          ==>>  [#]
%%   \cite[chap. 2]{key} ==>>  [#, chap. 2]
%%   \citet{key}         ==>>  Author [#]

%% References with bibTeX database:

% \clearpage
\section*{References}
\bibliographystyle{ieeetr}
\bibliography{viscoelastic_damper}

%% Authors are advised to submit their bibtex database files. They are
%% requested to list a bibtex style file in the manuscript if they do
%% not want to use model1-num-names.bst.

%% References without bibTeX database:

% \begin{thebibliography}{00}

%% \bibitem must have the following form:
%%   \bibitem{key}...
%%

% \bibitem{}

% \end{thebibliography}

% TABLES
% -----------------------------------------------------------------------------

% FIGURES
% -----------------------------------------------------------------------------

\end{document}